\documentclass[12pt]{article}
\usepackage{epsfig,amssymb,amsmath,psfrag,multirow,epstopdf,color}

\numberwithin{equation}{section}

\newcommand{\be}{\begin{equation}}
\newcommand{\ee}{\end{equation}}
\newcommand{\bea}{\begin{eqnarray}}
\newcommand{\eea}{\end{eqnarray}}

\def\Li{{\rm Li}}
\def\cN{{\mathcal N}}
\def\cO{{\mathcal O}}
\def\cL{{\mathcal L}}
\def\ws{{w^\ast}}

\font\cyr=wncyr8
\newcommand{\sha}{{\mbox{\cyr X}}}
\newfont{\scyr}{wncyr10 scaled 550}
\newcommand{\ssha}{\mbox{\bf \scyr X}}

\newcommand{\cS}{{\cal S}}

\def\beq{\begin{equation}}
\def\eeq{\end{equation}}

\def\bsp#1\esp{\begin{split}#1\end{split}}

\def\CC{{C\nolinebreak[4]\hspace{-.05em}\raisebox{.4ex}{\scriptsize{++}}}}

\newcommand{\rat}[2]{{#1\over#2}}

\newcommand{\LZeroMM}{L_0^-}
\newcommand{\LOnePP}{L_1^+}
\newcommand{\LTwoMM}{L_2^-}
\newcommand{\LThreePP}{L_3^+}
\newcommand{\LTwoOneMM}{L_{2,1}^-}
\newcommand{\LThreeOnePP}{L_{3,1}^+}
\newcommand{\LFourMM}{L_4^-}
\newcommand{\LTwoOneOneMM}{L_{2,1,1}^-}
\newcommand{\LFivePP}{L_5^+}
\newcommand{\LThreeOneOnePP}{L_{3,1,1}^+}

\newcommand{\LZeroM}[1]{[L_0^-]^#1}
\newcommand{\LOneP}[1]{[L_1^+]^#1}
\newcommand{\LTwoM}[1]{[L_2^-]^#1}

\newcommand{\LPA}[2]{[L^+_{#1}]^#2}

\newcommand{\LSPA}[1]{L^+_{#1}}
\newcommand{\LSPB}[2]{L^+_{#1,#2}}
\newcommand{\LSPC}[3]{L^+_{#1,#2,#3}}
\newcommand{\LSPD}[4]{L^+_{#1,#2,#3,#4}}
\newcommand{\LSPE}[5]{L^+_{#1,#2,#3,#4,#5}}

\newcommand{\LMA}[2]{[L^-_{#1}]^#2}
\newcommand{\LMB}[3]{[L^-_{#1,#2}]^#3}

\newcommand{\LSMA}[1]{L^-_{#1}}
\newcommand{\LSMB}[2]{L^-_{#1,#2}}
\newcommand{\LSMC}[3]{L^-_{#1,#2,#3}}
\newcommand{\LSMD}[4]{L^-_{#1,#2,#3,#4}}
\newcommand{\LSME}[5]{L^-_{#1,#2,#3,#4,#5}}

\newcommand{\zp}[0]{{\bar{z}}}

\newcommand{\Enun}[1]{{E_{\nu,n}^#1}}
\newcommand{\EnunOne}[0]{{E_{\nu,n}}}
\newcommand{\dE}[1]{{D_\nu^#1 E_{\nu,n}}}
\newcommand{\dEOne}[0]{{D_\nu E_{\nu,n}}}
\newcommand{\dEP}[2]{{[D_\nu^#1 E_{\nu,n}]^#2}}
\newcommand{\dEPOne}[1]{{[D_\nu E_{\nu,n}]^#1}}

\newcommand{\kd}[1]{{\delta_{0,n}/(i\nu)^#1}}
\newcommand{\Ffourtilde}[0]{{\tilde{F}_4}}
\newcommand{\Fsixatilde}[0]{{\tilde{F}_{6a}}}
\newcommand{\Fseventilde}[0]{{\tilde{F}_7}}

\newcommand{\gFL}[2]{{g_#2^{(#1)}(w,\ws)}}
\newcommand{\hFL}[2]{{h_#2^{(#1)}(w,\ws)}}

\DeclareMathOperator{\sgn}{sgn}

\newcommand{\Hb}[0]{{\overline{H}}}

\newcommand{\dnu}[0]{{D_{\nu}}}

\newenvironment{sloppyequation}[0]{\normalsize\sloppy\begin{flushleft}\hspace*{0.75cm}\(\displaystyle}{\)\end{flushleft}\fussy\normalsize}

\newcommand{\beqsloppy}{\begin{sloppyequation}}
\newcommand{\eeqsloppy}{\end{sloppyequation}}

\begin{document}

\catcode`\@=11
\font\manfnt=manfnt
\def\Watchout{\@ifnextchar [{\W@tchout}{\W@tchout[1]}}
\def\W@tchout[#1]{{\manfnt\@tempcnta#1\relax%
  \@whilenum\@tempcnta>\z@\do{%
    \char"7F\hskip 0.3em\advance\@tempcnta\m@ne}}}
\let\foo\W@tchout
\def\dubious{\@ifnextchar[{\@dubious}{\@dubious[1]}}
\let\enddubious\endlist
\def\@dubious[#1]{%
  \setbox\@tempboxa\hbox{\@W@tchout#1}
  \@tempdima\wd\@tempboxa
  \list{}{\leftmargin\@tempdima}\item[\hbox to 0pt{\hss\@W@tchout#1}]}
\def\@W@tchout#1{\W@tchout[#1]}
\catcode`\@=12


\thispagestyle{empty}

\begin{flushright}
SLAC--PUB--15132
\end{flushright}

\begingroup\centering
{\Large\bfseries\mathversion{bold}
Single-valued harmonic polylogarithms \\and the multi-Regge limit \par}%
\vspace{8mm}

\begingroup\scshape\large
Lance~J.~Dixon$^{(1)}$, Claude Duhr$^{(2)}$, Jeffrey Pennington$^{(1)}$\\
\endgroup
\vspace{6mm}
\begingroup\small
$^{(1)}$  \emph{SLAC National Accelerator Laboratory,
Stanford University, \\
Stanford, CA 94309, USA} \\
$^{(2)}$  \emph{Institut f\"{u}r Theoretische Physik, ETH Z\"{u}rich,\\
Wolfgang-Paulistrasse 27, CH-8093, Z\"{u}rich, Switzerland}
\endgroup

\vspace{0.6cm}
\begingroup\small
E-mails:\\
{\tt lance@slac.stanford.edu}, {\tt duhrc@itp.phys.ethz.ch}, 
{\tt jpennin@stanford.edu}\endgroup
\vspace{1.2cm}

\textbf{Abstract}\vspace{5mm}\par
\begin{minipage}{14.7cm}
We argue that the natural functions for describing the multi-Regge
limit of six-gluon scattering in planar $\cN=4$ super Yang-Mills
theory are the single-valued harmonic polylogarithmic functions
introduced by Brown.  These functions depend on a single complex
variable and its conjugate, $(w,\ws)$. Using these functions, and
formulas due to Fadin, Lipatov and Prygarin, we determine the
six-gluon MHV remainder function in the leading-logarithmic
approximation (LLA) in this limit through ten loops, and the
next-to-LLA (NLLA) terms through nine loops.  In separate work, we
have determined the symbol of the four-loop remainder function for
general kinematics, up to 113 constants.  Taking its multi-Regge limit
and matching to our four-loop LLA and NLLA results, we fix all but one
of the constants that survive in this limit.  The multi-Regge limit
factorizes in the variables $(\nu,n)$ which are related to $(w,\ws)$
by a Fourier-Mellin transform.  We can transform the single-valued
harmonic polylogarithms to functions of $(\nu,n)$ that incorporate
harmonic sums, systematically through transcendental weight six.
Combining this information with the four-loop results, we determine
the eigenvalues of the BFKL kernel in the adjoint representation to
NNLLA accuracy, and the MHV product of impact factors to N$^3$LLA
accuracy, up to constants representing beyond-the-symbol terms and the
one symbol-level constant.  Remarkably, only derivatives of the polygamma
function enter these results. Finally, the LLA approximation to the
six-gluon NMHV amplitude is evaluated through ten loops.
\end{minipage}\par
\endgroup

\newpage

\section{Introduction}

Enormous progress has taken place recently in unraveling the properties
of relativistic scattering amplitudes in four-dimensional gauge theories
and gravity.  Perhaps the most intriguing developments have been in
maximally supersymmetric $\cN=4$ Yang-Mills theory, in the planar
limit of a large number of colors.  Many lines of evidence suggest
that it should be possible to solve for the scattering amplitudes in 
this theory to all orders in perturbation theory.  There are also
semi-classical results based on the AdS/CFT duality to match to at
strong coupling~\cite{Alday2007hr}.  The scattering amplitudes 
in the planar theory can be expressed in terms of a set of dual (or region)
variables $x_i^\mu$, which are related to the usual external momentum
four-vectors $k_i^\mu$ by $k_i = x_i - x_{i+1}$.  Remarkably, the 
planar $\cN=4$ super-Yang-Mills amplitudes are governed by a
dual conformal symmetry acting on the
$x_i$~\cite{Alday2007hr,Drummond2006rz,Bern2006ew,Drummond2007aua,%
Brandhuber2007yx,Alday2007he,Drummond2007au}.
This symmetry can be extended to a dual superconformal
symmetry~\cite{Drummond2008vq}, which acts on supermultiplets of
amplitudes that are packaged together by using an $\cN=4$ on-shell
superfield and associated Grassmann
coordinates~\cite{Nair1988bq,ArkaniHamed2008gz,Brandhuber2008pf,%
Elvang2009wd}.

Due to infrared divergences, amplitudes are not invariant under
dual conformal transformations.  Rather, there is an anomaly, which
was first understood in terms of polygonal Wilson loops rather
than amplitudes~\cite{Drummond2007au}.  (For such Wilson loops the
anomaly is ultraviolet in nature.)   A solution to the anomalous
Ward identity for maximally-helicity violating (MHV) amplitudes
is to write them in terms of the BDS ansatz~\cite{Bern2005iz},
\be\label{eq:Rndef}
A_n^{\textrm{MHV}} = A_n^{\textrm{BDS}} \times \exp(R_n),
\ee
where $R_n$ is the so-called
{\it remainder function}~\cite{Bern2008ap,Drummond2008aq},
which is fully dual-conformally invariant.

For the four- and five-gluon scattering amplitudes, the only 
dual-conformally invariant functions are constants, and because
of this fact the BDS ansatz is exact and the remainder function
vanishes to all loop orders, $R_4 = R_5 = 0$.
For six-gluon amplitudes, dual conformal invariance restricts
the functional dependence to have the form $R_6(u_1,u_2,u_3)$,
where the $u_i$ are the unique invariant cross ratios constructed from
distances $x_{ij}^2$ in the dual space:
\bea\label{eq:cross_ratio_def}
u_1 = \frac{x_{13}^2 x_{46}^2}{x_{14}^2 x_{36}^2}
= \frac{ s_{12} s_{45} }{ s_{123} s_{345} }\,, \qquad
u_2 = \frac{x_{24}^2 x_{15}^2}{x_{25}^2 x_{14}^2}
= \frac{ s_{23} s_{56} }{ s_{234} s_{456} }\,, \qquad
u_3 = \frac{x_{35}^2 x_{26}^2}{x_{36}^2 x_{25}^2}
= \frac{ s_{34} s_{61} }{ s_{345} s_{561} }\,.
\eea
The need for a nonzero remainder function $R_n$ for Wilson loops was
first indicated by the strong-coupling behavior of polygonal loops
corresponding to amplitudes with a large number of gluons
$n$~\cite{Alday2007he}.  At the six-point level, investigation of the
multi-Regge limits of $2\to 4$ gluon scattering amplitudes led to the
conclusion that $R_6$ must be nonvanishing at two
loops~\cite{Bartels2008ce}.  Numerical evidence was found soon
thereafter for a nonvanishing two-loop coefficient $R_6^{(2)}$ for
generic nonsingular kinematics~\cite{Bern2008ap}, in agreement with
the numerical values found simultaneously for the corresponding
hexagonal Wilson loop~\cite{Drummond2008aq}.

Based on the Wilson line representation~\cite{Drummond2008aq}, and
using dual conformal invariance to take a quasi-multi-Regge limit and
simplify the integrals, an analytic result for $R_6^{(2)}$ was
derived~\cite{DelDuca2009au,DelDuca2010zg} in terms of
Goncharov's multiple polylogarithms~\cite{Gonchpoly}.  Making use of
properties of the
{\it symbol}~\cite{symbolsC,symbolsB,symbols,Goncharov2010jf,Duhr2011zq}
associated with iterated integrals, the analytic result for $R_6^{(2)}$ was
then simplified to just a few lines of classical
polylogarithms~\cite{Goncharov2010jf}.

A powerful constraint on the structure of the remainder function at 
higher loop order is provided by the operator product expansion (OPE) for
polygonal Wilson loops~\cite{Alday2010ku,Gaiotto2010fk,Gaiotto2011dt}.
At three loops, this constraint, together with symmetries,
collinear vanishing, and an assumption about the final entry of the symbol,
can be used to determine the symbol of $R_6^{(3)}$ up to just two constant
parameters~\cite{Dixon2011pw}.  Another powerful technique for
determining the remainder function is to exploit an infinite-dimensional
Yangian invariance~\cite{Drummond2009fd,Bargheer2009qu} which includes
the dual superconformal generators.  These symmetries are anomalous
at the loop level~(or alternatively one can say that the algebra has to be
deformed)~\cite{Beisert2010gn}.  However, the symmetries imply
a first order linear differential equation for the $\ell$-loop $n$-point
amplitude, and the anomaly dictates the inhomogenous term in the
differential equation, in terms of an integral over an $(\ell-1)$-loop
$(n+1)$-point amplitude~\cite{Bullimore2011kg,CaronHuot2011kk}.
Using this differential equation, a number of interesting results
were obtained in ref.~\cite{CaronHuot2011kk}.  In particular, the result
for the symbol of $R_6^{(3)}$ found in ref.~\cite{Dixon2011pw} was
recovered and the two previously-undetermined constants were fixed.

In principle, the method of refs.~\cite{Bullimore2011kg,CaronHuot2011kk}
works to arbitrary loop order.  However, it requires
knowing lower-loop amplitudes with an increasing number of external legs,
for which the number of kinematical variables (the dual conformal cross
ratios) steadily increases.  Although the symbol of the two-loop
remainder function $R_n^{(2)}$ is known for arbitrary
$n$~\cite{CaronHuot2011ky}, the same is not true of the three-loop
seven-point remainder function, which would feed into the four-loop
six-point remainder function --- one of the subjects of this paper.

In this article, we focus on features of the six-point kinematics
that allow us to push directly to higher loop orders for this
amplitude, without having to solve for amplitudes with more legs.
In fact, most of our paper is concerned with a special limit
of the kinematics in which we can make even more progress:
multi-Regge kinematics (MRK), a limit which has
already received considerable attention in the context
of $\cN=4$ super-Yang-Mills
theory~\cite{Bartels2008ce,Dixon2011pw,Bartels2008sc,Schabinger2009bb,Lipatov2010qg,%
Lipatov2010ad,Bartels2010tx,Fadin2011we,Prygarin2011gd,%
Bartels2011ge,Lipatov2012gk}.
In the MRK limit of $2\to4$ gluon scattering,
the four outgoing gluons are widely-spaced in rapidity.
In other words, two of the four gluons are emitted far forward,
with almost the same energies and directions of the two incoming gluons.
The other two outgoing gluons are also well-separated from each other,
and have smaller energies than the two far-forward gluons.

The MHV amplitude possesses a unique limit of this type.  For definiteness,
we will take legs 3 and 6 to be incoming, legs 1 and 2 to be the far-forward
outgoing gluons, and legs 4 and 5 to be the other two outgoing gluons. 
Neglecting power-suppressed terms, helicity must be conserved along the
high-energy lines.  In the usual all-outgoing convention for labeling
helicities, the helicity configuration can be taken to be
$({+}{+}{-}{+}{+}{-})$.
For generic $2\to4$ scattering in four dimensions there are
eight kinematic variables.  Dual conformal invariance reduces the eight
variables down to just the three dual conformal cross ratios $u_i$.
Taking the multi-Regge limit essentially reduces the amplitude to a
function of just two variables, $w$ and $\ws$,
which turn out to be the complex conjugates of each other.

We will argue that the function space relevant for this limit has been
completely characterized by Brown~\cite{BrownSVHPLs}.  We call the functions
{\it single-valued harmonic polylogarithms} (SVHPLs).  They 
are built from the analytic functions of a single complex
variable that are known as harmonic polylogarithms (HPLs) in the physics
literature~\cite{Remiddi1999ew}.  These functions have branch cuts at $w=0$
and $w=-1$.  However, bilinear combinations of HPLs in $w$
and in $\ws$ can be constructed~\cite{BrownSVHPLs} to cancel the branch cuts,
so that the resulting functions are single-valued in the $(w,\ws)$ plane.
The single-valued property matches perfectly a physical constraint on the
remainder function in the multi-Regge limit.  SVHPLs, like HPLs, are
equipped with an integer transcendental {\it weight}.  The required weight
increases with the loop order.  However, at any given weight there is only
a finite-dimensional vector space of available functions.  Thus, once we
have identified
the proper function space, the problem of solving for the remainder function
in MRK reduces simply to determining a set of rational numbers, namely 
the coefficients multiplying the allowed SVHPLs at a given weight.

In order to further appreciate the simplicity of the multi-Regge limit, we
recall that for generic six-point kinematics there are nine
possible choices for the entries in the
symbol for the remainder function
$R_6(u_1,u_2,u_3)$~\cite{Goncharov2010jf,Dixon2011pw}:
\be\label{nineentries}
\{ u_1, u_2, u_3, 1-u_1, 1-u_2, 1-u_3, y_1, y_2, y_3 \} \,,
\ee
where
\bea\label{eq:y_z_def}
y_i &=& \frac{u_i-z_+}{u_i-z_-} \,, \\
z_\pm &=& \frac{-1+u_1+u_2+u_3 \pm \Delta}{2} \,, \\
\Delta &=& (1-u_1-u_2-u_3)^2 - 4u_1u_2u_3 \,.
\eea
The first entry of the symbol
is actually restricted to the set $\{ u_1, u_2, u_3 \}$
due to the location of the amplitude's branch
cuts~\cite{Gaiotto2011dt}; the integrability of the symbol
restricts the second entry to the set
$\{u_i,1-u_i\}$~\cite{Gaiotto2011dt,Dixon2011pw};
and a ``final-entry condition''~\cite{Dixon2011pw,CaronHuot2011ky}
implies that there are only six, not nine, possibilities for
the last entry.  However, the remaining entries are unrestricted.
The large number of possible entries, and the fact that the $y_i$ variables
are defined in terms of square-root functions of the cross ratios
(although the $u_i$ can be written as rational functions of the
$y_i$~\cite{Dixon2011pw}), complicates the task of identifying the 
proper function space for this problem.

So in this paper we will solve a simpler problem.
The MRK limit consists of taking one of the $u_i$, say $u_1$, to unity,
and letting the other two cross ratios vanish at the same rate that
$u_1\to1$:  $u_2 \approx x(1-u_1)$ and $u_3 \approx y(1-u_1)$ for
two fixed variables $x$ and $y$.  To reach the Minkowski version
of the MRK limit, which is relevant for $2\to4$ scattering, it is necessary
to analytically continue $u_1$ from the Euclidean region according to 
$u_1 \to e^{-2\pi i} |u_1|$, before taking this limit~\cite{Bartels2008ce}.
Although the square-root variables $y_2$ and $y_3$ remain nontrivial in
the MRK limit, all of the square roots can be rationalized by a clever
choice of variables~\cite{Lipatov2010ad}. We define $w$ and $\ws$ by
\be\label{eq:xyw}
x\equiv {1\over (1+w)(1+\ws)}, \qquad y\equiv {w\,\ws\over (1+w)(1+\ws)}\,.
\ee
Then the MRK limit of the other variables is
\be\label{eq:y12lim}
u_1 \to 1, \qquad y_1 \to 1, \qquad y_2 \to \tilde{y}_2 = \frac{1+\ws}{1+w} \,,
\qquad y_3 \to \tilde{y}_3 = \frac{(1+w)\ws}{w(1+\ws)} \,.
\ee
Neglecting terms that vanish like powers of $(1-u_1)$,
we expand the remainder function in the multi-Regge limit in
terms of coefficients multiplying powers of the large logarithm $\log(1-u_1)$
at each loop order, following the conventions of ref.~\cite{Dixon2011pw}, 
\beq\label{eq:R6_MRK_intro}
R_6(u_1,u_2,u_3)|_{\textrm{MRK}}
\,=\, 2\pi i\,\sum_{\ell=2}^\infty\sum_{n=0}^{\ell-1}a^\ell\,\log^n(1-u_1)\,
\left[g_n^{(\ell)}(w,\ws) + 2\pi i\,h_n^{(\ell)}(w,\ws)\right]\,,
\eeq
where the coupling constant for planar $\cN=4$ super-Yang-Mills theory
is $a = g^2 N_c/(8\pi^2)$.

The remainder function $R_6$ is a transcendental function with weight $2\ell$
at loop order $\ell$.  Therefore the coefficient functions $g_n^{(\ell)}$
and $h_n^{(\ell)}$ have weight $2\ell-n-1$ and $2\ell-n-2$ respectively.
As a consequence of eqs.~(\ref{eq:xyw}) and (\ref{eq:y12lim}), their
symbols have only four possible entries,
\be\label{fourentries}
\{ w, 1+w, \ws, 1+\ws \} \,.
\ee
Furthermore, $w$ and $\ws$ are independent complex variables.  Hence
the problem of determining the coefficient functions
factorizes into that of determining functions of $w$ whose
symbol entries are drawn from $\{ w, 1+w \}$ --- a special class of HPLs
--- and the complex conjugate functions of $\ws$.

On the other hand, not every combination of HPLs in $w$ and HPLs in $\ws$
will appear.  When the symbol is expressed in terms of the original variables
$\{ x, y, \tilde{y}_2, \tilde{y}_3 \}$, the first entry must be either
$x$ or $y$, reflecting the branch-cut behavior and first-entry condition
for general kinematics.  Also, the full function must be a single-valued
function of $x$ and $y$, or equivalently a single-valued function of $w$
and $\ws$.  These conditions imply that the coefficient functions belong
to the class of SVHPLs defined by Brown~\cite{BrownSVHPLs}.

The MRK limit~(\ref{eq:R6_MRK_intro}) is organized hierarchically into
the leading-logarithmic approximation (LLA) with $n=\ell-1$, 
the next-to-leading-logarithmic approximation (NLLA) with $n=\ell-2$,
and in general the N$^k$LL terms with $n=\ell-k-1$.  Just as the problem
of DGLAP evolution in $x$ space is diagonalized by transforming to
the space of Mellin moments $N$, the MRK limit can be diagonalized
by performing a Fourier-Mellin transform from $(w,\ws)$ to a new
space labeled by $(\nu,n)$.  In fact, Fadin, Lipatov and 
Prygarin~\cite{Lipatov2010ad, Fadin2011we} have given an all-loop-order
formula for $R_6$ in the multi-Regge limit, in terms
of two functions of $(\nu,n)$:  The eigenvalue $\omega(\nu,n)$
of the BFKL kernel in the adjoint representation, and the (regularized)
MHV impact factor $\Phi_{\textrm{Reg}}(\nu,n)$.  Each function can be
expanded in $a$, and each successive order in $a$ corresponds to
increasing $k$ by one in the N$^k$LLA.  
It is possible that the assumption that was made in 
refs.~\cite{Lipatov2010ad,Fadin2011we},
of single Reggeon exchange through NLL, breaks down beyond that order,
due to Reggeon-number changing interactions or other possible
effects~\cite{GregoryAgustinPrivate}.
In this paper we will assume that it holds through N$^3$LL (for the impact
factor); the three quantities we extract beyond NLL could be affected if
this assumption is wrong.

The leading term in
the impact factor is just one, while the leading BFKL eigenvalue $E_{\nu,n}$
was found in ref.~\cite{Bartels2008sc}.  The NLL term in the
impact factor was found in ref.~\cite{Lipatov2010ad},
and the NLL contribution to the BFKL eigenvalue in ref.~\cite{Fadin2011we}.

With this information it is possible to compute
the LLA functions $g_{\ell-1}^{(\ell)}$, NLLA functions 
$g_{\ell-2}^{(\ell)}$ and $h_{\ell-2}^{(\ell)}$, and even the real part
at NNLLA, $h_{\ell-3}^{(\ell)}$.  All one needs to do is perform the
inverse Fourier-Mellin transform back to the $(w,\ws)$ variables.
At the three-loop level, this was carried out at LLA
for $g_2^{(3)}$ and $h_1^{(3)}$ in ref.~\cite{Lipatov2010ad},
and at NLLA for $g_1^{(3)}$ and $h_0^{(3)}$ in ref.~\cite{Fadin2011we}.
Here we will use the SVHPL basis to make this step very simple.
The inverse transform contains an explicit sum over $n$, and an integral
over $\nu$ which can be evaluated via residues in terms of a sum over
a second integer $m$.  For low loop orders we can perform the double sum
analytically using harmonic
sums~\cite{Euler_sum,Zagier_sum,Vermaseren1998uu,Blumlein1998if,%
Blumlein2009ta,Blumlein2009fz}.
For high loop orders, it is more efficient
to simply truncate the double sum.  In the $(w,\ws)$ plane this truncation
corresponds to truncating the power series expansion in $|w|$ around the
origin.  We know the answer is a linear combination of a finite number
of SVHPLs with rational-number coefficients.  In order to determine the
coefficients, we simply compute the power series expansion of the 
generic linear combination of SVHPLs and match it against the truncated
double sum over $m$ and $n$.  We can now perform the inverse Fourier-Mellin
transform, in principle to all orders, and in practice through
weight 10, corresponding to 10 loops for LLA and 9 loops for NLLA.

Furthermore, we can bring in additional information at fixed loop order,
in order to obtain more terms in the expansion of the BFKL eigenvalue
and the MHV impact factor.  In ref.~\cite{Fadin2011we}, the NLLA
results for $g_1^{(3)}$ and $h_0^{(3)}$ confirmed a previous
prediction~\cite{Dixon2011pw} based on an analysis of the multi-Regge
limit of the symbol for $R_6^{(3)}$.  In this limit, the two free symbol 
parameters mentioned above dropped out.  The symbol could be
integrated back up into a function, but a few more ``beyond-the-symbol''
constants entered at this stage.  One of the constants was fixed in
ref.~\cite{Fadin2011we} using the NLLA information.
As noted in ref.~\cite{Fadin2011we}, the result from
ref.~\cite{Dixon2011pw} for $g_0^{(3)}$ can be used to determine
the NNLLA term in the impact factor.  In this paper, we will use
our knowledge of the space of functions of $(w,\ws)$ (the SVHPLs) to build
up a dictionary of the functions of $(\nu,n)$ (special types
of harmonic sums) that are the Fourier-Mellin transforms of the SVHPLs.
From this dictionary and $g_0^{(3)}$ we will determine the NNLLA term
in the impact factor.

We can go further if we know the four-loop remainder function $R_6^{(4)}$.
In separate work~\cite{fourloop}, we have heavily constrained the
symbol of $R_6^{(4)}(u_1,u_2,u_3)$ for generic kinematics, using exactly
the same constraints used in ref.~\cite{Dixon2011pw}:
integrability of the symbol, branch-cut behavior, symmetries, the
final-entry condition, vanishing of collinear limits, and
the OPE constraints (which at four loops are a constraint on the
triple discontinuity).  Although there are millions of possible
terms before applying these constraints, afterwards the symbol contains
just 113 free constants (112 if we apply the overall normalization for the
OPE constraints).  Next we construct the multi-Regge limit of this symbol,
and apply all the information we have about this limit:
\begin{itemize}
\item Vanishing of the super-LLA terms $g_n^{(4)}$ and $h_n^{(4)}$
for $n=4,5,6,7$; 
\item LLA and NLLA predictions for $g_n^{(4)}$ and $h_n^{(4)}$ for $n=2,3$;
\item the NNLLA real part $h_1^{(4)}$, which is also predicted by the
NLLA formula;
\item a consistency condition between $g_1^{(4)}$ and $h_0^{(4)}$.
\end{itemize}
Remarkably, these conditions determine all but one of the symbol-level
parameters in the MRK limit.  (The one remaining free parameter seems
highly likely to vanish, given the complicated way it enters
various formulae, but we have not yet proven that to be the case.)

We then extract the remaining four-loop coefficient functions,
$g_1^{(4)}$, $h_0^{(4)}$ and $g_0^{(4)}$, introducing some additional
beyond-the-symbol parameters at this stage.  We use this information
to determine the NNLLA BFKL eigenvalue and the N$^3$LLA MHV impact
factor, up to these parameters.  Although our general dictionary of
functions of $(\nu,n)$ contains various multiple harmonic sums, we
find that the key functions entering the multi-Regge limit can all be
expressed just in terms of certain rational combinations of $\nu$
and $n$, together with the polygamma functions $\psi$, $\psi^\prime$,
$\psi^{\prime\prime}$, etc. (derivatives of the logarithm of the $\Gamma$
function) with arguments $1\pm i\nu+|n|/2$.

As a byproduct, we find that the SVHPLs also describe the multi-Regge
limit of the one remaining helicity configuration for six-gluon scattering
in $\cN=4$ super-Yang-Mills theory, namely the next-to-MHV (NMHV)
configuration with three negative and three positive gluon helicities.
It was shown recently~\cite{Lipatov2012gk}
that in LLA the NMHV and MHV remainder functions
are related by a simple integro-differential operator.  This operator
has a natural action in terms of the SVHPLs, allowing us to easily
extend the NMHV LLA results of ref.~\cite{Lipatov2012gk} from three loops
to 10 loops.

This article is organized as follows. In Section~\ref{sec:MRK} we review
the structure of the six-point MHV remainder function in the multi-Regge
limit.  Section~\ref{sec:SVHPLs} reviews Brown's construction of
single-valued harmonic polylogarithms.  In Section~\ref{sec:R6_LLA_NLLA}
we exploit the SVHPL basis to determine the functions
$g_n^{(\ell)}$ and $h_n^{(\ell)}$ at LLA through 10 loops and at NLLA
through 9 loops. Section~\ref{sec:NMHV} determines the NMHV remainder
function at LLA through 10 loops.  In Section~\ref{sec:nu_n}
we describe our construction of the functions of $(\nu,n)$ that
are the Fourier-Mellin transforms of the SVHPLs.
Section~\ref{sec:applications} applies this knowledge, plus
information from the four-loop remainder function~\cite{fourloop},
in order to determine the NNLLA MHV impact factor and BFKL eigenvalue,
and the N$^3$LLA MHV impact factor, in terms of a handful of (mostly)
beyond-the-symbol constants.  In Section~\ref{sec:Concl} we report
our conclusions and discuss directions for future research.

We include two appendices.  Appendix~\ref{app:svhpl}
collects expressions for the SVHPLs (after diagonalizing the
action of a $\mathbb{Z}_2\times\mathbb{Z}_2$ symmetry), in terms
of HPLs through weight 5.  It also gives expressions before diagonalizing
one of the two $\mathbb{Z}_2$ factors.  Appendix~\ref{app:blumlein}
gives a basis for the function space in $(\nu,n)$ through weight 5,
together with the Fourier-Mellin map to the SVHPLs.
In addition, for the lengthier formulae, we provide separate
computer-readable text files as ancillary material.  In particular, we include
files (in {\tt Mathematica} format) that contain the expressions for the 
SVHPLs  in terms of ordinary
HPLs up to weight six, decomposed into an eigenbasis of the
$\mathbb{Z}_2\times\mathbb{Z}_2$ symmetry, as well as the analytic results
 up to weight ten for the imaginary parts of the MHV 
remainder function at LLA and NLLA
and for the NMHV remainder function at LLA. Furthermore, we include 
the expressions for the NNLL BFKL eigenvalue and impact factor and the N$^3$LL 
impact factor in terms of the building blocks in the variables $(\nu,n)$ constructed in Section~\ref{sec:nu_n}, as well as a dictionary between these building blocks and the SVHPLs up to weight five.


\section{The six-point remainder function in the multi-Regge limit}
\label{sec:MRK}

The principal aim of this paper is to study the six-point MHV amplitude in
$\cN=4$ super Yang-Mills theory in multi-Regge kinematics.  This limit is
defined by the hierarchy of scales,
\beq
s_{12} \gg s_{345},\,s_{456} \gg s_{34},\, s_{45}\,,s_{56} 
      \gg s_{23},\, s_{61},\,s_{234}\,.
\eeq
In this limit the cross ratios~\eqref{eq:cross_ratio_def} behave as
\beq
1-u_1,\, u_2,\, u_3 \sim 0\,,
\eeq
together with the constraint that the following ratios are held fixed,
\beq\label{eq:xy_def}
x\equiv {u_2\over1-u_1} = \cO(1) {\rm~~and~~} 
y\equiv {u_3\over1-u_1} = \cO(1)\,.
\eeq
In the following it will be convenient~\cite{Lipatov2010ad}
to parametrize the dependence on $x$ and $y$ by a single complex variable $w$,
\beq\label{eq:xy_to_wws}
x\equiv {1\over (1+w)(1+\ws)} {\rm~~and~~} y\equiv {w\,\ws\over (1+w)(1+\ws)}\,.
\eeq
Any function of the three cross ratios can then develop large logarithms $\log(1-u_1)$ in the multi-Regge limit, and we can write generically,
\beq\label{eq:MRK_expansion}
F(u_1,u_2,u_3)  = \sum_i\log^i(1-u_1)\,f_i(w,\ws) + \cO(1-u_1)\,.
\eeq
Let us make at this point an important observation which will be a recurrent theme in the rest of the paper: If $F(u_1,u_2,u_3)$ represents a physical quantity like a scattering amplitude, then $F$ should only have cuts in physical channels, corresponding to branch cuts starting at points where one of the cross ratios vanishes.  Rotation around the origin in the complex $w$ plane,
i.e.~$(w,\ws)\to(e^{2\pi i}w,e^{-2\pi i}\ws)$, does not correspond to crossing any branch cut.  As a consequence, the functions $f_i(w,\ws)$ should not change under this operation.  More generally, the functions $f_i(w,\ws)$ must be \emph{single-valued} in the complex $w$ plane.

Let us start by reviewing the multi-Regge limit of the MHV remainder function $R(u_1,u_2,u_3)\equiv R_6(u_1,u_2,u_3)$ introduced in eq.~\eqref{eq:Rndef}. It can be shown that, while in the Euclidean region the remainder function vanishes in the multi-Regge limit, there is a Mandelstam cut such that we obtain a non-zero contribution in MRK after performing the analytic continuation~\cite{Bartels2008ce}
\beq\label{eq:MRK_anal_cont}
u_1 \to e^{-2\pi i}\,|u_1|\,.
\eeq
After this analytic continuation, the six-point remainder function can
be expanded into the form given in eq.~\eqref{eq:R6_MRK_intro}, which we
repeat here for convenience,
\beq\label{eq:R6_MRK}
R|_{\textrm{MRK}} = 2\pi i\,\sum_{\ell=2}^\infty\sum_{n=0}^{\ell-1}a^\ell\,\log^n(1-u_1)\,\left[g_n^{(\ell)}(w,\ws) + 2\pi i\,h_n^{(\ell)}(w,\ws)\right]\,.
\eeq
The functions $g_n^{(\ell)}(w,\ws)$ and $h_n^{(\ell)}(w,\ws)$ will in the following be referred to as the \emph{coefficient functions} for the logarithmic expansion in the MRK limit.  The imaginary part $g_n^{(\ell)}$ is associated
with a single discontinuity, and the real part $h_n^{(\ell)}$ with a double
discontinuity, although both functions also include information from 
higher discontinuities, albeit with accompanying explicit factors of $\pi^2$.

The coefficient functions are single-valued pure transcendental functions
in the complex variable $w$, of weight $2\ell-n-1$ for $g_n^{(\ell)}$
and weight $2\ell-n-2$ for $h_n^{(\ell)}$.  They are left invariant by a 
$\mathbb{Z}_2\times\mathbb{Z}_2$ symmetry acting via complex conjugation 
and inversion,
\beq\label{eq:Z2xZ2w}
w\leftrightarrow \ws {\rm~~and~~} (w,\ws) \leftrightarrow (1/w,1/\ws)\,.
\eeq
The complex conjugation symmetry arises because the MHV remainder function
has a parity symmetry, or invariance under $\Delta \to -\Delta$, which
inverts $\tilde{y}_2$ and $\tilde{y}_3$ in eq.~\eqref{eq:y12lim}.
The inversion symmetry is a consequence of the fact that the six-point
remainder function is a totally symmetric function of the three cross
ratios $u_1$, $u_2$ and $u_3$.  In particular, exchanging
$\tilde{y}_2 \leftrightarrow \tilde{y}_3$ is the product of conjugation
and inversion.  The inversion symmetry is sometimes referred to as 
target-projectile symmetry~\cite{Lipatov2010qg}.  Finally, the vanishing
of the six-point remainder function in the collinear limit implies the
vanishing of $g_n^{(\ell)}(w,\ws)$ and $h_n^{(\ell)}(w,\ws)$  in the limit where 
$(w,\ws) \to0$.  Clearly the functions $g_n^{(\ell)}$ and
$h_n^{(\ell)}$ are already highly constrained on general grounds.

In ref.~\cite{Lipatov2010ad, Fadin2011we} an all-loop integral formula for 
the six-point amplitude in MRK was presented\footnote{There is a difference
in conventions regarding the definition of the remainder function. What we
call $R$ is called $\log(R)$ in refs.~\cite{Lipatov2010ad, Fadin2011we}.
Apart from the zeroth order term, the first place this makes a difference
is at four loops, in the real part.},
\beq\label{eq:MHV_MRK}
e^{R+i\pi\delta}|_{\textrm{MRK}} = \cos\pi\omega_{ab} 
+ i \, {a\over 2} \sum_{n=-\infty}^\infty
(-1)^n\,\left({w\over \ws}\right)^{{n\over 2}}\int_{-\infty}^{+\infty}
{d\nu\over \nu^2+{n^2\over 4}}\,|w|^{2i\nu}\,\Phi_{\textrm{Reg}}(\nu,n)
\,\left(-{1\over \sqrt{u_2\,u_3}}\right)^{\omega(\nu,n)} .
\eeq
The first term is the Regge pole contribution, with
\beq
\omega_{ab} = {1\over 8}\,\gamma_K(a)\,\log{u_3\over u_2}
 = {1\over 8}\,\gamma_K(a)\,\log|w|^2\,,
\eeq
and $\gamma_K(a)$ is the cusp anomalous dimension, known to all orders in
perturbation theory~\cite{BES},
\be\label{eq:gamma_cusp}
\gamma_K(a) \, = \, \sum_{\ell=1}^\infty \gamma_K^{(\ell)} a^\ell
           \,  = \, 4\,a - 4\,\zeta_2 \, a^2 + 22 \, \zeta_4 \, a^3 
             - ( \textstyle{\frac{219}{2}} \, \zeta_6 + 4 \, \zeta_3^2 ) \, a^4
             + \cdots \,. 
\ee
The second term in eq.~(\ref{eq:MHV_MRK}) arises from a Regge cut
and is fully determined to all orders by the BFKL eigenvalue
$\omega(\nu,n)$ and the (regularized) impact factor
$\Phi_{\textrm{Reg}}(\nu,n)$.  The function $\delta$ appearing in the exponent
on the left-hand side is the contribution from a Mandelstam cut present in
the BDS ansatz, and is given to all loop orders by
\beq
\delta = {1\over 8}\,\gamma_K(a)\,\log{(xy)}
 = {1\over 8}\,\gamma_K(a)\,\log{|w|^2\over |1+w|^4}\,.
\eeq
In addition, we have
\beq\label{eq:removeu2u3}
{1\over \sqrt{u_2\,u_3}} = {1\over 1-u_1}\,{|1+w|^2\over |w|}\,.
\eeq
The BFKL eigenvalue and the impact factor can be expanded perturbatively,
\beq\bsp
\omega(\nu,n) &\,= 
- a \left(E_{\nu,n} + a\,E_{\nu,n}^{(1)}+ a^2\,E_{\nu,n}^{(2)}+\cO(a^3)\right)\,,\\
\Phi_{\textrm{Reg}}(\nu,n)&\, = 1 + a \, \Phi_{\textrm{Reg}}^{(1)}(\nu,n)
 + a^2 \, \Phi_{\textrm{Reg}}^{(2)}(\nu,n)
 + a^3 \, \Phi_{\textrm{Reg}}^{(3)}(\nu,n)+\cO(a^4)\,.
\esp\eeq
The BFKL eigenvalue is known to the first two orders in perturbation
theory~\cite{Fadin2011we,Bartels2008sc},
\bea\label{eq:E_0}
E_{\nu,n} &=& -{1\over2}\,{|n|\over \nu^2+{n^2\over 4}}
+\psi\left(1+i\nu+{|n|\over2}\right) +\psi\left(1-i\nu+{|n|\over2}\right) 
- 2\psi(1)\,,\\
E_{\nu,n}^{(1)} &=&-{1\over 4}\biggl[\psi''\left(1+i\nu+{|n|\over2}\right)
   +\psi''\left(1-i\nu+{|n|\over2}\right) \label{eq:E_1}\\
&&\hskip-0.7cm\null
- {2i\nu\over \nu^2+{n^2\over 4}}
\left(\psi'\left(1+i\nu+{|n|\over2}\right)
 -\psi'\left(1-i\nu+{|n|\over2}\right)\right)\biggr] - \zeta_2\,E_{\nu,n}
 - 3\zeta_3 - {1\over 4}\,{|n|\,\left(\nu^2-{n^2\over 4}\right)
          \over\left(\nu^2+{n^2\over 4}\right)^3}\,, \nonumber
\eea
where $\psi(z) = {d\over dz}\log\Gamma(z)$ is the digamma function, 
and $\psi(1)=-\gamma_E$ is the Euler-Mascheroni constant. The NLL
contribution to the impact factor is given by~\cite{Lipatov2010qg}
\beq\label{eq:Phi_1}
\Phi_{\textrm{Reg}}^{(1)}(\nu,n) 
= -\frac{1}{2}E_{\nu,n}^2 - {3\over 8}\,{n^2\over (\nu^2+{n^2\over 4})^2}
 - \zeta_2\,.
\eeq
The BFKL eigenvalues and impact factor in eqs.~\eqref{eq:E_0},
\eqref{eq:E_1} and \eqref{eq:Phi_1} are enough to compute the six-point
remainder function in the Regge limit in the leading and next-to-leading
logarithmic approximations (LLA and NLLA). Indeed, we can interpret
the integral in eq.~\eqref{eq:MHV_MRK} as a contour integral in the
complex $\nu$ plane and close the contour at infinity. By summing up
the residues we then obtain the analytic expression of the remainder
function in the LLA and NLLA in MRK.  This procedure will be discussed
in greater detail in Section~\ref{sec:R6_LLA_NLLA}.  Some comments are in
order about the integral in eq.~\eqref{eq:MHV_MRK}:
\begin{enumerate}
\item The contribution coming from $n=0$ seems ill-defined, as the integral
in eq.~\eqref{eq:MHV_MRK} diverges. After closing the contour at infinity, 
our prescription is to take only half of the residue at $\nu=n=0$ into
account.
\item We need to specify the Riemann sheet of the exponential factor 
in the right-hand side of eq.~\eqref{eq:MHV_MRK}. We find that the 
replacement
\beq\label{eq:ipi}
\left(-{1\over \sqrt{u_2\,u_3}}\right)^{\omega(\nu,n)}
\to e^{-i\pi\omega(\nu,n)}\,\left({1\over \sqrt{u_2\,u_3}}\right)^{\omega(\nu,n)}
\eeq
gives the correct result.
\end{enumerate}
The $i\pi$ factor in the right-hand side of eq.~\eqref{eq:ipi} generates the real parts $h_n^{(\ell)}$ in eq.~\eqref{eq:R6_MRK}.
It is easy to see that the $g_n^{(\ell)}$ and $h_n^{(\ell)}$ functions are not
independent, but they are related.  For example, at LLA and NLLA we have,
\beq\bsp\label{eq:htog}
h_{\ell-1}^{(\ell)}(w,\ws) &\,= 0\,, \\
h_{\ell-2}^{(\ell)}(w,\ws) &\,= 
{\ell-1\over2} \, g_{\ell-1}^{(\ell)}(w,\ws)
+ {1\over 16} \, \gamma_K^{(1)} \, g_{\ell-2}^{(\ell-1)}(w,\ws)
\,\log{|1+w|^4 \over|w|^2} \\
 &\quad - \frac{1}{2} \sum_{k=2}^{\ell-2} g_{k-1}^{(k)} g_{\ell-k-1}^{(\ell-k)} \,, 
\qquad \ell > 2,
\esp\eeq
where $\gamma_K^{(1)} = 4$ from eq.~(\ref{eq:gamma_cusp}).
(Note that the sum over $k$ in the formula for $h_{\ell-2}^{(\ell)}$ would not
have been present if we had used the convention for $R$ in 
refs.~\cite{Lipatov2010ad, Fadin2011we}.)
Similar relations can be derived beyond NLLA, i.e.~for $n<\ell-2$.

So far we have only considered $2\to4$ scattering. In
ref.~\cite{Bartels2010tx} it was shown that if the remainder function
is analytically continued to the region corresponding to $3\to3$
scattering, then it takes a particularly simple form. The analytic
continuation from $2\to4$ to $3\to3$ scattering can be obtained easily
by performing the replacement
\beq
\log(1-u_1) \to \log(u_1-1) -i\pi
\eeq
in eq.~\eqref{eq:MHV_MRK}.  After analytic continuation the real part of 
the remainder function only gets contributions from the Regge pole and
is given by~\cite{Bartels2010tx}
\beq\label{eq:3to3}
\textrm{Re}\left(e^{R_{3\to3}-i\pi\delta}\right) = \cos\pi\omega_{ab}\,.
\eeq
It is manifest from eq.~\eqref{eq:MHV_MRK} that eq.~\eqref{eq:3to3} is
automatically satisfied if the relations among the coefficient functions
derivable by tracking the $i\pi$ from eq.~\eqref{eq:ipi}
(e.g.~eq.~\eqref{eq:htog}) are satisfied in $2\to4$ kinematics.

So far we have only reviewed some general properties of the six-point
remainder function in MRK, but we have not yet given explicit analytic
expressions for the coefficient functions. The two-loop contributions
to eq.~\eqref{eq:MHV_MRK} in LLA and NLLA were computed in
refs.~\cite{Lipatov2010qg,Lipatov2010ad}, while the three-loop
contributions up to the NNLLA were found in
refs.~\cite{Dixon2011pw,Lipatov2010qg}. In all cases the results have been
expressed as combinations of classical polylogarithms in the complex
variable $w$ and its complex conjugate $\ws$, with potential
branching points at $w=0$ and $w=-1$. As discussed at the beginning of
this section, all the branch cuts in the complex $w$ plane must
cancel, i.e., the function must be single-valued in $w$. The class of
functions satisfying these constraints has been studied in full generality
in the mathematical literature, as will be reviewed in the next section.


\section{Harmonic polylogarithms and their single-valued analogues}
\label{sec:SVHPLs}
\subsection{Review of harmonic polylogarithms}
In this section we give a short review of the classical and harmonic polylogarithms, one of the main themes in the rest of this paper. 
The simplest possible polylogarithmic functions are the so-called \emph{classical} polylogarithms, defined inside the unit circle by a convergent power series,
\begin{equation}
\label{eq:lindef}
{\rm Li}_{m} (z) = \sum_{k=1}^{\infty} \frac{z^k}{k^m}\,, \quad\quad |z|<1\,.
\end{equation}
They can be continued to the cut plane $\mathbb{C}\backslash[1,\infty)$ by an iterated integral representation,
\begin{equation}
\label{eq:Li_int}
{\rm Li}_{m} (z) = \int_0^z dz'\; \frac{{\rm Li}_{m-1}(z')}{z'}\,.
\end{equation}
For $m=1$, the polylogarithm reduces to the ordinary logarithm, ${\rm Li}_1(z) = -\log(1-z)$, a fact that dictates the location of the branch cut for all $m$ (along the real axis for $z>1$). It also determines the discontinuity across the cut,
\begin{equation}\label{eq:lin_disc}
\Delta {\rm Li}_m(z) = 2\pi i\, \frac{\log^{m-1} z}{(m-1)!}\,.
\end{equation}

It is possible to define more general classes of polylogarithmic functions by allowing for different kernels inside the iterated integral in eq.~\eqref{eq:Li_int}. The \emph{harmonic} polylogarithms (HPLs)~\cite{Remiddi1999ew} are a special class of generalized polylogarithms whose properties and construction we review in the remainder of this section. To begin, let $w$ be a word formed from the letters $x_0$ and $x_1$, and let $e$ be the empty word. Then, for each $w$, define a function $H_w(z)$ which obeys the differential equations,
\begin{equation}\label{eq:HPLdef_1}
\frac{\partial}{\partial z}H_{x_0w}(z) = \frac{H_{w}(z)}{z} \quad\quad\text{and}\quad\quad\frac{\partial}{\partial z}H_{x_1w}(z) = \frac{H_{w}(z)}{1-z}\,,
\end{equation}
subject to the following conditions,
\begin{equation}
H_{e}(z)=1,\quad\quad H_{x_0^n}(z)=\frac{1}{n!}\log^nz,\quad\quad\text{and}\quad\quad \lim_{z\rightarrow 0}H_{w\neq x_0^n}(z)=0\,.
\end{equation}
There is a unique family of solutions to these equations, and it defines the HPLs. Note that we use the term ``HPL'' in a restricted sense\footnote{%
In the mathematical literature, these functions are sometimes referred to
as \emph{multiple polylogarithms in one variable}.}
-- we only consider poles in the differential equations~\eqref{eq:HPLdef_1} at $z=0$ and $z=1$.  (In our MRK application, we will let
$z=-w$, so that the poles are at $w=0$ and $w=-1$.)

The \emph{weight} of an HPL is the length of the word $w$, and its \emph{depth} is the number of $x_1$'s\footnote{For ease of notation, we will often impose the replacement $\{x_0 \rightarrow 0, x_1 \rightarrow 1\}$ in subscripts. In some cases, we will use the collapsed notation where a subscript $m$ denotes $m-1$ zeroes followed by a single $1$. For example, if $w=x_0x_0x_1x_0x_1$,
\begin{equation}
H_w(z)=H_{x_0x_0x_1x_0x_1}(z) = H_{0,0,1,0,1}(z) = H_{3,2}(z)\,.
\end{equation}
In the collapsed notation, the \emph{weight} is the sum of the indices, and the \emph{depth} is the number of nonzero indices.}. HPLs of depth one are simply the classical polylogarithms, $H_{n}(z) ={ \rm Li}_n(z)$.
Like the classical polylogarithms, the HPLs can be written as iterated integrals,
\begin{equation}\label{eq:HPL_integral}
H_{x_0 w}(z) = \int_0^z dz'\; \frac{H_w(z')}{z'} \quad\quad \text{and}\quad\quad H_{x_1w}= \int_0^z dz'\; \frac{H_w(z')}{1-z'}\,.
\end{equation}
The structure of the underlying iterated integrals endows the HPLs with an important property: they form a \emph{shuffle algebra}. The shuffle relations can be written,
\begin{equation}
\label{eq:HPL_shuffle}
H_{w_1}(z)\,H_{w_2}(z) = \sum_{{w}\in{w_1}\ssha {w_2}}H_{w}(z)\,,
\end{equation}
where ${w_1}\sha{w_2}$ is the set of mergers of the sequences $w_1$ and $w_2$ that preserve their relative ordering. Equation~(\ref{eq:HPL_shuffle}) may be used to express all HPLs of a given weight in terms of a relatively small set of basis functions and products of lower-weight HPLs.
 One convenient such basis~\cite{Blumlein2003gb} of irreducible functions is the \emph{Lyndon} basis, defined by $\{H_w(z): w\in {\rm Lyndon}(x_0,x_1)\}$. The Lyndon words Lyndon($x_0,x_1$) are those words $w$ such that for every decomposition into two words $w=uv$, the left word is lexicographically smaller than the right, $u<v$. Table \ref{tab:Lyndon_table} gives the first few examples of Lyndon words.
\begin{table}[!t]
\centering
\begin{tabular}[t]{c|c|c}
\hline\hline
\textrm{Weight} & \textrm{Lyndon words} & \textrm{Dimension} \\
\hline
1 & 0, 1 & 2\\
2 & 01 & 1 \\
3& 001, 011 & 2 \\
4&0001, 0011, 0111 & 3\\
5&00001, 00011, 00101, 00111, 01011, 01111&6\\
\hline\hline
\end{tabular}
\caption{ All Lyndon words ${\rm Lyndon}(x_0,x_1)$ through weight five}
\label{tab:Lyndon_table}
\end{table}

All HPLs are real whenever the argument $z$ is less than 1, and so, in
particular, the HPLs are analytic in a neighborhood of $z=0$. The Taylor
expansion around $z=0$ is particularly simple and involves only a
special class of harmonic numbers~\cite{Remiddi1999ew,Vermaseren1998uu}
(hence the name \emph{harmonic} polylogarithm),
\beq\label{eq:HPL_series}
H_{m_1,\ldots,m_k}(z) = \sum_{l=1}^\infty{z^l\over l^{m_1}}Z_{m_2,\ldots,m_k}(l-1)\,,
\qquad m_i>0\,,
\eeq
where $Z_{m_1,\ldots,m_k}(n)$ denote the so-called Euler-Zagier
sums~\cite{Euler_sum,Zagier_sum}, defined recursively by
\beq\label{eq:Euler-Zagier}
Z_{m_1}(n) = \sum_{l=1}^n{1\over l^{m_1}} {\rm~~and~~}
Z_{m_1,\ldots,m_k}(n) = \sum_{l=1}^n{1\over l^{m_1}}Z_{m_2,\ldots,m_k}(l-1)\,.
\eeq
Note that the indexing of the weight vectors $m_1,\ldots,m_k$ in
eqs.~(\ref{eq:HPL_series}) and (\ref{eq:Euler-Zagier}) is in the
collapsed notation.

Another important property of HPLs is that they are closed under
certain transformations of the arguments~\cite{Remiddi1999ew}. In
particular, using the integral representation~\eqref{eq:HPL_integral},
it is easy to show that the set of all HPLs is closed under the
following transformations,
\beq\label{eq:S3_map}
z\mapsto 1-z,\quad z\mapsto 1/z, \quad z\mapsto1/(1-z),
\quad z\mapsto1-1/z, \quad z\mapsto z/(z-1)\,.
\eeq
If we add to these mappings the identity map $z\mapsto z$, we can
identify the transformations in eq.~\eqref{eq:S3_map} as forming a
representation of the symmetric group $S_3$. In other words, the
vector space spanned by all HPLs is endowed with a natural action of
the symmetric group $S_3$.

Finally, it is evident from the iterated integral representation~\eqref{eq:HPL_integral} that HPLs can have branch cuts starting at $z=0$ and/or $z=1$, i.e., HPLs define in general multi-valued functions on the complex plane. In the next section we will define analogues of HPLs without any branch cuts, thus obtaining a single-valued version of the HPLs.

\subsection{Single-valued harmonic polylogarithms}
Before reviewing the definition of single-valued harmonic polylogarithms in general, let us first review the special case of single-valued classical polylogarithms.
The knowledge of the discontinuities of the classical polylogarithms, eq.~\eqref{eq:lin_disc}, can be leveraged to construct a sequence of real analytic functions on the punctured plane $\mathbb{C}\backslash\{0,1\}$. The idea is to consider linear combinations of (products of) classical polylogarithms and ordinary logarithms such that all the branch cuts cancel.  Although the space of single-valued functions is unique, the choice of basis is not unique, and there have been several versions proposed in the literature. As an illustration, consider the functions of Zagier~\cite{Zagier},
\beq\label{eq:D_m_def}
D_m(z) =
\mathfrak{R}_m \left\{\sum_{k=1}^m \frac{(-\log |z|)^{m-k}}{(m-k)!} {\rm Li}_k(z)+\frac{\log^m|z|}{2\ m!}\right\}\,,
\eeq
where $\mathfrak{R}_m$ denotes the imaginary part for $m$ even and the real part for $m$ odd.
The discontinuity of the function inside the curly brackets is given by
\begin{equation}\label{eq:D_m_disc}
2\pi i \sum_{k=1}^m \frac{(-\log |z|)^{m-k}}{(m-k)!}\frac{\log^{k-1} z}{(k-1)!} = 2\pi \frac{ i^m}{(m-1)!}(\arg z)^{m-1}\,.
\end{equation}
Since eq.~\eqref{eq:D_m_disc} is real for even $m$ and pure imaginary for odd $m$, $D_m(z)$ is indeed single-valued. For the special case $m=2$, we reproduce the famous Bloch-Wigner dilogarithm~\cite{Bloch},
\begin{equation}
D_2(z) = {\rm Im} \{{\rm Li}_2(z) \}+ \arg(1-z)\log|z|\,.
\end{equation}

Just as there have been numerous proposals in the literature for single-valued versions of the classical polylogarithms, there are many potential choices of bases for single-valued HPLs. On the other hand, if we choose to demand some reasonable properties, it turns out that a unique set of functions emerges.
Following ref.~\cite{BrownSVHPLs}, we require the single-valued HPLs to be built entirely from holomorphic and anti-holomorphic HPLs. Specifically, they should be a linear combination of terms of the form $H_{w_1}(z)H_{w_2}(\bar{z})$, where $w_1$ and $w_2$ are words in $x_0$ and $x_1$ or the empty word $e$. The single-valued classical polylogarithms obey an analogous property, and it can be understood as the condition that the single-valued functions are the proper extensions of the original functions. The remaining requirements are simply the analogues of the conditions used to construct the ordinary HPLs.

Define a function $\cL_w(z)$, which is a linear combination of functions $H_{w_1}(z)H_{w_2}(\bar{z})$ and which obeys the differential equations
\begin{equation}\label{eq:Lzdiffeq}
\frac{\partial}{\partial z}\cL_{x_0w}(z) = \frac{\cL_{w}(z)}{z} \quad\quad\text{and}\quad\quad\frac{\partial}{\partial z}\cL_{x_1w}(z) = \frac{\cL_{w}(z)}{1-z}\,,
\end{equation}
subject to the conditions,
\begin{equation}
\cL_{e}(z)=1\,,\quad\quad \cL_{x_0^n}(z)=\frac{1}{n!}\log^n|z|^2\quad\quad\text{and}\quad\quad \lim_{z\rightarrow 0}\cL_{w\neq x_0^n}(z)=0\,.
\end{equation}
In ref.~\cite{BrownSVHPLs} Brown showed that there is a unique family of solutions to these equations that is single-valued in the complex $z$ plane, and it defines the single-valued HPLs (SVHPLs). The functions $\cL_w(z)$ are linearly independent and span the space. That is to say, every single-valued linear combination of functions of the form $H_{w_1}(z)H_{w_2}(\bar{z})$ can be written in terms of the $\cL_w(z)$. In ref.~\cite{BrownSVHPLs} an algorithm was presented that allows for the explicit construction of all SVHPLs as linear combinations of (products of) ordinary HPLs. We present a short review of this algorithm in Section~\ref{sec:SVHPL_construction}.

The SVHPLs of ref.~\cite{BrownSVHPLs} share all the nice features of their multi-valued analogues. First, like the ordinary HPLs, they  obey shuffle relations,
\begin{equation}
\label{SVeq:HPL_shuffle}
\cL_{w_1}(z)\,\cL_{w_2}(z) = \sum_{{w}\in{w_1}\ssha {w_2}}\cL_{w}(z),
\end{equation}
where again ${w_1}\sha {w_2}$ represents the {shuffles} of $w_1$ and $w_2$. As a consequence, we may again choose to solve eq.~\eqref{SVeq:HPL_shuffle} in terms of a Lyndon basis.  It follows that if we want the full list of all SVHPLs of a given weight, it is enough to know the corresponding Lyndon basis up to that weight.

Furthermore, the space of SVHPLs is also closed under the $S_3$ action defined by eq.~\eqref{eq:S3_map}. Indeed, if we extend the action to the complex conjugate variable $\bar{z}$, then the closure of the space of all ordinary HPLs implies the closure of the space spanned by all products of the form $H_{w_1}(z)H_{w_2}(\bar{z})$, and, in particular, the closure of the subspace of SVHPLs. For the SVHPLs, it is possible to enlarge the symmetry group to $\mathbb{Z}_2\times S_3$, where the $\mathbb{Z}_2$ subgroup acts by complex conjugation, $z \leftrightarrow \bar{z}$.

It turns out that the functions $\cL_w(z)$ can generically be decomposed as
\begin{equation}
\cL_w(z)=\left( H_w(z) - (-1)^{|w|} H_w(\bar{z})\right)+ \textrm{[products of lower weight]}\,,
\end{equation}
where $|w|$ denotes the weight. As such, it is convenient to consider the even and odd projections, i.e., the decomposition into eigenfunctions of the $\mathbb{Z}_2$ action, 
\begin{equation}
\label{eq:Lplusminus}
L_w(z) =\frac{1}{2}\left( \cL_w(z) - (-1)^{|w|}\,\cL_w(\bar{z})\right) \quad\quad \textrm{and}\quad\quad\overline{L}_w(z) =\frac{1}{2}\left( \cL_w(z)+(-1)^{|w|}\cL_w(\bar{z})\right)\,.
\end{equation}
The basis defined by $\cL_w(z)$ was already complete, and yet here we have doubled the number of potential basis functions. Therefore $L_w(z)$ and $\overline{L}_w(z)$ must be related to one another. Writing $L_w(z) =\mathfrak{R}_{|w|}(\cL_w(z))$, we see that it has the same parity as Zagier's single-valued versions of the classical polylogarithms given in eq.~\eqref{eq:D_m_def}. Therefore we might expect the $L_w(z)$ to form a complete basis on their own. Indeed this turns out to be the case, and the $\overline{L}_w(z)$ can be expressed as products of the functions $L_w(z)$,
\begin{equation}
\overline{L}_w(z)= \textrm{[products of lower weight $L_{w'}(z)$]}\,.
\end{equation}
Hence we will not consider the functions $\overline{L}_w(z)$ any further and will concentrate solely on the functions $L_{w}(z)$.

The functions $L_{w}(z)$ do not automatically form simple representations of the $S_3$ symmetry. For the current application, we will mostly be concerned with the $\mathbb{Z}_2 \subset S_3$ subgroup generated by inversions $z\leftrightarrow1/z$. The functions $L_{w}(z)$ can easily be decomposed into eigenfunctions of this $\mathbb{Z}_2$, and, furthermore, these eigenfunctions form a basis for the space of all SVHPLs. The latter follows from the observation that,
\beq
L_{w}(z) - (-1)^{|w|+d_w}L_{w}\Big(\frac{1}{z}\Big) =  \textrm{[products of lower weight]},
\eeq
where $|w|$ is the weight and $d_w$ is the depth of the word $w$. We will denote these eigenfunctions of $\mathbb{Z}_2\times\mathbb{Z}_2$ by,
\beq
L_{w}^\pm(z) \equiv {1\over 2}\,\left[L_w(z)\pm L_w\Big({1\over z}\Big)\right] ,
\eeq
and present most of our results in terms of this convenient basis. For low weights, Appendix \ref{app:svhpl} gives explicit representations of these basis functions in terms of HPLs. The expressions through weight six can be found in the ancillary files.

We have seen in the previous section that in the multi-Regge limit the six-point amplitude is described to all loop orders by single-valued functions of a single complex variable $w$ satisfying certain reality and inversion properties. It turns out that the SVHPLs we just defined are particularly well-suited to describe these multi-Regge limits. This description will be the topic of the rest of this paper.

\subsection{Explicit construction}
\label{sec:SVHPL_construction}

The explicit construction of the functions $\cL_w(z)$ is somewhat involved so we take a brief detour to describe the details. Let $X^*$ be the set of words in the alphabet $\{x_0,x_1\}$, along with the empty word $e$. Define  the Drinfel'd associator $Z(x_0,x_1)$ as the generating series,
\begin{equation}
Z(x_0,x_1) = \sum_{w\in X^*} \zeta(w)w,
\end{equation}
where $\zeta(w)=H_w(1)$ for $w\neq x_1$ and $\zeta(x_1)=0$.  The $\zeta(w)$ are regularized by the shuffle algebra.  Using the collapsed notation for $w$, these $\zeta(w)$ are the familiar multiple zeta values.

Next, define an alphabet $\{y_0,y_1\}$ (and a set of words $Y^*$) and a map  $^ \sim: Y^*\rightarrow Y^*$ as the operation that reverses words. The alphabet $\{y_0, y_1\}$ is related to the alphabet $\{x_0,x_1\}$  by the following relations:
\beq\bsp
\label{y_alph}
y_0&\,=\,x_0\\
\tilde{Z}(y_0,y_1)y_1\tilde{Z}(y_0,y_1)^{-1} &\,=\, Z(x_0,x_1)^{-1}x_1 Z(x_0,x_1).
\esp\eeq
The inversion operator is to be understood as a formal series expansion in the weight $|w|$. Solving eq.~\eqref{y_alph} iteratively in the weight yields a series expansion for $y_1$. The first few terms are,
\beq\bsp\label{eq:y1_ser}
&y_1=x_1-\zeta_3\left(2x_0x_0x_1x_1-4x_0x_1x_0x_1 +2x_0x_1x_1x_1 + 4x_1x_0x_1x_0 \right.\\
&\quad\quad\quad\quad\quad\quad\left.-6x_1x_0x_1x_1-2x_1x_1x_0x_0+6x_1x_1x_0x_1-2x_1x_1x_1x_0\right)  + \ldots
\esp\eeq
Letting $\phi:Y^*\rightarrow X^*$ be the map that renames $y$ to $x$, i.e. $\phi(y_0)=x_0$ and $\phi(y_1)=x_1$,  define the generating functions
\beq\label{eq:LXY0}
L_X(z)\,=\,\sum_{w\in X^*}H_w(z)w \,, \qquad
\tilde{L}_Y(\bar{z})\,=\,\sum_{w\in Y^*}H_{\phi(w)}(\bar{z})\tilde{w} \,.
\eeq
In the following, we use a condensed notation for the HPL arguments, 
in order to improve the readability of explicit formulas:
\beq\label{eq:Hwcompact}
H_w \equiv H_w(z) {\rm~~and~~}\Hb_w \equiv H_w(\zp)\,.
\eeq
Then we can write 
\beq\bsp\label{eq:LX}
L_X(z)\,=\;&1+H_{0} x_0+H_{1} x_1  \\
& + \,H_{0,0} x_0x_0+H_{0,1} x_0x_1+H_{1,0}  x_1x_0+H_{1,1}  x_1 x_1\\
& + \,H_{0,0,0} x_0x_0x_0+H_{0,0,1} x_0x_0x_1+H_{0,1,0}  x_0x_1x_0+H_{0,1,1}  x_0x_1 x_1\\
& + \,H_{1,0,0} x_1x_0x_0+H_{1,0,1} x_1x_0x_1+H_{1,1,0}  x_1x_1x_0+H_{1,1,1}  x_1x_1 x_1\,+\,\ldots\,,\\
\esp\eeq
and
\beq\bsp\label{eq:LYtilde}
\tilde{L}_Y(\bar{z})\,=\;&
1 + \Hb_{0} y_0 + \Hb_{1} y_1  \\
& + \,\Hb_{0,0} y_0y_0+\Hb_{0,1} y_1y_0+\Hb_{1,0}  y_0y_1+\Hb_{1,1}  y_1 y_1\\
& + \,\Hb_{0,0,0} y_0y_0y_0+\Hb_{0,0,1} y_1y_0y_0+\Hb_{0,1,0}  y_0y_1y_0+\Hb_{0,1,1}  y_1y_1 y_0\\
& + \,\Hb_{1,0,0} y_0y_0y_1+\Hb_{1,0,1} y_1y_0y_1+\Hb_{1,1,0}  y_0y_1y_1+\Hb_{1,1,1}  y_1y_1 y_1\,+\,\ldots\,\\
=\;&1+\Hb_{0} x_0+\Hb_{1} x_1  \\
& + \,\Hb_{0,0} x_0x_0+\Hb_{0,1} x_1x_0+\Hb_{1,0}  x_0x_1+\Hb_{1,1}  x_1 x_1\\
& + \,\Hb_{0,0,0} x_0x_0x_0+\Hb_{0,0,1} x_1x_0x_0+\Hb_{0,1,0}  x_0x_1x_0+\Hb_{0,1,1}  x_1x_1 x_0\\
& + \,\Hb_{1,0,0} x_0x_0x_1+\Hb_{1,0,1} x_1x_0x_1+\Hb_{1,1,0}  x_0x_1x_1+\Hb_{1,1,1}  x_1x_1 x_1\,+\,\ldots\; .\\
\esp\eeq
In the last step of eq.~\eqref{eq:LYtilde} we used $y_0=x_0$ and $y_1=x_1$. Note that the latter only holds through weight three, as is clear from eq.~\eqref{eq:y1_ser}.
Finally, we are able to construct the SVHPLs as a generating series,
\begin{equation}
\cL(z)=L_X(z)\tilde{L}_Y(\bar{z}) \equiv \sum_{w\in X^*} \cL_w(z)w.
\end{equation}
Indeed, taking the product of eq.~\eqref{eq:LX} with eq.~\eqref{eq:LYtilde} and keeping terms through weight three, we obtain,
\beq\bsp
\sum_{w\in X^*} &\cL_w(z)w \;=\;1+\;\cL_{0}(z)\,x_0+\cL_{1}(z)\,x_1\\
&+\cL_{0,0}(z)\,x_0x_0+\cL_{0,1}(z)\,x_0x_1+\cL_{1,0}(z)\,x_1x_0 +\cL_{1,1}(z)\,x_1x_1 \\
&+\cL_{0,0,0}(z)\,x_0x_0x_0 +\cL_{0,0,1}(z)\,x_0x_0x_1 +\cL_{0,1,0}(z)\,x_0x_1x_0 +\cL_{0,1,1}(z)\,x_0x_1x_1\\
&+\cL_{1,0,0}(z)\,x_1x_0x_0 +\cL_{1,0,1}(z)\,x_1x_0x_1+\cL_{1,1,0}(z)\,x_1x_1x_0 +\cL_{1,1,1}(z)\,x_1x_1x_1 + \ldots \; ,
\esp\eeq
where the SVHPL's of weight one are,
\beq
\cL_{0}(z) \;=\; H_{0} + \Hb_{0}\, ,\quad 
\cL_{1}(z)\;=\;H_{1} + \Hb_{1}, \\
\eeq
the SVHPL's of weight two are,
\beq\bsp
\cL_{0,0}(z)&\;=\;H_{0,0} + \Hb_{0,0} + H_{0} \Hb_{0} \,,\\
\cL_{0,1}(z)&\;=\;H_{0,1} + \Hb_{1,0} + H_{0} \Hb_{1} \,,\\
\cL_{1,0}(z)&\;=\;H_{1,0} + \Hb_{0,1} + H_{1} \Hb_{0} \,,\\
\cL_{1,1}(z)&\;=\;H_{1,1} + \Hb_{1,1} + H_{1} \Hb_{1} \,,\\
\esp\eeq
and the SVHPL's of weight three are,
\beq\bsp
\cL_{0,0,0}(z)&\;=\;H_{0,0,0}+\Hb_{0,0,0}+H_{0,0}\Hb_{0}+H_{0}\Hb_{0,0}\,,\\
\cL_{0,0,1}(z)&\;=\;H_{0,0,1}+\Hb_{1,0,0}+H_{0,0}\Hb_{1}+H_{0}\Hb_{1,0}\,,\\
\cL_{0,1,0}(z)&\;=\;H_{0,1,0}+\Hb_{0,1,0}+H_{0,1}\Hb_{0}+H_{0}\Hb_{0,1}\,,\\
\cL_{0,1,1}(z)&\;=\;H_{0,1,1}+\Hb_{1,1,0}+H_{0,1}\Hb_{1}+H_{0}\Hb_{1,1}\,,\\
\cL_{1,0,0}(z)&\;=\;H_{1,0,0}+\Hb_{0,0,1}+H_{1,0}\Hb_{0}+H_{1}\Hb_{0,0}\,,\\
\cL_{1,0,1}(z)&\;=\;H_{1,0,1}+\Hb_{1,0,1}+H_{1,0}\Hb_{1}+H_{1}\Hb_{1,0}\,,\\
\cL_{1,1,0}(z)&\;=\;H_{1,1,0}+\Hb_{0,1,1}+H_{1,1}\Hb_{0}+H_{1}\Hb_{0,1}\,,\\
\cL_{1,1,1}(z)&\;=\;H_{1,1,1}+\Hb_{1,1,1}+H_{1,1}\Hb_{1}+H_{1}\Hb_{1,1}\,.\\
\esp\eeq
\\
The $y$ alphabet differs from the $x$ alphabet starting at weight four. Referring to eq.~\eqref{eq:y1_ser}, we expect the difference to generate factors of $\zeta_3$. To illustrate this effect, we list here the subset of weight-four SVHPLs with explicit $\zeta$ terms:
\beq\bsp
\cL_{0,0,1,1}(z)&\;=\;H_{0,0,1,1}+\Hb_{1,1,0,0}+H_{0,0,1}\Hb_{1}+H_{0}\Hb_{1,1,0}+H_{0,0}\Hb_{1,1}-2\zeta_3\,\Hb_{1}\,,\\
\cL_{0,1,0,1}(z)&\;=\;H_{0,1,0,1}+\Hb_{1,0,1,0}+H_{0,1,0}\Hb_{1}+H_{0}\Hb_{1,0,1}+H_{0,1}\Hb_{1,0}+4\zeta_3\,\Hb_{1}\,,\\
\cL_{0,1,1,1}(z)&\;=\;H_{0,1,1,1}+\Hb_{1,1,1,0}+H_{0,1,1}\Hb_{1}+H_{0}\Hb_{1,1,1}+H_{0,1}\Hb_{1,1}-2\zeta_3\,\Hb_{1}\,,\\
\cL_{1,0,1,0}(z)&\;=\;H_{1,0,1,0}+\Hb_{0,1,0,1}+H_{1,0,1}\Hb_{0}+H_{1}\Hb_{0,1,0}+H_{1,0}\Hb_{0,1}-4\zeta_3\,\Hb_{1}\,,\\
\cL_{1,0,1,1}(z)&\;=\;H_{1,0,1,1}+\Hb_{1,1,0,1}+H_{1,0,1}\Hb_{1}+H_{1}\Hb_{1,1,0}+H_{1,0}\Hb_{1,1}+6\zeta_3\,\Hb_{1}\,,\\
\cL_{1,1,0,0}(z)&\;=\;H_{1,1,0,0}+\Hb_{0,0,1,1}+H_{1,1,0}\Hb_{0}+H_{1}\Hb_{0,0,1}+H_{1,1}\Hb_{0,0}+2\zeta_3\,\Hb_{1}\,,\\
\cL_{1,1,0,1}(z)&\;=\;H_{1,1,0,1}+\Hb_{1,0,1,1}+H_{1,1,0}\Hb_{1}+H_{1}\Hb_{1,0,1}+H_{1,1}\Hb_{1,0}-6\zeta_3\,\Hb_{1}\,,\\
\cL_{1,1,1,0}(z)&\;=\;H_{1,1,1,0}+\Hb_{0,1,1,1}+H_{1,1,1}\Hb_{0}+H_{1}\Hb_{0,1,1}+H_{1,1}\Hb_{0,1}+2\zeta_3\,\Hb_{1}\,.\\
\esp\eeq
\\
Finally, we remark that the generating series $\cL(z)$ provides a convenient way to represent the differential equations~\eqref{eq:Lzdiffeq}. Together with the $y$ alphabet, it also allows us to write down the differential equations in $\bar{z}$,
\beq\label{eq:dLzz}
\frac{\partial}{\partial z} \cL(z) = \left(\frac{x_0}{z}+\frac{x_1}{1-z}\right)\cL(z) \quad\quad \textrm{and}\quad\quad\frac{\partial}{\partial \bar{z}} \cL(z) = \cL(z)\left(\frac{y_0}{\bar{z}}+\frac{y_1}{1-\bar{z}}\right).
\eeq
These equations will be particularly useful in Section~\ref{sec:NMHV} when we study the multi-Regge limit of the ratio function of the six-point NMHV amplitude.

\section{The six-point remainder function in LLA and NLLA}
\label{sec:R6_LLA_NLLA}
In Section~\ref{sec:MRK}, we showed that in MRK the remainder function is fully determined by the coefficient functions $g_n^{(\ell)}(w,\ws)$ and $h_n^{(\ell)}(w,\ws)$ in the logarithmic expansion of its real and imaginary part in eq.~\eqref{eq:R6_MRK}. We further argued that these functions are single-valued in the complex $w$ plane, and suggested that they can be computed explicitly by interpreting the $\nu$-integral in eq.~\eqref{eq:MHV_MRK} as a contour integral and summing the residues. In this section, we describe how knowledge about the space of SVHPLs can be used to facilitate this calculation. In particular, we present results for LLA through ten loops and for NLLA through nine loops.

The main integral we consider is eq.~\eqref{eq:MHV_MRK}, which we reproduce here
for clarity, rewriting the last factor to take into account
eqs.~\eqref{eq:removeu2u3} and \eqref{eq:ipi},
\bea
e^{R+i\pi\delta}|_{\textrm{MRK}}
 &=& \cos\pi\omega_{ab}
 + i \, {a\over 2}\sum_{n=-\infty}^\infty(-1)^n
 \,\left({w\over \ws}\right)^{{n\over 2}}\int_{-\infty}^{+\infty}
 {d\nu\over \nu^2+{n^2\over 4}}\,|w|^{2i\nu}
  \,\Phi_{\textrm{Reg}}(\nu,n) \nonumber\\
&&\hskip2cm\null
 \times \exp\left[ -\omega(\nu,n) \left( \log(1-u_1) + i\pi
   + \frac{1}{2} \log \frac{|w|^2}{|1+w|^4} \right) \right] \,.
\label{eq:MHV_MRK_2}
\eea
The integrand depends on the BFKL eigenvalue and impact factor,
which are known through order $a^2$ and are given in
eqs.~\eqref{eq:E_0}, \eqref{eq:E_1} and \eqref{eq:Phi_1}.  These
functions can be written as rational functions of $\nu$ and $n$, and
polygamma functions ($\psi$ and its derivatives) 
with arguments $1 \pm i\nu+|n|/2$. Recalling that
the polygamma functions have poles at the non-positive integers, it is
easy to see that all poles are found in the complex $\nu$ plane at
values $\nu = -i(m+{|n|\over2})$, $m\in\mathbb{N}$, $n\in\mathbb{Z}$.
When the integral is performed by summing residues, the result will be
of the form,
\beq\label{eq:double_sum}
\sum_{m,n}\,a_{m,n}\,w^{m+n}\,\ws^m\,.
\eeq
Because residues of the polygamma functions are rational numbers,
and because polygamma functions evaluate to Euler-Zagier sums for
positive integers, the coefficients $a_{m,n}$ are combinations of
\begin{enumerate}
\item rational functions in $m$ and $n$,
\item Euler-Zagier sums of the form $Z_{\vec \imath}(m)$, 
$Z_{\vec \imath}(n)$ and $Z_{\vec \imath}(m+n)$,
\item $\log|w|$, arising from taking residues at multiple poles.
\end{enumerate}
Identifying $(z,\bar{z}) \equiv(-w,-\ws)$, and comparing the double sum~\eqref{eq:double_sum} to the formal series expansion of the HPLs around $z=0$, eq.~\eqref{eq:HPL_series}, we conclude that the double sums will evaluate to linear combinations of terms of the form $H_{w_1}(-w) H_{w_2}(-\ws)$. Moreover, as discussed above, this combination should be single-valued. Therefore, based on the discussion in Section~\ref{sec:SVHPLs}, we expect $g_n^{(\ell)}(w,\ws)$ and $h_n^{(\ell)}(w,\ws)$ to belong to the space spanned by the SVHPLs.

Furthermore, we know that $g_n^{(\ell)}(w,\ws)$ and $h_n^{(\ell)}(w,\ws)$ are invariant under the action of the $\mathbb{Z}_2\times\mathbb{Z}_2$ transformations of eq.~\eqref{eq:Z2xZ2w}. In terms of SVHPLs, this symmetry is just an (abelian) subgroup of the larger $\mathbb{Z}_2\times S_3$ symmetry, where the $\mathbb{Z}_2$ is complex conjugation and the $S_3$ action is given in eq.~\eqref{eq:S3_map}. As such, we do not expect an arbitrary linear combination of SVHPLs, but only those that are eigenfunctions with eigenvalue $(+,+)$ of the $\mathbb{Z}_2\times\mathbb{Z}_2$ symmetry.

Putting everything together, and taking into account that scattering amplitudes in $\cN=4$ SYM are expected to have uniform transcendentality, we are led to conjecture that, to all loop orders, $g_n^{(\ell)}(w,\ws)$ and $h_n^{(\ell)}(w,\ws)$ should be expressible as a linear combination of SVHPLs in $(z,\bar{z})=(-w,-\ws)$ of uniform transcendental weight, with eigenvalue $(+,+)$ under the $\mathbb{Z}_2\times\mathbb{Z}_2$ symmetry.  Inspecting eq.~\eqref{eq:R6_MRK}, the weight should be $2\ell-n-1$ for $g_n^{(\ell)}$ and $2\ell-n-2$ for $h_n^{(\ell)}$.  Our conjecture allows us to predict \emph{a priori} the set of functions that can appear at a given loop order, and in practice this set turns out to be rather small. Knowledge of this set of functions can be used to facilitate the evaluation of eq.~\eqref{eq:MHV_MRK_2}. We outline two strategies to achieve this:

\begin{enumerate}
\item Evaluate the double sum~\eqref{eq:double_sum} with the summation algorithms of ref.~\cite{Moch2001zr}. The result is a complicated expression involving multiple polylogarithms which can be matched to a combination of SVHPLs and zeta values by means of the
symbol~\cite{symbolsC,symbolsB,symbols,Goncharov2010jf,Duhr2011zq}
and coproduct~\cite{GoncharovGalois,Brown2011ik,Duhr2012fh}.

\item The double sum~\eqref{eq:double_sum} should be equal to the formal series expansion of some linear combination of SVHPLs and zeta values. The unknown coefficients of this combination can be fixed by matching the two expressions term by term.
\end{enumerate}
To see how this works, we calculate the two-loop remainder function in MRK. Expanding eq.~\eqref{eq:MHV_MRK_2} to two loops, we find,
\beq\bsp\label{eq:R62}
a^2 R^{(2)} \simeq 2\pi i\, &\left\{a\left[-\frac{1}{2}L_1^{+}+\frac{1}{4} \mathcal{I}[1] \right] \right.\\
&\; \null + a^2\left[\log(1-u_1)\,\frac{1}{4}\,\mathcal{I}[E_{\nu,n}] + \Big(\frac{1}{2}\zeta_2L_1^{+}+\frac{1}{4}\,\mathcal{I}[\Phi_{\textrm{Reg}}^{(1)}(\nu,n)]+\frac{1}{4}L_1^{+}\,\mathcal{I}[E_{\nu,n}]\Big) \right.\\
&\quad\quad\left.\left.+2\pi i\,\Big(\frac{1}{32}\,[L_0^{-}]^2+\frac{1}{8}\,[L_1^{+}]^2-\frac{1}{8}\,L_1^{+}\,\mathcal{I}[1]+\frac{1}{8}\,\mathcal{I}[E_{\nu,n}]\Big) \right] \right\},
\esp\eeq
where we have introduced the notation,
\beq\label{eq:integral_transform}
\mathcal{I}[\mathcal{F}(\nu,n)] = \frac{1}{\pi}\sum_{n=-\infty}^\infty(-1)^n\,\left({w\over \ws}\right)^{{n\over 2}}\int_{-\infty}^{+\infty}{d\nu\over \nu^2+{n^2\over 4}}\,|w|^{2i\nu}\,\mathcal{F}(\nu,n) \, .
\eeq
Explicit expressions for the functions $L_w^\pm$ for low weights are provided
in Appendix~\ref{app:svhpl}.
Equation~\eqref{eq:R62} is consistent only if the term of order $a$ vanishes. Indeed this is the case,
\beq\bsp\label{eq:I[1]}
\mathcal{I}[1] &\,= {1\over\pi}\sum_{n=-\infty}^\infty(-1)^n\,\left({w\over \ws}\right)^{{n\over 2}}\int_{-\infty}^{+\infty}
{d\nu\over \nu^2+{n^2\over 4}}\,|w|^{2i\nu}\\
&\,=\log|w|^2+2\sum_{n=1}^\infty{(-w)^n\over n}+2\sum_{n=1}^\infty{(-\ws)^n\over n}\\
&\,=\log|w|^2-2\log|1+w|^2\\
&\,=2L_1^+\,.
\esp\eeq
As previously mentioned, we only take half of the residue at $\nu=n=0$. 

Moving on to the terms of order $a^2$, we refer to eq.~\eqref{eq:R6_MRK} and extract from eq.~\eqref{eq:R62} the expressions for the coefficient functions,
\beq\bsp\label{eq:R62_gh}
g^{(2)}_1(w,\ws) &= \frac{1}{4}\,\mathcal{I}[E_{\nu,n}] \\
g^{(2)}_0(w,\ws) &= \frac{1}{2}\zeta_2L_1^{+}+\frac{1}{4}\,\mathcal{I}[\Phi_{\textrm{Reg}}^{(1)}(\nu,n)]+\frac{1}{4}L_1^{+}\,\mathcal{I}[E_{\nu,n}]\\
h^{(2)}_0(w,\ws) &= \frac{1}{32}\,[L_0^{-}]^2+\frac{1}{8}\,[L_1^{+}]^2-\frac{1}{8}\,L_1^{+}\,\mathcal{I}[1]+\frac{1}{8}\,\mathcal{I}[E_{\nu,n}]\,.
\esp\eeq
Note that $h^{(2)}_1 =0$, in accordance with the general expectation that $h^{(l)}_{l-1}=0$. Proceeding onwards, we have to calculate $\mathcal{I}[E_{\nu,n}]$,
\beq\bsp\label{eq:I[E]}
\mathcal{I}[E_{\nu,n}] &= \frac{1}{\pi}\sum_{n=-\infty}^\infty(-1)^n\,\left({w\over \ws}\right)^{{n\over 2}}\int_{-\infty}^{+\infty}{d\nu\over \nu^2+{n^2\over 4}}\,|w|^{2i\nu}\bigg\{ 2\gamma_E+\frac{|n|}{2(\nu^2+\frac{n^2}{4})}\\
&\quad\quad\quad\quad\quad +\psi\left(i\nu+\frac{|n|}{2}\right) + \psi\left(-i\nu+\frac{|n|}{2}\right)\bigg\}\\
&=\sum_{m=1}^\infty\left\{2{|w|^{2m}\over m^2} -2{(-w)^m+(-\ws)^m\over m^2}
+[\log|w|^2+2Z_1(m)]{(-w)^m+(-\ws)^m\over m}\right\}\\
&\quad+2\,\sum_{n=1}^\infty\sum_{m=1}^\infty{(-1)^n\over m(m+n)}\,\left\{w^{m+n}\ws^m+w^{m}\ws^{m+n}\right\}\,.
\esp\eeq
The single sum in the first line immediately evaluates to polylogarithms,
\beq\bsp
\sum_{m=1}^\infty&\left\{2{|w|^{2m}\over m^2} -2{(-w)^m+(-\ws)^m\over m^2}+[\log|w|^2+2Z_1(m)]{(-w)^m+(-\ws)^m\over m}\right\}\\
&=\sum_{m=1}^\infty\left\{2{|w|^{2m}\over m^2} +[\log|w|^2+2Z_1(m-1)]{(-w)^m+(-\ws)^m\over m}\right\}\\
&=\log|w|^2\left[ H_{1}(-w)+H_{1}(-\ws)\right]
  + 2 H_{0,1}(|w|^2) + 2 H_{1,1}(-w) + 2 H_{1,1}(-\ws)\,.
\esp\eeq
Next we transform the double sum into a nested sum by shifting the summation variables by $n=N-m$,
\beq\bsp\label{eq:I[E]_dbl_sum}
\sum_{n=1}^\infty&\sum_{m=1}^\infty{(-1)^n\over m(m+n)}\,\left\{w^{m+n}\ws^m+w^{m}\ws^{m+n}\right\}=\sum_{N=1}^\infty\sum_{m=1}^{N-1}\left\{{(-w)^{N}(-\ws)^m\over N\,m}+{(-w)^{m}(-\ws)^N\over N\,m}\right\}\\
&=\textrm{Li}_{1,1}(-w,-\ws)+\textrm{Li}_{1,1}(-\ws,-w)\\
&= H_{1}(-w)\,H_{1}(-\ws) - H_{0,1}(|w|^2)\,,
\esp\eeq
where the last step follows from a stuffle identity
among multiple polylogarithms~\cite{Borwein1999js}.
Putting everything together, we obtain
\beq\bsp
\mathcal{I}[E_{\nu,n}] &\,= \log|w|^2\left[ H_{1}(-w) + H_{1}(-\ws)\right]
 + 2 H_{1,1}(-w) + 2 H_{1,1}(-\ws) + 2 H_{1}(-w)\,H_{1}(-\ws) \\
&\,=[L_1^+]^2-{1\over 4}[L_0^-]^2\,.
\esp\eeq
Referring to eqs.~\eqref{eq:I[1]} and~\eqref{eq:R62_gh}, we can now write down the results,
\beq\bsp\label{eq:g21}
g_1^{(2)}(w,\ws) \,&= {1\over4}\,\LOneP{2}  -{1\over16}\, \LZeroM{2} \,, \\
h_0^{(2)}(w,\ws) \,&=0\,.
\esp\eeq

For higher weights the nested double sums can be more complicated, but they are always of a form that can be performed using the algorithms of ref.~\cite{Moch2001zr}. These algorithms will in general produce complicated multiple polylogarithms that, unlike in eq.~\eqref{eq:I[E]_dbl_sum}, cannot in general be reduced to HPLs by the simple application of stuffle identities. In this case we can use symbols~\cite{symbols,Goncharov2010jf,Duhr2011zq} and the coproduct on multiple polylogarithms~\cite{GoncharovGalois,Brown2011ik,Duhr2012fh} to perform this reduction.

The above strategy becomes computationally taxing for high weights. For this reason, we also employ an alternative strategy, based on matching series expansions, which is computationally simpler. We demonstrate this method in the computation of $g^{(2)}_0$, for which the only missing ingredient in eq.~\eqref{eq:R62_gh} is $\mathcal{I}[\Phi_{\textrm{Reg}}^{(1)}(\nu,n)]$, where $\Phi_{\textrm{Reg}}^{(1)}(\nu,n)$ is defined in eq.~\eqref{eq:Phi_1}. To proceed, we write the $\nu$-integral as a sum of residues, and truncate the resulting double sum to some finite order,
\beq\bsp\label{eq:I[Phi]_ser}
\mathcal{I}[\Phi_{\textrm{Reg}}^{(1)}(\nu,n)] &= 
\frac{1}{\pi}\sum_{n=-\infty}^\infty(-1)^n\,\left({w\over \ws}\right)^{{n\over 2}}
\int_{-\infty}^{+\infty}{d\nu\over \nu^2+{n^2\over 4}}\,|w|^{2i\nu}
\Bigg\{-\zeta_2 - {3\over 8}\,{n^2\over (\nu^2+{n^2\over 4})^2} \\
&\quad -\frac{1}{2}\left(2\gamma_E+\frac{|n|}{2(\nu^2+\frac{n^2}{4})}
+\psi\left(i\nu+\frac{|n|}{2}\right)
+ \psi\left(-i\nu+\frac{|n|}{2}\right)\right)^2 \Bigg\}
\esp\eeq
\beq\bsp\nonumber
\phantom{\mathcal{I}[\Phi_{\textrm{Reg}}^{(1)}(\nu,n)]}&=-\zeta_2 \, \log|w|^2 -\left( \log|w|^2  \right)|w|^2 -\left(1+\frac{1}{4}\log|w|^2\right) |w|^4+ \ldots\\
&\quad+(w+\ws)\left[2\zeta_2
+\left(4-2\log|w|^2+\frac{1}{2}\log^2|w|^2\right) 
+\left(1+\frac{1}{2}\log|w|^2\right) |w|^2+ \ldots \right]\\
&\quad+(w^2+\ws^2)\left[-\zeta_2-\left(\frac{1}{2}+\frac{1}{4}\log^2|w|^2\right) +\left(-1-\frac{1}{3}\log|w|^2\right)|w|^2+ \ldots\right]\\
&\quad + \ldots \,.
\esp\eeq
Here we show on separate lines the contributions to the sum from $n=0$, $n=\pm1$, and $n=\pm2$. Next, we construct an ansatz of SVHPLs whose series expansion we attempt to match to the above expression. We expect the result to be a weight-three SVHPL with parity $(+,+)$ under conjugation and inversion. Including zeta values, there are five functions satisfying these criteria, and we can write the ansatz as,
\beq
\mathcal{I}[\Phi_{\textrm{Reg}}^{(1)}(\nu,n)] =c_1\, \LThreePP + c_2\,\LZeroM{2}\LOnePP + c_3\,\LOneP{3} + c_4\,\zeta_2\,\LOnePP +c_5\, \zeta_3\,.
\eeq
Using the series expansions of the constituent HPLs~\eqref{eq:HPL_series}, it is straightforward to produce the series expansion of this ansatz,
\beq\bsp\label{eq:Phi_ser}
\mathcal{I}[\Phi_{\textrm{Reg}}^{(1)}(\nu,n)]
&=\left(\frac{c_1}{12}+\frac{c_2}{2}+\frac{c_3}{8}\right)\log^3|w|^2 
+ \frac{1}{2}c_4\zeta_2\log|w|^2 + c_5\,\zeta_3  
+ 3\,c_3\left( \log|w|^2\right)|w|^2 + \ldots \\
&+ (w+\ws) \left[ - \zeta_2 c_4
  + \left(-c_1+\frac{1}{2}c_1\,\log|w|^2\right)
  + \left(-\frac{c_1}{4}-c_2-\frac{3c_3}{4}\right)\,\log^2|w|^2
  + \ldots\right]  \\
& + \ldots\, .
\esp\eeq
We have only listed the terms necessary to fix the undetermined constants. In practice we generate many more terms than necessary to cross-check the result. Consistency of eqs.~\eqref{eq:I[Phi]_ser} and~\eqref{eq:Phi_ser} requires,
\beq
c_1=-4,\quad c_2=\frac{3}{4},\quad c_3=-\frac{1}{3},\quad c_4=-2,\quad c_5=0\,,
\eeq
which gives,
\beq
\mathcal{I}[\Phi_{\textrm{Reg}}^{(1)}(\nu,n)] = -4\, \LThreePP + \frac{3}{4}\,\LZeroM{2}\LOnePP -\frac{1}{3}\,\LOneP{3} -2\,\zeta_2\,\LOnePP\,.
\eeq
Finally, putting everything together in eq.~\eqref{eq:R62_gh},
\beq
g_0^{(2)}(w,\ws) \,=-L_3^{+}+{1\over 6}\left[L_1^+\right]^3+{1\over 8}\LZeroM{2}\,L_1^+\, .
\eeq
This completes the two-loop calculation, and we find agreement with~\cite{Lipatov2010qg,Lipatov2010ad}. Moving on to three loops, we can proceed in exactly the same way, and we reproduce the LLA~\cite{Lipatov2010ad} and NLLA results~\cite{Dixon2011pw,Fadin2011we} for the imaginary parts of the coefficient functions,
\beq\bsp
g_2^{(3)}(w,\ws) \,&=-{1\over 8}L_3^{+}+{1\over 12}\left[L_1^+\right]^3\,,\\
g_1^{(3)}(w,\ws) \,&={1\over8}\LZeroMM\,L_{2,1}^- - {5\over8}L_1^+\,L_{3}^+
+{5\over48}[L_1^+]^4+{1\over16}[\LZeroMM]^2\,[L_1^+]^2-{5\over 768}\LZeroM{4}\\
&\,-{\pi^2\over12}[L_1^+]^2+{\pi^2\over48}\LZeroM{2}+{1\over 4}\zeta_3\,L_1^+\,.
\esp\eeq
(The result for $g_1^{(3)}$ agrees with that in ref.~\cite{Dixon2011pw}
once the constants are fixed to $c=0$ and
$\gamma' = -9/2$~\cite{Fadin2011we}.)
The real parts are given by,
\beq\bsp
h_2^{(3)}(w,\ws) \,&=0\,,\\
h_1^{(3)}(w,\ws) \,&=-{1\over8}L_3^+-{1\over24}[L_1^+]^3+{1\over32}\LZeroM{2}\,L_1^+\,,
\esp\eeq
in agreement with ref.~\cite{Lipatov2010ad}. Using the fact that 
\beq
L_1^+ = \frac{1}{2}\log{|w|^2\over|1+w|^4}\,,
\eeq
it is easy to check that $h_1^{(3)}(w,\ws)$ satisfies eq.~\eqref{eq:htog} for $\ell=3$.

\begin{figure}[t]
\centering
\includegraphics[height=100mm]{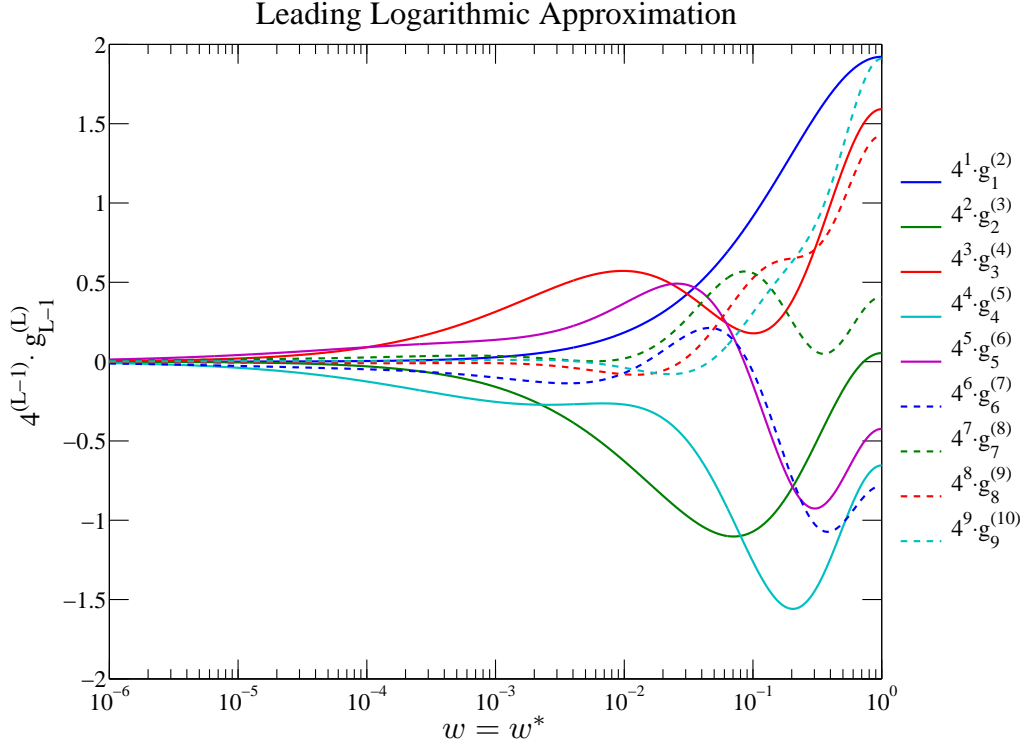}
\caption{Imaginary parts $g_{\ell-1}^{(\ell)}$ of the MHV remainder function
in MRK and LLA through 10 loops, on the line segment with $w=\ws$ running 
from 0 to 1.  The functions have been rescaled by powers of 4 so that they
are all roughly the same size.}
\label{fig:LLA_fig}
\end{figure}

\begin{figure}[t]
\centering
\includegraphics[height=100mm]{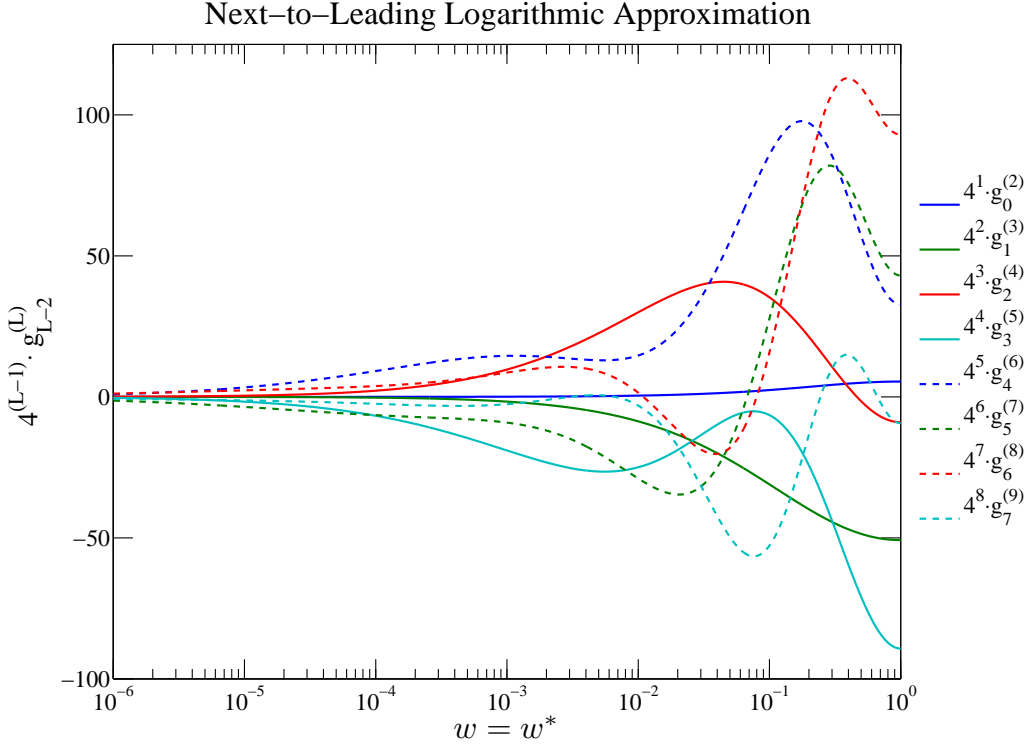}
\caption{Imaginary parts $g_{\ell-2}^{(\ell)}$ of the MHV remainder function
in MRK and NLLA through 9 loops.}
\label{fig:NLLA_fig}
\end{figure}

It is straightforward to extend these methods to higher loops. We
have produced results for all functions with weight less than or equal to
10, which is equivalent to 10 loops in the LLA, and 9 loops in
the NLLA.  Using the \CC~symbolic computation framework
GiNaC~\cite{Bauer2000cp}, which allows for the efficient numerical
evaluation of HPLs to high precision~\cite{Vollinga2004sn}, we can
evaluate these functions numerically. Figures~\ref{fig:LLA_fig}
and~\ref{fig:NLLA_fig} show the
functions plotted on the line segment for which $w=\ws$ and
$0<w<1$. Here we also show the analytical results through six loops.
We provide a separate computer-readable text file, compatible with
the {\tt Mathematica} package {\tt HPL}~\cite{Maitre2005uu,Maitre2007kp},
which contains all the expressions through weight 10.


Up to six loops, we find,
\begin{eqnarray}
g_3^{(4)}(w,\ws) &=&  \frac{1}{48}\,\LMA{2}{2}+\frac{1}{48}\,\LMA{0}{2}\,\LPA{1}{2}+\frac{7}{2304}\,\LMA{0}{4}+\frac{1}{48}\,\LPA{1}{4}-\frac{1}{16}\,\LSMA{0}\,\LSMB{2}{1}\\
\nonumber&&-\frac{5}{48}\,\LSPA{1}\,\LSPA{3}-\frac{1}{8}\,\LSPA{1}\,\zeta_3\,,\\
g_2^{(4)}(w,\ws) &=& \frac{3}{64}\,\LMA{0}{2}\,\LPA{1}{3}+\frac{1}{128}\,\LSPA{1}\,\LMA{0}{4}-\frac{3}{32}\,\LSPA{3}\,\LMA{0}{2}+\frac{1}{8}\,\LMA{0}{2}\,\zeta_3\\
\nonumber&&-\frac{1}{8}\,\LPA{1}{2}\,\zeta_3+\frac{3}{80}\,\LPA{1}{5}-\frac{\pi ^2}{24}\,\LPA{1}{3}-\frac{1}{16}\,\LSMA{0}\,\LSMB{2}{1}\,\LSPA{1}+\frac{13}{16}\,\LSPA{5}\\
\nonumber&&+\frac{3}{8}\,\LSPC{3}{1}{1}+\frac{1}{4}\,\LSPC{2}{2}{1}-\frac{5}{16}\,\LSPA{3}\,\LPA{1}{2}+\frac{\pi ^2}{16}\,\LSPA{3}\,,
\end{eqnarray}
\begin{eqnarray}
g_4^{(5)}(w,\ws) &=&\frac{1}{96}\,\LMA{0}{2}\,\LPA{1}{3}+\frac{17}{9216}\,\LSPA{1}\,\LMA{0}{4}-\frac{5}{384}\,\LSPA{3}\,\LMA{0}{2}+\frac{1}{24}\,\LMA{0}{2}\,\zeta_3\\
\nonumber&&-\frac{1}{12}\,\LPA{1}{2}\,\zeta_3+\frac{1}{240}\,\LPA{1}{5}-\frac{1}{24}\,\LSMA{0}\,\LSMB{2}{1}\,\LSPA{1}+\frac{43}{384}\,\LSPA{5}+\frac{1}{8}\,\LSPC{3}{1}{1}+\frac{1}{12}\,\LSPC{2}{2}{1}\\
\nonumber&&-\frac{1}{24}\,\LSPA{3}\,\LPA{1}{2}\,,\\
g_3^{(5)}(w,\ws) &=& -\frac{1}{384}\,\LMA{2}{2}\,\LMA{0}{2}+\frac{5}{64}\,\LMA{2}{2}\,\LPA{1}{2}-\frac{\pi ^2}{72}\,\LMA{2}{2}+\frac{1}{384}\,\LMA{0}{4}\,\LPA{1}{2}-\frac{7}{48}\,\zeta_3^2\\
\nonumber&&+\frac{5}{144}\,\LMA{0}{2}\,\LPA{1}{4}-\frac{\pi ^2}{72}\,\LMA{0}{2}\,\LPA{1}{2}-\frac{31}{1152}\,\LSMB{2}{1}\,\LMA{0}{3}-\frac{11}{384}\,\LSPA{1}\,\LSPA{3}\,\LMA{0}{2}\\
\nonumber&&-\frac{7}{48}\,\LSPA{1}\,\LMA{0}{2}\,\zeta_3+\frac{31}{69120}\,\LMA{0}{6}-\frac{7 \pi ^2}{3456}\,\LMA{0}{4}+\frac{7}{48}\,\LMB{2}{1}{2}-\frac{31}{192}\,\LSMA{0}\,\LSMB{2}{1}\,\LPA{1}{2}\\
\nonumber&&-\frac{65}{576}\,\LSPA{3}\,\LPA{1}{3}-\frac{13}{96}\,\LPA{1}{3}\,\zeta_3+\frac{7}{720}\,\LPA{1}{6}-\frac{\pi ^2}{72}\,\LPA{1}{4}+\frac{1}{48}\,\LPA{3}{2}+\frac{5}{96}\,\LSMA{4}\,\LSMA{2}\\
\nonumber&&-\frac{7}{24}\,\LSMA{2}\,\LSMC{2}{1}{1}+\frac{1}{192}\,\LSMA{0}\,\LSMB{4}{1}+\frac{1}{16}\,\LSMA{0}\,\LSMB{3}{2}+\frac{\pi ^2}{24}\,\LSMA{0}\,\LSMB{2}{1}+\frac{9}{16}\,\LSMA{0}\,\LSMD{2}{1}{1}{1}\\
\nonumber&&+\frac{33}{64}\,\LSPA{5}\,\LSPA{1}+\frac{5 \pi ^2}{72}\,\LSPA{1}\,\LSPA{3}-\frac{7}{48}\,\LSPA{1}\,\LSPC{3}{1}{1}+\frac{25}{32}\,\LSPA{1}\,\zeta_5+\frac{\pi ^2}{12}\,\LSPA{1}\,\zeta_3-\frac{5}{32}\,\LSPA{3}\,\zeta_3\,,
\end{eqnarray}
\begin{eqnarray}
g_5^{(6)}(w,\ws) &=& \frac{103}{15360}\,\LMA{2}{2}\,\LMA{0}{2}-\frac{1}{64}\,\LMA{2}{2}\,\LPA{1}{2}+\frac{1}{576}\,\LMA{0}{2}\,\LPA{1}{4}+\frac{1}{720}\,\LMA{0}{4}\,\LPA{1}{2}\\
\nonumber&&+\frac{29}{9216}\,\LSMB{2}{1}\,\LMA{0}{3}-\frac{77}{5120}\,\LSPA{1}\,\LSPA{3}\,\LMA{0}{2}+\frac{29}{512}\,\LSPA{1}\,\LMA{0}{2}\,\zeta_3+\frac{73}{1382400}\,\LMA{0}{6}\\
\nonumber&&-\frac{1}{48}\,\LMB{2}{1}{2}-\frac{1}{192}\,\LSMA{0}\,\LSMB{2}{1}\,\LPA{1}{2}-\frac{7}{576}\,\LSPA{3}\,\LPA{1}{3}-\frac{1}{32}\,\LPA{1}{3}\,\zeta_3+\frac{1}{1440}\,\LPA{1}{6}\\
\nonumber&&+\frac{43}{3840}\,\LPA{3}{2}-\frac{29}{960}\,\LSMA{4}\,\LSMA{2}+\frac{1}{24}\,\LSMA{2}\,\LSMC{2}{1}{1}-\frac{25}{768}\,\LSMA{0}\,\LSMB{4}{1}-\frac{3}{128}\,\LSMA{0}\,\LSMB{3}{2}\\
\nonumber&&-\frac{1}{16}\,\LSMA{0}\,\LSMD{2}{1}{1}{1}+\frac{301}{3840}\,\LSPA{5}\,\LSPA{1}+\frac{7}{48}\,\LSPA{1}\,\LSPC{3}{1}{1}+\frac{1}{12}\,\LSPA{1}\,\LSPC{2}{2}{1}-\frac{3}{128}\,\LSPA{1}\,\zeta_5\\
\nonumber&&+\frac{3}{128}\,\LSPA{3}\,\zeta_3+\frac{1}{48}\,\zeta_3^2\,,\\
g_4^{(6)}(w,\ws) &=&  \frac{5}{1536}\,\LSPA{1}\,\LMA{2}{2}\,\LMA{0}{2}+\frac{1}{48}\,\LMA{2}{2}\,\LPA{1}{3}-\frac{1}{48}\,\LMA{2}{2}\,\zeta_3-\frac{101}{3072}\,\LSPA{3}\,\LMA{0}{2}\,\LPA{1}{2}\\
\nonumber&&+\frac{89}{1536}\,\LMA{0}{2}\,\LPA{1}{2}\,\zeta_3+\frac{59}{5760}\,\LMA{0}{2}\,\LPA{1}{5}+\frac{85}{18432}\,\LMA{0}{4}\,\LPA{1}{3}-\frac{5 \pi ^2}{576}\,\LMA{0}{2}\,\LPA{1}{3}\\
\nonumber&&-\frac{317}{9216}\,\LSMB{2}{1}\,\LSPA{1}\,\LMA{0}{3}-\frac{43}{768}\,\LSPA{5}\,\LMA{0}{2}+\frac{77}{221184}\,\LSPA{1}\,\LMA{0}{6}-\frac{85 \pi ^2}{55296}\,\LSPA{1}\,\LMA{0}{4}\\
\nonumber&&+\frac{65}{9216}\,\LSPA{3}\,\LMA{0}{4}+\frac{25 \pi ^2}{2304}\,\LSPA{3}\,\LMA{0}{2}-\frac{1}{128}\,\LSPC{2}{2}{1}\,\LMA{0}{2}+\frac{1}{768}\,\LMA{0}{4}\,\zeta_3\\
\nonumber&&
-\frac{17}{192}\,\LMA{0}{2}\,\zeta_5-\frac{5 \pi ^2}{144}\,\LMA{0}{2}\,\zeta_3-\frac{1}{24}\,\LSPA{1}\,\LMB{2}{1}{2}-\frac{3}{64}\,\LSMA{0}\,\LSMB{2}{1}\,\LPA{1}{3}+\frac{205}{768}\,\LSPA{5}\,\LPA{1}{2}\\
\nonumber&&-\frac{17}{576}\,\LSPA{3}\,\LPA{1}{4}+\frac{5 \pi ^2}{144}\,\LSPA{3}\,\LPA{1}{2}-\frac{1}{48}\,\LSPC{3}{1}{1}\,\LPA{1}{2}+\frac{1}{24}\,\LSPC{2}{2}{1}\,\LPA{1}{2}-\frac{7}{96}\,\LPA{1}{4}\,\zeta_3\\
\nonumber&&+\frac{65}{128}\,\LPA{1}{2}\,\zeta_5+\frac{5 \pi ^2}{72}\,\LPA{1}{2}\,\zeta_3+\frac{1}{504}\,\LPA{1}{7}-\frac{\pi ^2}{288}\,\LPA{1}{5}+\frac{7}{192}\,\LSPA{1}\,\LPA{3}{2}\\
\nonumber&&-\frac{5}{192}\,\LSMA{4}\,\LSMA{2}\,\LSPA{1}+\frac{11}{192}\,\LSMA{2}\,\LSMA{0}\,\LSPB{3}{1}-\frac{1}{6}\,\LSMA{2}\,\LSMC{2}{1}{1}\,\LSPA{1}-\frac{5}{768}\,\LSMA{0}\,\LSMB{4}{1}\,\LSPA{1}\\
\nonumber&&-\frac{13}{384}\,\LSMA{0}\,\LSMB{3}{2}\,\LSPA{1}+\frac{5 \pi ^2}{144}\,\LSMA{0}\,\LSMB{2}{1}\,\LSPA{1}+\frac{23}{384}\,\LSMA{0}\,\LSMB{2}{1}\,\LSPA{3}-\frac{21}{64}\,\LSMA{0}\,\LSMB{2}{1}\,\zeta_3\\
\nonumber&&+\frac{3}{16}\,\LSMA{0}\,\LSMD{2}{1}{1}{1}\,\LSPA{1}-\frac{215 \pi ^2}{2304}\,\LSPA{5}-\frac{29}{384}\,\LSPA{1}\,\LSPA{3}\,\zeta_3-\frac{19}{192}\,\LSPA{1}\,\zeta_3^2+\frac{1}{16}\,\LSPA{7}\\
\nonumber&&-\frac{151}{128}\,\LSPC{5}{1}{1}-\frac{3}{32}\,\LSPC{4}{1}{2}-\frac{27}{64}\,\LSPC{4}{2}{1}-\frac{5 \pi ^2}{48}\,\LSPC{3}{1}{1}-\frac{7}{64}\,\LSPC{3}{3}{1}-\frac{5 \pi ^2}{72}\,\LSPC{2}{2}{1}\\
\nonumber&&+\frac{13}{4}\,\LSPE{3}{1}{1}{1}{1}+\frac{1}{2}\,\LSPE{2}{1}{2}{1}{1}+\frac{3}{2}\,\LSPE{2}{2}{1}{1}{1}\,.
\end{eqnarray}

We present only the imaginary parts, as the real parts are determined
by eq.~\eqref{eq:htog}. However, as a cross-check of our result, we
computed the $h_n^{(\ell)}$ explicitly and checked that
eq.~\eqref{eq:htog} is satisfied. Furthermore, we checked that in the
collinear limit $w\to0$ our results agree with the all-loop prediction
for the six-point MHV amplitude in the double-leading-logarithmic
(DLL) and next-to-double-leading-logarithmic (NDLL) approximations of
ref.~\cite{Bartels2011xy},
\beq\bsp
e^{R_{\textrm{DLLA}}} \,&= i\pi\,a\,(w+\ws)
\,\left[1-I_0\left(2\sqrt{a\log|w|\log(1-u_1)}\right)\right]\,,\\
\textrm{Re}\left(e^{R_{\textrm{NDLLA}}}\right) \,&
= 1 \, + \, \pi^2a^{3/2}(w+\ws)\,\sqrt{\log|w|}\,
{I_1\left(2\sqrt{a\log|w|\log(1-u_1)}\right)\over\sqrt{\log(1-u_1)}}\\
&\,- \pi^2a^2 \, (w+\ws) \, \log|w| \, 
I_0\left(2\sqrt{a\log|w|\log(1-u_1)}\right) \,,
\esp\eeq
where $I_0(z)$ and $I_1(z)$ denote modified Bessel functions.

Let us conclude this section with an observation: All the results for
the six-point remainder function that we computed only involve
ordinary $\zeta$ values of depth one ($\zeta_k$ for some $k$), 
despite the fact that multiple $\zeta$ values are expected to appear
starting from weight eight. In addition, the LLA results only involve
odd $\zeta$ values -- even $\zeta$ values never appear.


\section{The six-point NMHV amplitude in MRK}
\label{sec:NMHV}
So far we have only discussed the multi-Regge limit of the six-point amplitude in an MHV helicity configuration. In this section we extend the discussion to the second independent helicity configuration for six points, the NMHV configuration.  We will see that the SVHPLs provide the natural function space for describing this case as well.

The NMHV case was recently analyzed in the LLA~\cite{Lipatov2012gk}.
It was shown that the two-loop expression agrees with the limit of the 
analytic formula for the NMHV amplitude for general
kinematics~\cite{Dixon2011nj}, and the three-loop result was also obtained.
Here we will extend these results to 10 loops.

Due to helicity conservation along the high-energy line, the only difference between the MHV and NMHV configurations is a flip in helicity of one of the lower energy external gluons (labeled by 4 and 5).  Instead of the MHV helicity configuration $({+}{+}{-}{+}{+}{-})$, we consider $({+}{+}{-}{-}{+}{-})$.
The tree amplitudes for MHV and NMHV become identical in
MRK~\cite{Lipatov2012gk}.  In this limit, we can define the NMHV remainder
function $R_{\textrm{NMHV}}$ in the same way as in the MHV case~(\ref{eq:Rndef}),
\be\label{eq:RdefNMHV}
A_6^{\textrm{NMHV}}|_{\textrm{MRK}} = A_6^{\textrm{BDS}} \times \exp(R_{\textrm{NMHV}}) \,.
\ee
Recall the LLA version\footnote{The distinction between $R$ and $\exp(R)$
is irrelevant at LLA, because the LLA has one fewer logarithm than the
loop order, so the square of an LL term has two fewer logarithms and is NLL.}
of eq.~\eqref{eq:MHV_MRK}:
\beq\label{eq:MHV_MRK_LLA}
R_{\textrm{MHV}}^{\textrm{LLA}} =  i{a\over 2}
\sum_{n=-\infty}^\infty(-1)^n\,
\int_{-\infty}^{+\infty}
{d\nu \, w^{i\nu+n/2} \, \ws^{i\nu-n/2} \over (i\nu+{n\over 2})(-i\nu+{n\over 2})}
\, \Bigl[ (1-u_1)^{a\,E_{\nu,n}} - 1 \Bigr] \,.
\eeq
At LLA, the effect of changing the impact factor for emitting gluon 4
with positive helicity to the one for a negative-helicity emission is simply
to perform the replacement
\beq
{1\over-i\nu+{n\over2}}\to-{1\over i\nu+{n\over2}}
\eeq
in eq.~\eqref{eq:MHV_MRK_LLA}, obtaining~\cite{Lipatov2012gk}
\beq\label{eq:NMHV_MRK}
R_{\textrm{NMHV}}^{\textrm{LLA}} \simeq -{ia\over2}\,\sum_{n=-\infty}^\infty(-1)^n
\,\int_{-\infty}^{+\infty}
{d\nu  \, w^{i\nu+n/2} \, \ws^{i\nu-n/2} \over (i\nu+{n\over 2})^2}
\, \Bigl[ (1-u_1)^{a\,E_{\nu,n}} - 1 \Bigr] \,.
\eeq
The NMHV ratio function is normally defined in terms of 
the ratio of NMHV to MHV superamplitudes ${\cal A}$,
\be\label{ratiogeneral}
\mathcal{P}_{\textrm{NMHV}} = 
\frac{{\cal A}_{\textrm{NMHV}}}{{\cal A}_{\textrm{MHV}}} \,.
\ee
However, in MRK, because the tree amplitudes become identical, it suffices
to consider the ordinary ratio, which in LLA becomes
\be\label{ratiomrk}
\mathcal{P}_{\textrm{NMHV}}^{\textrm{LLA}} = 
\frac{A_{\textrm{NMHV}}^{\textrm{LLA}}}{A_{\textrm{MHV}}^{\textrm{LLA}}} 
= \exp(R_{\textrm{NMHV}}^{\textrm{LLA}} - R_{\textrm{MHV}}^{\textrm{LLA}})\,.
\ee
Thus eq.~\eqref{eq:NMHV_MRK}, together with eq.~\eqref{eq:MHV_MRK_LLA},
is sufficient to generate both the remainder function and the ratio function
in LLA.

Comparing eq.~\eqref{eq:NMHV_MRK} to eq.~\eqref{eq:MHV_MRK}, we see that in MRK the MHV and NMHV remainder functions are related by
\beq\label{eq:NMHV_integral}
R_{\textrm{NMHV}}^{\textrm{LLA}} = 
\int dw\,{\ws\over w}\,{\partial\over\partial \ws }R_{\textrm{MHV}}^{\textrm{LLA}}\,.
\eeq
It is convenient to write this equation slightly differently. First, define a sequence of single-valued functions $f^{(l)}(w,\ws)$ in analogy with eq.~\eqref{eq:R6_MRK}\footnote{Ref.~\cite{Lipatov2012gk} defines a similar set of functions,
$f_l$, which are related to ours by $f_2=-\frac{1}{4} f^{(2)}$,
$f_3=\frac{1}{8} f^{(3)}$, etc.}
\beq
R_{\textrm{NMHV}}^{\textrm{LLA}} = 2\pi i\,\sum_{l=2}^\infty a^l \log^{l-1}(1-u_1)\left[\frac{1}{1+\ws}f^{(l)}(w,\ws) + \frac{\ws}{1+\ws} f^{(l)}\Big(\frac{1}{w},\frac{1}{\ws}\Big)\right].
\eeq
Then eq.~\eqref{eq:NMHV_integral} can be used to relate $f^{(l)}(w,\ws)$ to $g^{(l)}_{l-1}(w,\ws)$,
\beq\label{eq:g_int_f}
\int dw\,{\ws\over w}\,{\partial\over\partial \ws }g^{(l)}_{l-1}(w,\ws) = 
\frac{1}{1+\ws}f^{(l)}(w,\ws)
+ \frac{\ws}{1+\ws} f^{(l)}\Big(\frac{1}{w},\frac{1}{\ws}\Big)\,.
\eeq
In Section~\ref{sec:R6_LLA_NLLA} we computed the MHV remainder function in the LLA in the multi-Regge limit up to ten loops. Using these results and eq.~\eqref{eq:g_int_f}, we can immediately obtain NMHV expressions through ten loops as well. Indeed, $g^{(l)}_{l-1}(w,\ws)$ is a sum of SVHPLs, so the differentiation $\frac{\partial}{\partial \ws}$ can be performed with the aid of eq.~\eqref{eq:dLzz}. The result is again a sum of SVHPLs with rational coefficients $1/(1+\ws)$ and $\ws/(1+\ws)$. As such, the differential equations~\eqref{eq:dLzz} also uniquely determine the result of the $w$-integral as a sum of SVHPLs, up to an undetermined function $F(\ws)$. This function can be at most a constant in order to preserve the single-valuedness condition. It turns out that to respect the vanishing of the remainder function in the collinear limit, $F(\ws)$ must actually be zero.

To see how this works, consider the two loop case. From eq.~\eqref{eq:g21},
\beq
g_1^{(2)}(w,\ws) \,= {1\over4}\,\LOneP{2}  -{1\over16}\, \LZeroM{2}
 = \frac{1}{2}\cL_{1,1} + \frac{1}{4} \cL_{0,1}+\frac{1}{4}\cL_{1,0}.
\eeq
Recalling that $(w,\ws) = (-z,-\bar{z})$, first use the second
eq.~\eqref{eq:dLzz} to take the $\ws$ derivative, which clips off the last
index in the SVHPL, with a different prefactor depending on whether 
it is a `0' or a `1' (and with corrections due to the $y$ alphabet
at higher weights):
\beq\bsp
\ws \frac{\partial}{\partial \ws} g_1^{(2)}(w,\ws) &= 
-\frac{1}{2}\left(\frac{\ws}{1+\ws}\right)\cL_{1}
-\frac{1}{4}\left(\frac{\ws}{1+\ws}\right)\cL_{0}
+\frac{1}{4}\cL_{1}\\
&= \frac{\ws}{1+\ws}\left[-\frac{1}{4}\,\cL_{1}
-\frac{1}{4}\,\cL_{0}\right]+\frac{1}{1+\ws}\left[\frac{1}{4}\,\cL_{1}\right]\,.
\esp\eeq
Next, use the first eq.~\eqref{eq:dLzz} to perform the $w$-integration.
In practice, this amounts to prepending a `0' to the weight vector of
each SVHPL,
\beq\bsp\label{eq:identify_f2}
\int dw\,{\ws\over w}\,{\partial\over\partial \ws }g^{(2)}_{1} & = 
\frac{\ws}{1+\ws}\left[-\frac{1}{4}\,\cL_{0,1}-\frac{1}{4}\,\cL_{0,0}\right]
+\frac{1}{1+\ws}\left[\frac{1}{4}\,\cL_{0,1}\right]\\
&= \frac{1}{1+\ws}f^{(2)}(w,\ws)
 + \frac{\ws}{1+\ws} f^{(2)}\Big(\frac{1}{w},\frac{1}{\ws}\Big)\,,
\esp\eeq
where
\beq\bsp
f^{(2)}(w,\ws)&= \frac{1}{4}\,\cL_{0,1}\\
&= \frac{1}{4}\,L_2 + \frac{1}{8}\,L_0\,L_1\\
&= - \frac{1}{4} \left(\log|w|^2\log(1+\ws) - \Li_2(-w)+\Li_2(-\ws)\right)\,.
\esp\eeq
This result agrees with the one presented in ref.~\cite{Lipatov2012gk}.
Furthermore, we can check that the inversion property implicit
in eq.~\eqref{eq:identify_f2} is satisfied,
\beq\bsp
f^{(2)}\Big(\frac{1}{w},\frac{1}{\ws}\Big)&=
-\frac{1}{4}\left[-\log|w|^2\log\left(1+\frac{1}{\ws}\right)
  - \Li_2\left(-\frac{1}{w}\right)+\Li_2\left(-\frac{1}{\ws}\right)\right]\\
&=-\frac{1}{4}\left[\frac{1}{2}\log^2|w|^2-\log|w|^2\log(1+\ws)
  +\Li_2(-w)-\Li_2(-\ws)\right]\\
&= -\frac{1}{4}\,\cL_{0,1}-\frac{1}{4}\,\cL_{0,0} \, .
\esp\eeq
Moving on to three loops, we start with the MHV LLA term,
\beq\bsp
g_2^{(3)}(w,\ws) \,&=-{1\over 8}L_3^{+}+{1\over 12}\left[L_1^+\right]^3 \\
&=\frac{1}{16}\cL_{0,0,1}+\frac{1}{8}\cL_{0,1,0}+\frac{1}{4}\cL_{0,1,1}
+\frac{1}{16}\cL_{1,0,0}+\frac{1}{4}\cL_{1,0,1}+\frac{1}{4}\cL_{1,1,0}
+\frac{1}{2}\cL_{1,1,1}\, .
\esp\eeq
As before, we can take derivatives and integrate using eq.~\eqref{eq:dLzz},
\beq\bsp
\int dw\,{\ws\over w}\,{\partial\over\partial \ws }g^{(3)}_{2} &
= \frac{\ws}{1+\ws}\left[-\frac{1}{16}\,\cL_{0,0,0}-\frac{1}{8}\cL_{0,0,1}
           -\frac{3}{16}\cL_{0,1,0}-\frac{1}{4}\cL_{0,1,1}\right] \\
&\quad+\frac{1}{1+\ws}\left[\frac{1}{8}\,\cL_{0,0,1}+\frac{1}{16}\,\cL_{0,1,0}
                +\frac{1}{4}\, \cL_{0,1,1}\right]\,,
\esp\eeq
and we find,
\beq\bsp
f^{(3)}(w,\ws)&=\frac{1}{8}\,\cL_{0,0,1}+\frac{1}{16}\,\cL_{0,1,0}
+\frac{1}{4}\, \cL_{0,1,1}\\
&=\frac{1}{4}\,L_{2,1}+\frac{1}{8}\,L_1\,L_2+\frac{1}{16}\,L_0\,L_2
+\frac{1}{32}\,L_0^2\,L_1\\
&=\frac{1}{8}\,\Big[-\frac{1}{2}\log^2|w|^2\log(1+\ws)
 + \log(-w)\Big(\log^2(1+\ws)-\log^2(1+w)\Big) \\
&\quad\quad\quad+2\,\zeta_2\log|1+w|^2
+\frac{1}{2}\log|w|^2\,\Big(\,\Li_2(-w)-\Li_2(-\ws)\,\Big)\\ 
&\quad\quad\quad-2\log|1+w|^2\,\Li_2(-w)-2\,\Li_3(1+w)-2\,\Li_3(1+\ws)
+4\,\zeta_3\;\Big]\,.
\esp\eeq
The last form agrees with the one given in ref.~\cite{Lipatov2012gk},
up to the sign of the second term, which we find must be 
$+1$ for the function to be single-valued.

Continuing on to higher loops, we find,
\begin{align}
f^{(4)}(w,\ws)
&=-\frac{1}{8}\,L_1\,\zeta_3+\frac{1}{4}\,L_{2,1,1}-\frac{1}{8}\,L_{3,1}+\frac{1}{32}\,L_2^2-\frac{1}{32}\,L_4+\frac{1}{8}\,L_1\,L_{2,1}-\frac{1}{16}\,L_1\,L_3\\
\nonumber&-\frac{1}{96}\,L_0\,L_1^3+\frac{1}{96}\,L_0^2\,L_2-\frac{1}{192}\,L_0\,L_3+\frac{1}{256}\,L_0^3\,L_1+\frac{3}{128}\,L_0^2\,L_1^2\\
\nonumber&+\frac{1}{16}\,L_0\,L_1\,L_2\,,\\
f^{(5)}(w,\ws)
&=
-\frac{1}{96}\,L_2\,\zeta_3-\frac{1}{24}\,L_0\,L_1\,\zeta_3+\frac{1}{4}\,L_{2,1,1,1}-\frac{1}{8}\,L_{2,2,1}+\frac{1}{32}\,L_{4,1}+\frac{1}{48}\,L_{3,2}\\
\nonumber&+\frac{1}{8}\,L_1\,L_{2,1,1}+\frac{1}{16}\,L_0\,L_{2,1,1}-\frac{1}{16}\,L_1\,L_{3,1}+\frac{1}{32}\,L_1\,L_2^2-\frac{1}{64}\,L_1\,L_4-\frac{1}{96}\,L_0^2\,L_{2,1}\\
\nonumber&-\frac{1}{96}\,L_1^3\,L_2+\frac{1}{192}\,L_0\,L_2^2-\frac{1}{256}\,L_0\,L_4-\frac{1}{384}\,L_0^2\,L_1^3+\frac{1}{1152}\,L_0^3\,L_2-\frac{1}{1536}\,L_0^2\,L_3\\
\nonumber&+\frac{5}{768}\,L_0^3\,L_1^2+\frac{5}{18432}\,L_0^4\,L_1-\frac{7}{192}\,L_0\,L_{3,1}+\frac{1}{16}\,L_0\,L_1\,L_{2,1}-\frac{1}{48}\,L_0\,L_1\,L_3\\
\nonumber&+\frac{1}{64}\,L_0\,L_1^2\,L_2+\frac{11}{768}\,L_0^2\,L_1\,L_2-\frac{3}{8}\,L_{3,1,1}\,,\\
f^{(6)}(w,\ws)
=&\frac{1}{4}\,L_{2,1,1,1,1}-\frac{1}{8}\,L_{3,1,1,1}+\frac{1}{12}\,L_{3,2,1}-\frac{1}{32}\,L_{2,1}^2+\frac{1}{48}\,L_{5,1}+\frac{1}{288}\,L_2^3+\frac{1}{384}\,L_3^2\\
\nonumber&+\frac{1}{768}\,L_6-\frac{1}{768}\,L_{4,2}+\frac{7}{32}\,L_{4,1,1}+\frac{1}{8}\,L_1\,L_{2,1,1,1}-\frac{1}{16}\,L_1\,L_{3,1,1}+\frac{1}{16}\,L_2\,L_{2,1,1}\\
\nonumber&+\frac{1}{24}\,L_1\,L_{3,2}+\frac{1}{32}\,L_3\,L_{2,1}-\frac{1}{32}\,L_2\,L_{3,1}+\frac{1}{96}\,L_0^2\,L_{2,1,1}-\frac{1}{96}\,L_1^3\,L_{2,1}+\frac{1}{96}\,L_1^3\,\zeta_3\\
\nonumber&-\frac{1}{128}\,L_1^2\,L_2^2-\frac{1}{192}\,L_0\,L_{3,1,1}-\frac{1}{192}\,L_1\,\zeta_5+\frac{1}{192}\,L_1^3\,L_3-\frac{1}{256}\,L_0^2\,L_1^4+\frac{1}{384}\,L_3\,\zeta_3\\
\nonumber&-\frac{1}{512}\,L_0\,L_{3,2}-\frac{1}{768}\,L_0\,L_{4,1}+\frac{1}{960}\,L_0\,L_1^5-\frac{1}{2560}\,L_0^2\,L_4+\frac{1}{7680}\,L_0\,L_5\\
\nonumber&-\frac{1}{18432}\,L_0^3\,L_3+\frac{1}{73728}\,L_0^5\,L_1+\frac{5}{96}\,L_{2,1}\,\zeta_3+\frac{5}{384}\,L_1\,L_5+\frac{5}{2048}\,L_0^2\,L_2^2\\
\nonumber&+\frac{5}{4096}\,L_0^4\,L_1^2+\frac{7}{64}\,L_1\,L_{4,1}+\frac{7}{1536}\,L_0^3\,L_1^3-\frac{11}{1536}\,L_0^2\,L_{3,1}-\frac{11}{1536}\,L_2\,L_4\\
\nonumber&+\frac{11}{184320}\,L_0^4\,L_2-\frac{19}{9216}\,L_0^3\,L_{2,1}+\frac{1}{16}\,L_0\,L_1\,L_{2,1,1}-\frac{1}{24}\,L_1\,L_2\,\zeta_3\\
\nonumber&-\frac{1}{32}\,L_0\,L_1\,L_{3,1}+\frac{1}{32}\,L_0\,L_1^2\,L_{2,1}-\frac{1}{48}\,L_0\,L_{2,1}\,L_2-\frac{1}{48}\,L_1\,L_3\,L_2\\
\nonumber&+\frac{1}{96}\,L_0^2\,L_1^2\,L_2-\frac{1}{192}\,L_0\,L_1^3\,L_2+\frac{1}{384}\,L_0\,L_1\,L_2^2-\frac{3}{256}\,L_0^2\,L_1\,L_{2,1}\\
\nonumber&-\frac{3}{512}\,L_0^2\,L_1\,\zeta_3-\frac{5}{96}\,L_0\,L_1^2\,\zeta_3-\frac{5}{768}\,L_0\,L_2\,\zeta_3-\frac{11}{1536}\,L_0\,L_1\,L_4\\
\nonumber&-\frac{11}{2048}\,L_0^2\,L_1\,L_3-\frac{19}{768}\,L_0\,L_1^2\,L_3+\frac{49}{18432}\,L_0^3\,L_1\,L_2\,.
\end{align}
The remaining expressions through 10 loops can be found in computer-readable 
format in a separate file attached to this article.


\section{Single-valued HPLs and Fourier-Mellin transforms}
\label{sec:nu_n}
\subsection{The multi-Regge limit in $(\nu,n)$ space}
So far we have only used the machinery of SVHPLs in order to obtain compact analytic expressions for
the six-point MHV amplitude in the LL and NLL approximation.  However, this was only possible because we knew \emph{a priori} the 
BFKL eigenvalues and the impact factor to the desired order in perturbation theory. Going beyond NLLA requires higher-order corrections 
to the BFKL eigenvalues and the impact factor which, by the same logic, can be computed if the corresponding amplitude is known. In other words, if we are given the functions $g_n^{(\ell)}(w,\ws)$ up to some loop order, we can use them to extract the corresponding impact factors and BFKL eigenvalues by transforming the expression from $(w,\ws)$ space back to $(\nu,n)$ space. The impact factors and BFKL eigenvalues obtained in this way can then be used to compute the six-point amplitude to any loop order for a given logarithmic accuracy.

In ref.~\cite{Dixon2011pw} the three-loop six point amplitude was computed up to next-to-next-to-leading logarithmic accuracy (NNLLA),
\beq\bsp\label{eq:g30}
g_0^{(3)}(w,\ws) &\,= \rat{27}{8}\,\LFivePP+\rat{3}{4}\,\LThreeOneOnePP-\rat{1}{2}\,\LThreePP\,\LOneP{2}-\rat{15}{32}\,\LThreePP\,\LZeroM{2}-\rat{1}{8}\,\LOnePP\,\LTwoOneMM\,\LZeroMM\\
&\,+\rat{3}{32}\,\LZeroM{2}\,\LOneP{3}+\rat{19}{384}\,\LOnePP\,\LZeroM{4}+\rat{3}{8}\,\LOneP{2}\,\zeta_3-\rat{5}{32}\,\LZeroM{2}\,\zeta_3+\rat{\pi ^2}{96}\,\LOneP{3}\\
&\,-\rat{\pi ^2}{384}\,\LOnePP\,\LZeroM{2}-\rat{3}{4}\,\zeta_5-{\pi^2\over 6}\,\gamma''\,\Big\{L_3^+-{1\over 6}\LOneP{3}-{1\over 8}\LZeroM{2}\,\LOnePP\Big\}\\
&\,+{1\over 4}\,d_1\,\zeta_3\Big\{[L_1^+]^2-{1\over 4}\LZeroM{2}\Big\}-{\pi^2\over 3}d_2\,L_1^+\Big\{[L_1^+]^2-{1\over 4}\LZeroM{2}\Big\}+\rat{1}{30}\,\LOneP{5}\,,\\
h_0^{(3)}(w,\ws) &\,=\rat{3}{16}\,\LOnePP\,\LThreePP+\rat{1}{16}\,\LTwoOneMM\,\LZeroMM-\rat{1}{32}\,\LOneP{4} -\rat{1}{32}\,\LZeroM{2}\,\LOneP{2}\\
&\,-\rat{5}{1536}\,\LZeroM{4}+\rat{1}{8}\,\LOnePP\,\zeta_3\,,
\esp\eeq
where $d_1$, $d_2$ and $\gamma''$ are some undetermined rational numbers. 
(To obtain eq.~\eqref{eq:g30} from ref.~\cite{Dixon2011pw}
one also needs the value for another constant, $\gamma^\prime=-9/2$,
or equivalently $\gamma^{\prime\prime\prime}=0$, which was obtained in
ref.~\cite{Fadin2011we} using the MRK limit at NLLA.)

These functions can be used to extract the NNLLA correction to the impact factor\footnote{In principle we should expect the amplitude to NNLLA to depend on both the NNLL impact factor and BFKL eigenvalue. The NNLL BFKL eigenvalue however only enters at four loops, see Section~\ref{sec:4loop}.}. Indeed, the NNLL impact factor has already been expressed~\cite{Fadin2011we} as an integral over the complex $w$ plane,
\beq\label{eq:FL_integral}
\Phi_{\text{Reg}}^{(2)}(\nu,n)=(-1)^n\,\left(\nu^2+{n^2\over4}\right)\,\int{d^2w\over \pi}\,\rho(w,\ws)\,|w|^{-2i\nu-2}\,\left({\ws\over w}\right)^{n\over 2}\,,
\eeq
where the kernel $\rho(w,\ws)$ is related to the three-loop amplitude in MRK,
\beq\bsp\label{eq:rho_def}
\rho(w,\ws) &\,= 2\left[g_0^{(3)}(w,\ws)+\log{|1+w|^2\over |w|}\,g_1^{(3)}(w,\ws)+\left(\log^2{|1+w|^2\over |w|}+\pi^2\right)\,g_2^{(3)}(w,\ws)\right]\\
&\,+\log{|1+w|^2\over |w|}\left(\zeta_2\,\log^2{|1+w|^2\over |w|}-{11\over2}\,\zeta_4\right)\,.
\esp\eeq
However, no analytic expression for $\Phi_{\text{Reg}}^{(2)}(\nu,n)$ is yet known.
Indeed, an explicit evaluation of the integral~\eqref{eq:FL_integral} would require a detailed study of the integrand's branch structure, a task which, if feasible in this case, does not seem particularly amenable to generalization.

Here we propose an alternative to evaluating the integral explicitly. The basic idea is to write down an ansatz for the function in $(\nu,n)$ space, and then perform the inverse transform to fix the unknown coefficients. The inverse transform is easly performed using the methods outlined in Section~\ref{sec:R6_LLA_NLLA}, so we are left only with the task of writing down a suitable ansatz. To be precise, consider the inverse Fourier-Mellin transform defined in eq.~\eqref{eq:integral_transform}. Our goal is to find a set of linearly independent functions $\{\mathcal{F}_i\}$ defined in $(\nu,n)$ space such that their transforms $\{\mathcal{I}[\mathcal{F}_i]\}$:
\begin{enumerate}
\item are combinations of HPLs of uniform weight,
\item are single-valued in the complex $w$ plane,
\item have a definite parity under $\mathbb{Z}_2\times\mathbb{Z}_2$ transformations in $(w,\ws)$ space,
\item span the whole space of SVHPLs.
\end{enumerate}
Through weight six, we find empirically that this problem has a unique solution, the construction of which we present in the remainder of this section. In particular, we will be led to extend the action of the $\mathbb{Z}_2\times\mathbb{Z}_2$ symmetry and the notion of uniform transcendentality to $(\nu,n)$ space.
\subsection{Symmetries in $(\nu,n)$ space}
Let us start by analyzing the $\mathbb{Z}_2\times\mathbb{Z}_2$ symmetry in $(\nu,n)$ space. It is easy to see from eq.~\eqref{eq:integral_transform} that
\beq \bsp
\mathcal{I}[\mathcal{F}(\nu,n)](\ws,w) &\,= \mathcal{I}[\mathcal{F}(\nu,-n)](w,\ws)\,,\\
\mathcal{I}[\mathcal{F}(\nu,n)]\left({1\over w},{1\over \ws}\right) &\,= \mathcal{I}[\mathcal{F}(-\nu,-n)](w,\ws)\,.
\esp\eeq
In other words, the $\mathbb{Z}_2\times\mathbb{Z}_2$ of conjugation and inversion acts on the $(\nu,n)$ space via $[n\leftrightarrow-n]$ and $[\nu\leftrightarrow-\nu,\ n\leftrightarrow-n]$, respectively.  Hence, in order that the functions in $(w,\ws)$ space have definite parity under conjugation and inversion, $\mathcal{F}(\nu,n)$ should have definite parity under $n\leftrightarrow-n$ and $\nu\leftrightarrow-\nu$. Our experience shows that the $n$- and $\nu$-symmetries manifest themselves differently: the $\nu$-symmetry appears as an explicit symmetrization or anti-symmetrization, whereas the $n$-symmetry requires the introduction of an overall factor of $\sgn(n)$. For example, suppose the target function in $(w,\ws)$ space is odd under conjugation, and even under inversion. This implies that the function in $(\nu,n)$ space must be odd under $n\leftrightarrow -n$ and odd under $\nu \leftrightarrow -\nu$. Such a function will decompose as follows,
\beq\label{eq:nu_n_decomposition}
\mathcal{F}(\nu,n) = \frac{1}{2}\,\sgn(n)\left[f(\nu,|n|)-f(-\nu,|n|)\right]\,,
\eeq
for some suitable function $f$. See Table \ref{tab:Z2xZ2_F_table} for the typical decomposition in all four cases.
\begin{table}[!t]
\centering
\begin{tabular}[t]{c|c|c}
\hline\hline
$(w \leftrightarrow \ws, w \leftrightarrow 1/w)$ & $(\nu \leftrightarrow -\nu, n \leftrightarrow -n)$ &$\mathcal{F}(\nu,n)$\\
\hline
$(+,+)$ &$[+,+]$ & $1/2\,\left[f(\nu,|n|)+f(-\nu,|n|)\right]$ \\
$(+,-)$ &$[-,+]$ & $1/2\,\left[f(\nu,|n|)-f(-\nu,|n|)\right]$ \\
$(-,+)$ &$[-,-]$  & $1/2\,\sgn(n)\left[f(\nu,|n|)-f(-\nu,|n|)\right]$\\
$(-,-)$ &$[+,-]$ & $1/2\,\sgn(n)\left[f(\nu,|n|)+f(-\nu,|n|)\right]$\\
\hline\hline
\end{tabular}
\caption{Decomposition of functions in $(\nu,n)$ space into eigenfunctions of the $\mathbb{Z}_2\times\mathbb{Z}_2$ action. Note the use of brackets rather than parentheses to denote the parity under $(\nu,n)$ transformations.}
\label{tab:Z2xZ2_F_table}
\end{table}
Furthermore, in the cases we have studied so far, the constituents $f(\nu,|n|)$ can always be expressed as sums of products of single-variable functions with arguments $\pm i\nu+|n|/2$,
\beq
\label{f_eqn}
f(\nu,|n|) = \sum_j c_j \prod_k f_{j,k}(\delta_k i \nu+|n|/2),
\eeq
where $c_j$ are constants, $\delta_k\in\{+1,-1\}$, and the $f_{j,k}(z)$ are single-variable functions that we now describe.
\subsection{General construction}
The functional form of $\mathcal{F}_i(\nu,n)$ can be further restricted by demanding that the integral~\eqref{eq:integral_transform} evaluate to a combination of HPLs. To see how, consider closing the $\nu$-contour in the lower half plane and summing residues at poles with $\textrm{Im}(\nu)<0$. A necessary condition for the result to yield HPLs is that the residues evaluate exclusively to rational functions and generalized harmonic numbers, e.g., the Euler-Zagier sums defined in eq.~\eqref{eq:Euler-Zagier}. This condition will clearly be satisfied if the $f_{j,k}(z)$ are purely rational functions of $z$. Less obviously, it is also satisfied by polygamma functions. Indeed, the polygamma functions evaluate to ordinary (depth one) harmonic numbers at integer values,
\beq
\psi(1+n)= -\gamma_E+Z_1(n) {\rm~~and~~} 
\psi^{(k)}(1+n) = (-1)^{k+1}k!\,(\zeta_{k+1}-Z_{k+1}(n))\,,
\eeq
where $\psi^{(1)}=\psi^{\prime}$, $\psi^{(2)}=\psi^{\prime\prime}$, etc. 
Referring to eq.~\eqref{eq:HPL_series}, we see that all HPLs through weight three can be constructed using ordinary harmonic numbers\footnote{Harmonic numbers of depth greater than one do appear at weight three; however, after applying the stuffle algebra relations for Euler-Zagier sums, they all can be rewritten in terms of ordinary harmonic numbers of depth one, namely $Z_{1,1}(k-1) = {1\over2}\,Z_{1}(k-1)^2 - {1\over2}\,Z_{2}(k-1)$.}. 

We therefore expect the $f_{j,k}(z)$ to be rational functions or polygamma functions through weight three. Starting at weight four, however, ordinary harmonic numbers are insufficient to cover all possible HPLs. Indeed, at weight four, the HPL
\beq
H_{1,2,1}(z) =\sum_{k=1}^\infty{z^k\over k}\,Z_{2,1}(k-1)
\eeq
requires a depth-two sum\footnote{Another depth-two sum appears in $H_{1,1,2}(x) =\sum_{k=1}^\infty{x^k\over k}\,Z_{1,2}(k-1)$ but the two are related by a stuffle identity, $Z_{2,1}(k-1)+ Z_{1,2}(k-1) = Z_{2}(k-1)\,Z_{1}(k-1) - Z_{3}(k-1)$.}, $Z_{2,1}(k-1)$. A meromorphic function that generates $Z_{2,1}(k-1)$ was presented in ref.~\cite{Blumlein2009ta}. It can be written as a Mellin transform, 
\beq\label{eq:F4}
F_4(N) = {\bf M}\left[\left({\textrm{Li}_2(x)\over 1-x}\right)_+\right](N)\,,\quad N\in\mathbb{C}\,,
\eeq
where the Mellin transform ${\bf M}$ is defined by
\beq
{\bf M}[(f(x))_+](N) \equiv \int_0^1dx\,(x^N-1)\,f(x)\,.
\eeq
If $N$ is a positive integer, then $F_4(N)$ evaluates to harmonic numbers of depth two,
\beq
F_4(N) = Z_{2,1}(N)+Z_3(N) - \zeta_2\,Z_1(N)\,, \quad N\in\mathbb{N}\,.
\eeq
Going to higher weight, new harmonic sums will be necessary to construct the full space of HPLs, and, correspondingly, new meromorphic functions will be necessary to give rise to those sums. The analysis of refs.~\cite{Blumlein1998if,Blumlein2009ta,Blumlein2009fz} uncovers precisely the functions we need\footnote{Actually, in refs.~\cite{Blumlein1998if,Blumlein2009ta,Blumlein2009fz} a more general class of functions is defined. It involves generic HPLs that are singular at $x=-1$ as well as at $x=0$ and 1. As we never encounter these HPLs in our present context, we do not discuss these functions any further.}. They are summarized in Appendix \ref{app:blumlein}. Through weight five, three new functions are necessary: $F_4$, $F_{6a}$ and $F_7$.

There is one final special case that deserves attention. Unlike the other SVHPLs, the pure logarithmic functions $[L_0^-]^k$ diverge as $|w|\to 0$. These functions have special behavior in $(\nu,n)$ space as well, requiring a Kronecker delta function:
\beq\label{eq:special_delta}
\mathcal{I}[\delta_{n,0}/(i \nu)^k] = {1\over\pi}\sum_{n=-\infty}^\infty(-1)^n\,\left({w\over \ws}\right)^{{n\over 2}}\int_{-\infty}^{+\infty}
{d\nu\over \nu^2+{n^2\over 4}}\,|w|^{2i\nu}{\delta_{n,0}\over(i \nu)^{k}}=\,{[L_0^-]^{k+1}\over (k+1)!}\,.
\eeq
Altogether, we find that the following functions of $z = \pm i\nu + |n|/2$ 
are sufficient to construct all the remaining SVHPLs through weight five:
\beq\label{eq:z_basis}
f_{j,k}(z) \in \left\{1,\frac{1}{z},\psi(1+z),\psi^{\prime}(1+z),
\psi^{\prime\prime}(1+z),\psi^{\prime\prime\prime}(1+z),F_4(z),F_{6a}(z),F_7(z)\right\}\,.
\eeq
However, as we will see, not all combinations of elements
in the list~\eqref{eq:z_basis} lead to functions of $(w,\ws)$
that are both single-valued and of definite transcendental weight.
Instead we will construct a smaller set of {\it building blocks} that do
have this property.
\subsection{Examples}
Let us see how to use the elements in the list~\eqref{eq:z_basis} to construct SVHPLs. The simplest case is $f(\nu,|n|)=1$. Referring to Table \ref{tab:Z2xZ2_F_table}, only two of the four sectors yield non-zero choices for $\mathcal{F}$. One of these, $\mathcal{F}=\sgn(n)$, produces something proportional to $H_1 - \Hb_1$, which is not single-valued. This leaves $\mathcal{F}=1$, which should produce a function in the $(+,+)$ sector.  Closing the $\nu$-contour in the lower half plane, and summing up the residues at $\nu=-i|n|/2$, we obtain the integral of eq.~\eqref{eq:I[1]},
\beq\bsp
\mathcal{I}[1] &=2L_1^+\,,
\esp\eeq
indeed a function in the $(+,+)$ sector. Including the special case $\LZeroMM$ from eq.~\eqref{eq:special_delta}, this completes the analysis at weight one.

The next simplest element is $1/z$, yielding $f(\nu,|n|)=1/(i\nu+|n|/2)$. It generates two single-valued functions, one in the $(+,-)$ sector and one in the $(-,-)$ sector (using the $(w,\ws)$ labeling in the first column of Table~\ref{tab:Z2xZ2_F_table}). Symmetrizing as indicated in Table~\ref{tab:Z2xZ2_F_table}, the two functions in $(\nu,n)$ space are $\mathcal{F}=-V$ and $\mathcal{F}=N/2$, with the useful shorthands
\beq\bsp
V&\equiv -\frac{1}{2}\left[\frac{1}{i\nu+\frac{|n|}{2}}-\frac{1}{-i\nu+\frac{|n|}{2}}\right] = \frac{i \nu}{\nu^2+\frac{|n|^2}{4}},\\
N&\equiv \sgn(n)\left[\frac{1}{i\nu+\frac{|n|}{2}}+\frac{1}{-i\nu+\frac{|n|}{2}}\right] = \frac{n}{\nu^2+\frac{|n|^2}{4}}\,.
\esp\eeq
The transforms of these functions yield two of the four SVHPLs of weight two.
\beq\bsp\label{eq:I[V]_I[N]_building_block}
\mathcal{I}[V] \,&=\, - \LZeroMM \, L_1^+ \,,\\
\mathcal{I}[N]\,&= 4\,L_2^-\, .
\esp\eeq
A third weight-two function is the pure logarithmic function $[L_0^-]^2$, a special case already considered. To find the fourth weight-two function, we turn to the next element in the list~\eqref{eq:z_basis}, $\psi(1+z)$. On its own, it does not generate any single-valued functions; however, a particular linear combination of $\{1,1/z,\psi(1+z)\}$ indeed produces such a function.
Specifically, $f(\nu,|n|) = 2\psi(1+i\nu+|n|/2)+2\gamma_E-1/(i\nu+|n|/2)$ generates the last weight-two SVHPL, which transforms in the $(+,+)$ sector. The function in $(\nu,n)$ space is actually the leading-order BFKL eigenvalue, $E_{\nu,n}$,
\beq
\mathcal{F} = \psi\left(1+i\nu+\frac{|n|}{2}\right) + \psi\left(1-i\nu+\frac{|n|}{2}\right) + 2\gamma_E-\frac{\sgn(n)N}{2}  = E_{\nu,n}\,,
\eeq
and its transform is the last SVHPL of weight two,
\beq\label{eq:I[E]_building_block}
\mathcal{I}[E_{\nu,n}] \,= \,[L_1^+]^2-\frac{1}{4}\,\LZeroM{2}\,.
\eeq

The next element in the list (\ref{eq:z_basis}) is $\psi^{\prime}(1+z)$.
Like $\psi(1+z)$, $\psi^{\prime}(1+z)$ does not by itself generate any single-valued functions; however, there is a particular linear combination that does, and it is given by $f(\nu,|n|) = 2\psi^{\prime}(1+i\nu+|n|/2)+1/(i\nu+|n|/2)^2$. Notice that, for the first time, the product in eq.~(\ref{f_eqn}) extends over more than one term (in this case, $f_{1,1}=f_{1,2}=1/(i\nu+|n|/2)$, but in general the $f_{j,k}$ will be different). The function in $(\nu,n)$ space is,
\beq
\mathcal{F} = \psi^{\prime}\left(1+i\nu+\frac{|n|}{2}\right)-\psi^{\prime}\left(1-i\nu+\frac{|n|}{2}\right) - \sgn(n) N V  = \dnu E_{\nu,n}\,,
\eeq
where $\dnu\equiv-i\partial_\nu \equiv -i\,\partial/\partial\nu$.
The main observation is that the basis in eq.~(\ref{eq:z_basis}) can be modified to consistently generate single-valued functions:  $1/z$ is replaced by $V$ and $N$, $\psi$ is replaced by $E_{\nu,n}$, and $\psi^{(k)}$ is replaced by $\dE{k}$. 

Furthermore, as mentioned previously, the basis at weight four requires a new function $F_4(z)$ that is outside the class of polygamma functions. Like the polygamma functions, $F_4(z)$ does not by itself generate a single-valued function; it too requires additional terms. We denote the resulting basis element by $\tilde{F}_4$. It is related to the function $F_4(z)$ in eq.~\eqref{eq:F4} by,
\beq\bsp\label{eq:f4_tilde}\tilde{F}_4 \,=\,&\sgn(n) \left\{F_4\Big(i\nu+\frac{|n|}{2}\Big) + F_4\Big(-i\nu+\frac{|n|}{2}\Big) - \frac{1}{4}\dE{2}-\frac{1}{8}N^2 E_{\nu,n}-\frac{1}{2}V^2 E_{\nu,n}\right.\\
&\left.\phantom{\sgn(n)}+\frac{1}{2}\,\Big(\psi_{-}+V\Big)\dEOne + \zeta_2 E_{\nu,n} -4\,\zeta_3\right\} + N \left\{\frac{1}{2}\,V \psi_{-} + \frac{1}{2}\zeta_2 \right\}\,,
\esp\eeq
where
\beq\bsp
\psi_{-}&\equiv\psi\Big(1+i\nu+\frac{|n|}{2}\Big)
              -\psi\Big(1-i\nu+\frac{|n|}{2}\Big) \,.
\esp\eeq
Appendix \ref{app:blumlein} contains further details about the functions in $(\nu,n)$ space, including the basis through weight five and expressions for the building blocks $\tilde{F}_{6a}$ and $\tilde{F}_7$ generated by the functions $F_{6a}(z)$ and $F_7(z)$.

Finally, we describe a heuristic method for assembling the basis in $(\nu,n)$ space. The idea is to start with the building blocks,
\beq\label{mod_basis}
\{1,N,V,E_{\nu,n},\tilde{F}_4,\tilde{F}_{6a},\tilde{F}_7\},
\eeq
and piece them together with multiplication and $\nu$-differentiation. These two operations do not always produce independent functions. For example,
\beq\label{eq:NV_der}
\dnu N=2\,NV {\rm~~and~~}\dnu V = \frac{1}{4}N^2+V^2\,.
\eeq
The building blocks have definite parity under $\nu\leftrightarrow-\nu$ and $n\leftrightarrow-n$ which helps determine which combinations appear in which sector. Additionally, we observe that they can be assigned a transcendental weight, which further assists in the classification. The weight in $(w,\ws)$ space is found by calculating the total weight of the constituent building blocks in $(\nu,n)$ space, and then adding one (to account for the increase in weight due to the integral transform itself). The relevant properties of the basic building blocks are summarized in Table~\ref{tab:NVE}.

As an example, let us consider the function $N \dnu E_{\nu,n}$. Referring to Table~\ref{tab:NVE}, the transcendental weight is $1+1+1=3$ in $(\nu,n)$ space, or $3+1=4$ in $(w,\ws)$ space. Under $[\nu\leftrightarrow-\nu,n\leftrightarrow-n]$, $N$ has parity $[+,-]$, $\dnu$ has parity $[-,+]$, and $E_{\nu,n}$ has parity $[+,+]$, so  $N \dnu E_{\nu,n}$ has parity $[-,-]$. We therefore expect this function to transform into a weight four function of $(w,\ws)$, with parity $(-,+)$ under $(w\leftrightarrow\ws,w\leftrightarrow1/w)$ (see Table~\ref{tab:Z2xZ2_F_table}). Indeed this turns out to be the case. A complete basis through weight four is presented in Table~\ref{tab:nu_n_table}.
\renewcommand{\arraystretch}{1.2}
\begin{table}[!t]
\centering
\begin{tabular}[t]{c|c|c}
\hline\hline
& weight & $(\nu \leftrightarrow -\nu, n \leftrightarrow -n)$\\
\hline
$1$ & $0$ &$[+,+]$  \\ 
$\dnu$ & $1$&$[-,+]$  \\ 
$V$ & $1$&$[-,+]$ \\ 
$N$ & $1$&$[+,-]$  \\ \hline\hline
\end{tabular}
\quad
\begin{tabular}[t]{c|c|c}
\hline\hline
& weight & $(\nu \leftrightarrow -\nu, n \leftrightarrow -n)$\\
\hline
$E_{\nu,n}$ & $1$&$[+,+]$  \\ 
$\tilde{F}_{4}$ & $3$&$[+,-]$  \\ 
$\tilde{F}_{6a}$ & $4$&$[-,-]$ \\ 
$\tilde{F}_{7}$ & $4$&$[-,+]$\\ \hline\hline
\end{tabular}
\caption{Properties of the building blocks for the basis in $(\nu,n)$ space.}
\label{tab:NVE}
\end{table}
\renewcommand{\arraystretch}{1.2}
\begin{table}[!t]
\centering
\begin{tabular}[t]{c|c|c|c|c}
\hline\hline
 weight & $\mathbb{Z}_2\times\mathbb{Z}_2$ & $(w,\ws)$ basis & $ (\nu,n)$ basis &dimension\\
\hline\hline
\multirow{4}{*}{1}& $(+,+)$ & $\LOnePP$ & $1$ & 1\\ \cline{2-5}
& $(+,-)$ & $\LZeroMM$ & $\delta_{n,0}$ & 1\\  \cline{2-5}
& $(-,+)$ &$-$&$-$&0\\  \cline{2-5}
& $(-,-)$ &$-$&$-$&0\\  \cline{2-5}
\hline\hline
\multirow{4}{*}{2}& $(+,+)$ & $\LOneP{2},\, \LZeroM{2}$ & $\delta_{n,0}/(i\nu),\, E_{\nu,n}$ & 2\\  \cline{2-5}
& $(+,-)$ & $\LZeroMM \LOnePP$ & $V$ & $1$ \\  \cline{2-5}
& $(-,+)$ &$-$&$-$& $0$ \\  \cline{2-5}
& $(-,-)$ & $\LTwoMM$ & $N$ & $1$ \\  \cline{2-5}
\hline\hline
\multirow{4}{*}{3}& $(+,+)$ & $\LOneP{3},\, \LZeroM{2} \LOnePP,\, \LThreePP$ & $V^2,\,N^2,\,E_{\nu,n}^2$ & 3\\  \cline{2-5}
& $(+,-)$ & $\LZeroM{3},\, \LZeroMM\LOneP{2},\,\LTwoOneMM$ & $\delta_{n,0}/(i\nu)^2,\, V E_{\nu,n},\, \dnu E_{\nu,n}$ & $3$ \\  \cline{2-5}
& $(-,+)$ & $\LZeroMM \LTwoMM$ & $V N$ & $1$ \\  \cline{2-5}
& $(-,-)$ & $\LOnePP \LTwoMM$ & $N E_{\nu,n}$ & $1$ \\  \cline{2-5}
\hline\hline
\multirow{5}{*}{4}& \multirow{2}{*}{$(+,+)$} & $\LZeroM{4},\,\LOneP{4},\,\LZeroM{2}\LOneP{2},\,$ & $\delta_{n,0}/(i\nu)^3,\, E_{\nu,n}^3,\, N^2 E_{\nu,n},$ & \multirow{2}{*}{$6$} \\  
& & $\LTwoM{2},\,\LZeroMM\LTwoOneMM,\,\LOnePP\LThreePP$ & $V^2 E_{\nu,n},\, V \dnu E_{\nu,n},\, \dE{2}$ & \\     \cline{2-5}
& $(+,-)$ & $\LZeroMM\LOneP{3},\,\LZeroM{3}\LOnePP,\,\LZeroMM\LThreePP,\,\LOnePP\LTwoOneMM$ & $V^3,\,N^2 V,\, V E_{\nu,n}^2,\, E_{\nu,n} \dnu E_{\nu,n}$ & 4\\     \cline{2-5}
& $(-,+)$ & $\LZeroMM\LOnePP\LTwoMM,\, \LThreeOnePP$ & $N V E_{\nu,n},\, N \dnu E_{\nu,n}$ & 2\\  \cline{2-5}
& $(-,-)$ & $\LZeroM{2}\LTwoMM,\, \LOneP{2}\LTwoMM,\, \LFourMM,\,\LTwoOneOneMM$ & $N^3,\, N V^2,\, N E_{\nu,n}^2,\, \tilde{F}_4$ & 4\\  \cline{2-5}
\hline\hline
\end{tabular}
\caption{Basis of SVHPLs in $(w,\ws)$ and $(\nu,n)$ space through weight four. Note that at each weight we can also add the product of zeta values with lower-weight entries.}
\label{tab:nu_n_table} 
\end{table}


\section{Applications in $(\nu,n)$ space:
the BFKL eigenvalues and impact factor}
\label{sec:applications}
\subsection{The impact factor at NNLLA}
\label{sec:Phi2}
In this section we report results for $g^{(4)}_1$ and $g^{(4)}_0$ and discuss how to transform these functions to $(\nu,n)$ space using the basis constructed in the previous section. We then give our results for the new data for the MRK logarithmic expansion:  $\Phi^{(2)}_{\textrm{Reg}}$, $\Phi^{(3)}_{\textrm{Reg}}$, and $E^{(2)}_{\nu,n}$.

Before discussing the case of the higher-order corrections to the BFKL eigenvalue and the impact factor, let us review how the known results for $E_{\nu,n}$, $E^{(1)}_{\nu,n}$ and $\Phi^{(1)}_{\textrm{Reg}}$ fit into the framework for $(\nu,n)$ space that we have developed in the previous section.
First, we have already seen in Section~\ref{sec:nu_n} that the LL BFKL eigenvalue is one of our basis elements of weight one in $(\nu,n)$ space (see Table~\ref{tab:NVE}). Next, we know that the first time the NLL impact factor $\Phi^{(1)}_{\textrm{Reg}}$ appears is in the NLLA of the two-loop amplitude, $g_0^{(2)}(w,\ws)$, which is a pure single-valued function of weight three. Following our analysis from the previous section, it should then be possible to express $\Phi^{(1)}_{\textrm{Reg}}$ as a pure function of weight two in $(\nu,n)$ space with the correct symmetries. Indeed, we can easily recast eq.~\eqref{eq:Phi_1} in terms of the basis elements shown in Table~\ref{tab:NVE},
\beq
\Phi^{(1)}_{\textrm{Reg}}(\nu,n) = - {1\over2}E_{\nu,n}^2 - {3\over8} N^2 - \zeta_2\,.
\eeq
Similarly, the NLL BFKL eigenvalue can be written as a linear combination of weight three of the basis elements in Table~\ref{tab:NVE},
\beq
E^{(1)}_{\nu,n} = - {1\over4} \, \dE{2}
 + {1\over2} \, V \, \dnu E_{\nu,n} - \zeta_2 \, E_{\nu,n} - 3 \, \zeta_3 \,.
\eeq
This completes the data for the MRK logarithmic expansion that can be extracted through two loops.

Now we proceed to three loops.
By expanding eq.~\eqref{eq:MHV_MRK_2} to order $a^3$, we obtain the following relation for the NNLLA correction to the impact factor, $\Phi^{(2)}_{\textrm{Reg}}(\nu,n)$,
\beq\bsp\label{eq:I[Phi_2]}
\mathcal{I}\left[\Phi^{(2)}_{\textrm{Reg}}(\nu,n)\right] &\,= 
4 \, g^{(3)}_2(w,\ws) \,\left\{[L_1^+]^2+\pi ^2\right\}
- 4 \, g^{(3)}_1(w,\ws) \,L_1^+ + 4 \, g^{(3)}_0(w,\ws)\\
&\,-4 \pi ^2 g^{(2)}_1(w,\ws) \,L_1^+
+ \frac{\pi^2}{180} L_1^+ 
 \left\{ -45 \, [L_0^-]^2 + 120 \, [L_1^+]^2 + 22 \, \pi^2\right \}\,.
\esp\eeq
This expression is exactly $2\,\rho(w,\ws)$, where $\rho$ was given
in eq.~\eqref{eq:rho_def} and in ref.~\cite{Fadin2011we}.  (The factor of two just has to do with our normalization of the Fourier-Mellin transform.)

To invert eq.~\eqref{eq:I[Phi_2]} and obtain $\Phi^{(2)}_{\textrm{Reg}}(\nu,n)$, we begin by observing that the right-hand side is a pure function of weight five in $(w,\ws)$ space. Moreover, it is an eigenfunction with eigenvalue $(+,+)$ under the $\mathbb{Z}_2\times\mathbb{Z}_2$ symmetry. Following the analysis of Section~\ref{sec:nu_n}, and using the results at the end of Appendix~\ref{app:blumlein}, we are led to make the following ansatz,
\beq\bsp\label{eq:Phi_2_ansatz}
\Phi^{(2)}_{\textrm{Reg}}(\nu,n) & = \alpha_1\,E^4_{\nu,n} + \alpha_2\,N^2 E^2_{\nu,n} + 
\alpha_3\,N^4 + \alpha_4\,V^2\, E^2_{\nu,n} + \alpha_5\,N^2 V^2 + \alpha_6\,V^4 \\
&\,+\alpha_7\,\EnunOne\,V\dEOne +\alpha_8\, \dEPOne{2} + \alpha_9\,\EnunOne\,\dE{2} + \alpha_{10}\,\Ffourtilde\, N \\
&\,+ \alpha_{11}\,\zeta_2 E^2_{\nu,n}+\alpha_{12}\,\zeta_2 N^2 + \alpha_{13}\, \zeta_2 V^2 + \alpha_{14}\, \zeta_3 E_{\nu,n} + \alpha_{15}\, \zeta_3\,[\delta_{n,0}/(i\nu)] + \alpha_{16}\,\zeta_4\,.
\esp\eeq
The $\alpha_i$ are rational numbers that can be determined by computing the integral transform to $(w,\ws)$ space of eq.~\eqref{eq:Phi_2_ansatz} (see Appendix~\ref{app:blumlein}) and then matching the result to the right-hand side of eq.~\eqref{eq:I[Phi_2]}. We find
\beq\bsp
\Phi^{(2)}_{\textrm{Reg}}(\nu,n) &\,=
{1\over2}\left[\Phi^{(1)}_{\textrm{Reg}}(\nu,n)\right]^2 - E^{(1)}_{\nu,n} \, \EnunOne
+ \frac{1}{8}\,\dEPOne{2} + \frac{5 \pi^2}{16} \, \Enun{2}
- \frac{1}{2}\,\zeta_3 \,\EnunOne + \frac{5}{64}\,N^4 \\
& \, + \frac{5}{16}\,N^2\,V^2 - \frac{5 \pi^2}{64}\,N^2
- \frac{\pi^2}{4}\,V^2 + \frac{17 \pi^4}{360}
+ d_1\,\zeta_3\,\EnunOne
- d_2\,{\pi^2\over6} \, \left[ 12\,\Enun{2} + N^2 \right]\\
& \, + \gamma''\,{\pi^2\over6} \, \left[\Enun{2} - \frac{1}{4}\,N^2 \right]\,.
\esp\eeq
Here $d_1$, $d_2$ and $\gamma''$ are the (not yet determined) rational numbers that appear in eq.~\eqref{eq:g30}.
We emphasize that the expression for $\Phi^{(2)}_{\textrm{Reg}}(\nu,n)$ does not involve the basis element $N\,{\tilde F}_4$ (see eq.~\eqref{eq:NF4tilde}).
That is, $\Phi^{(2)}_{\textrm{Reg}}(\nu,n)$ can be written purely in terms of $\psi$ functions (and their derivatives).

To determine the six-point remainder function in MRK to all loop orders in the NNLL approximation, we must apply some additional information beyond $\Phi^{(2)}_{\textrm{Reg}}(\nu,n)$. In particular, at four loops and higher, the second-order correction to the BFKL eigenvalue, $E^{(2)}_{\nu,n}$, is necessary. In the next section, we will show how to use information from the symbol of the four-loop remainder function to determine $E^{(2)}_{\nu,n}$. We will also derive the next correction to the impact factor, $\Phi^{(3)}_{\textrm{Reg}}(\nu,n)$, which enters the N$^3$LL approximation.

\subsection{The four-loop remainder function in the multi-Regge limit}
\label{sec:4loop}

In order to compute the next term in the perturbative expansion of the BFKL eigenvalue and the impact factor, we need the analytic expressions for the four-loop six-point remainder function in the multi-Regge limit. In an independent work, the symbol of the four-loop six-point remainder function has been heavily constrained~\cite{fourloop}. In ref.~\cite{fourloop} the symbol of $R_6^{(4)}$ is written in the form
\beq\label{eq:4-loop-symbol}
\cS(R_6^{(4)}) = \sum_{i=1}^{113}\alpha_i\,S_i\,,
\eeq
where $\alpha_i$ are undetermined rational numbers. The $S_i$ denote integrable tensors of weight eight satisfying the first- and final-entry conditions mentioned in the introduction, such that:
\begin{enumerate}
\item All entries in the symbol are drawn from the set  $\{u_i,1-u_i,y_i\}_{i=1,2,3}$, where the $y_i$'s are defined in eq.~\eqref{eq:y_z_def}.
\item The symbol is integrable.
\item The tensor is totally symmetric in $u_1$, $u_2$, $u_3$. Note that under a permutation $u_i\to u_{\sigma(i)}$, $\sigma\in S_3$, the $y_i$ variables transform as $y_i\to1/y_{\sigma(i)}$.
\item The tensor is invariant under the transformation $y_i\to 1/y_i$. 
\item The tensor vanishes in all simple collinear limits.
\item The tensor is in agreement with the prediction coming from the collinear OPE of ref.~\cite{Alday2010ku}. We implement this condition on the leading singularity exactly as was done at three loops~\cite{Dixon2011pw}.
\end{enumerate}
In Section~\ref{sec:R6_LLA_NLLA}, we presented analytic expressions for the four-loop remainder function in the LLA and NLLA of MRK. We can use these results to obtain further constraints on the free coefficients $\alpha_i$ appearing in eq.~\eqref{eq:4-loop-symbol}. In order to achieve this, we first have to understand how to write the symbol~\eqref{eq:4-loop-symbol} in MRK. 
In the following we give very brief account of this procedure. 

To begin, recall that the remainder function is non-zero in MRK only after performing the analytic continuation~\eqref{eq:MRK_anal_cont}, $u_1 \to e^{-2\pi i}\,|u_1|$. The function can then be expanded as in eq.~\eqref{eq:R6_MRK},
\beq
R_6^{(4)}|_{\textrm{MRK}} = 2\pi i\,\sum_{n=0}^{3}\,\log^n(1-u_1)\,\left[g_n^{(4)}(w,\ws) + 2\pi i\,h_n^{(4)}(w,\ws)\right]\,.
\eeq
The symbols of the imaginary and real parts can be extracted by taking single and double discontinuities,
\beq\bsp\label{eq:gh_symb}
2\pi i\,\sum_{n=0}^{3}\cS\left[\log^n(1-u_1)\,g^{(4)}_n(w,\ws)\right] &\;=\; \cS(\Delta_{u_1}R_6^{(4)})|_{\textrm{MRK}}\\
& \;=\; -2\pi i\,\sum_{i=1}^{113}\alpha_i\,\Delta_{u_1}(S_i)|_{\textrm{MRK}}\\
(2\pi i)^2\,\sum_{n=0}^{3}\cS\left[\log^n(1-u_1)\,h^{(4)}_n(w,\ws)\right] &\;=\; \cS(\Delta^2_{u_1}R_6^{(4)})|_{\textrm{MRK}}\\
& \;=\; (-2\pi i)^2\,\sum_{i=1}^{113}\alpha_i\,\Delta^2_{u_1}(S_i)|_{\textrm{MRK}}\,,
\esp\eeq
where the discontinuity operator $\Delta$ acts on symbols via,
\begin{eqnarray}
\Delta_{u_1}(a_1\otimes a_2\otimes \ldots\otimes a_n) &=& \left\{\begin{array}{cc}
a_2\otimes \ldots\otimes a_n\,, & \textrm{if } a_1 = u_1\,,\\
0\,, &\textrm{otherwise.}
\end{array}\right.\\
\Delta^2_{u_1}(a_1\otimes a_2\otimes \ldots\otimes a_n) &=& \left\{\begin{array}{cc}
\frac{1}{2}\,(a_3\otimes \ldots\otimes a_n)\,, & \textrm{if } a_1 = a_2 = u_1\,,\\
0\,, &\textrm{otherwise.}
\end{array}\right.
\end{eqnarray}
As indicated in eq.~\eqref{eq:gh_symb}, we need to evaluate the symbols $S_i$ in MRK, which we do by taking the multi-Regge limit of each entry of the symbol. This can be achieved by replacing $u_2$ and $u_3$ by the variables $x$ and $y$, defined in eq.~\eqref{eq:xy_def} (which we then write in terms of $w$ and $\ws$ using eq.~\eqref{eq:xy_to_wws}), while the $y_i$'s are replaced by their limits in MRK~\cite{Dixon2011pw},
\beq
y_1\to1\,,\qquad y_2\to {1+\ws\over1+w}\,,\qquad y_3\to {\ws(1+w)\over w(1+\ws)}\,.
\eeq 
Finally, we drop all terms in $\Delta_{u_1}^k(S_i)$, $k=1,2$, that have an entry corresponding to $u_1$, $y_1$, $1-u_2$ or $1-u_3$, since these quantities approach unity in MRK. In the end, the resulting tensors have entries drawn from the set $\{1-u_1, w,\ws,1+w,1+\ws\}$. The $1-u_1$ entries come from factors of $\log(1-u_1)$ and can be shuffled out, so that we can write eq.~\eqref{eq:gh_symb} as,
\beq\bsp\label{eq:gh_GH}
\sum_{n=0}^{3}\cS\left[\log^n(1-u_1)\right]\sha \cS\left[g^{(4)}_n(w,\ws))\right] &\;=\; \sum_{i=1}^{113}\sum_{n=0}^{7}\alpha_i\,\cS\left[\log^n(1-u_1)\right]\sha G_{i,n}\\
\sum_{n=0}^{3}\cS\left[\log^n(1-u_1)\right]\sha \cS\left[h^{(4)}_n(w,\ws))\right] &\;=\; \sum_{i=1}^{113}\sum_{n=0}^{6}\alpha_i\,\cS\left[\log^n(1-u_1)\right]\sha H_{i,n}\, ,
\esp\eeq
for some suitable tensors $G_{i,n}$ of weight $(7-n)$ and $H_{i,n}$ of weight $(6-n)$.  The sums on the right-hand side of eq.~\eqref{eq:gh_GH} turn out to extend past $n=3$.  Because the sums on the left-hand side do not, we immediately obtain homogeneous constraints on the $\alpha_i$ for the cases $n=4,5,6,7$. Furthermore, since the quantities on the left-hand side of eq.~\eqref{eq:gh_GH} are known for $n=3$ and $n=2$, we can use this information to further constrain the $\alpha_i$. Finally, there is a consistency condition which relates the real and imaginary parts,
\beq\bsp
h_1^{(4)}(w,\ws) &\,= g^{(4)}_2(w,\ws)
+\frac{\pi^2}{12}\, g^{(2)}_1(w,\ws) \,L_1^+
-\frac{1}{2} \, g^{(3)}_1(w,\ws)\,L_1^+ - g_1^{(2)}(w,\ws)\,g_0^{(2)}(w,\ws)\,,\\
h_0^{(4)}(w,\ws)&\,=  \frac{1}{2} \, g^{(4)}_1(w,\ws) + \pi ^2\, g^{(4)}_3(w,\ws)
-\pi ^2\,g^{(3)}_2(w,\ws)\,L_1^+
-\frac{1}{2} \, g^{(3)}_0(w,\ws) \, L_1^+\\
&\,
+\frac{\pi^2}{2} \, g^{(2)}_1(w,\ws)\,[L_1^+]^2
+\frac{\pi^2}{12} \, g^{(2)}_0(w,\ws)\,L_1^++
\frac{\pi^2}{64} \,[L_0^-]^2\,[L_1^+]^2
-\frac{\pi^2}{1536}\,[L_0^-]^4\\
&\,
+\frac{3}{640} \pi ^4\,[L_0^-]^2
-\frac{5}{96} \pi ^2\,[L_1^+]^4-\frac{3}{160} \pi ^4\,[L_1^+]^2
- {1\over2} \, [g_0^{(2)}(w,\ws)]^2\,.
\esp\eeq
In total, these constraints allow us to fix all of the coefficients $\alpha_i$ that survive in the multi-Regge limit, except for a single parameter which we will refer to as $a_0$. 

The results of the above analysis are expressions for the symbols of the functions $g^{(4)}_1$ and $g^{(4)}_0$. We would like to use this information to calculate new terms in the perturbative expansions of the BFKL eigenvalue $\omega(\nu,n)$ and the MHV impact factor $\Phi_{\textrm{Reg}}(\nu,n)$. For this purpose, we actually need the functions $g^{(4)}_1$ and $g^{(4)}_0$, and not just their symbols. Thankfully, using our knowledge of the space of SVHPLs, it is easy to integrate these symbols. We can constrain the beyond-the-symbol ambiguities by demanding that the function vanish in the collinear limit $(w,\ws)\to0$, and that it be invariant under conjugation and inversion of the $w$ variables.
Putting everything together, we find the following expressions for $g^{(4)}_1$ and $g^{(4)}_0$,
\beq\bsp
g_1^{(4)}(w,\ws) &\,= \frac{3}{128}\,\LMA{2}{2}\,\LMA{0}{2}-\frac{3}{32}\,\LMA{2}{2}\,\LPA{1}{2}+\frac{19}{384}\,\LMA{0}{2}\,\LPA{1}{4}+\frac{73}{1536}\,\LMA{0}{4}\,\LPA{1}{2}\\
&\,+\frac{1}{96}\,\LSMB{2}{1}\,\LMA{0}{3}-\frac{29}{64}\,\LSPA{1}\,\LSPA{3}\,\LMA{0}{2}-\frac{11}{30720}\,\LMA{0}{6}-\frac{1}{8}\,\LMB{2}{1}{2}-\frac{17}{48}\,\LSPA{3}\,\LPA{1}{3}\\
&\,+\frac{23}{12}\,\LPA{1}{3}\,\zeta_3+\frac{11}{480}\,\LPA{1}{6}+\frac{5}{32}\,\LPA{3}{2}-\frac{1}{4}\,\LSMA{4}\,\LSMA{2}+\frac{1}{4}\,\LSMA{2}\,\LSMC{2}{1}{1}+\frac{1}{4}\,\LSMA{0}\,\LSMB{4}{1}\\
&\,-\frac{3}{4}\,\LSMA{0}\,\LSMD{2}{1}{1}{1}+\frac{19}{8}\,\LSPA{5}\,\LSPA{1}+\frac{5}{4}\,\LSPA{1}\,\LSPC{3}{1}{1}+\frac{1}{2}\,\LSPA{1}\,\LSPC{2}{2}{1}-\frac{3}{2}\,\LSPA{1}\,\zeta_5+\frac{1}{8}\,\zeta_3^2\\
&\,+a_0\Bigg\{
\frac{1027}{2}\,\LMA{2}{2}\,\LMA{0}{2}+\frac{417}{8}\,\LMA{0}{2}\,\LPA{1}{4}+\frac{431}{24}\,\LMA{0}{4}\,\LPA{1}{2}+\frac{3155}{48}\,\LSMB{2}{1}\,\LMA{0}{3}\\
&\,\qquad-\frac{1581}{16}\,\LSPA{1}\,\LSPA{3}\,\LMA{0}{2}+\frac{9823}{1152}\,\LMA{0}{6}-\frac{871}{4}\,\LSMA{0}\,\LSMB{2}{1}\,\LPA{1}{2}-\frac{709}{4}\,\LSPA{3}\,\LPA{1}{3}\\
&\,\qquad+\frac{2223}{2}\,\LSPA{5}\,\LSPA{1}-157\,\LMA{2}{2}\,\LPA{1}{2}-256\,\LMB{2}{1}{2}+1593\,\LPA{1}{3}\,\zeta_3\\
&\,\qquad+681\,\LPA{3}{2}-1606\,\LSMA{4}\,\LSMA{2}+512\,\LSMA{2}\,\LSMC{2}{1}{1}-3371\,\LSMA{0}\,\LSMB{4}{1}\\
&\,\qquad-1730\,\LSMA{0}\,\LSMB{3}{2}-299\,\LSMA{0}\,\LSMD{2}{1}{1}{1}+2127\,\LSPA{1}\,\LSPC{3}{1}{1}+744\,\LSPA{1}\,\LSPC{2}{2}{1}\\
&\,\qquad+5489\,\LSPA{1}\,\zeta_5+256\,\zeta_3^2\Bigg\}
+a_1\,\pi ^2\,\gFL{3}{1} + a_2\,\pi ^2\,\gFL{4}{3} \\
&\,+ a_3\,\pi ^2\,[\gFL{2}{1}]^2 +a_4\,\pi ^2\,\hFL{4}{2} + a_5\,\pi ^2\,\hFL{3}{0}\\
&\,+a_6\,\pi ^4\,\gFL{2}{1}+a_7\,\zeta_3\,\gFL{2}{0}+a_8\,\zeta_3\,\gFL{3}{2}\,.
\esp\eeq

\beq\bsp
g_0^{(4)}(w,\ws) &\,=\frac{5}{64}\,\LSPA{1}\,\LMA{2}{2}\,\LMA{0}{2}-\frac{1}{16}\,\LMA{2}{2}\,\LPA{1}{3}-\frac{21}{64}\,\LSPA{3}\,\LMA{0}{2}\,\LPA{1}{2}+\frac{7}{144}\,\LMA{0}{4}\,\LPA{1}{3}\\
&\,+\frac{9}{320}\,\LMA{0}{2}\,\LPA{1}{5}-\frac{7}{192}\,\LSMB{2}{1}\,\LSPA{1}\,\LMA{0}{3}+\frac{129}{64}\,\LSPA{5}\,\LMA{0}{2}+\frac{1007}{46080}\,\LSPA{1}\,\LMA{0}{6}\\
&\,-\frac{5}{24}\,\LSPA{3}\,\LMA{0}{4}+\frac{3}{32}\,\LSPC{3}{1}{1}\,\LMA{0}{2}-\frac{1}{16}\,\LSPC{2}{2}{1}\,\LMA{0}{2}+\frac{7}{16}\,\LMA{0}{2}\,\zeta_5\\
&\,-\frac{1}{16}\,\LSMA{0}\,\LSMB{2}{1}\,\LPA{1}{3}+\frac{25}{16}\,\LSPA{5}\,\LPA{1}{2}-\frac{7}{48}\,\LSPA{3}\,\LPA{1}{4}+\frac{7}{8}\,\LSPC{3}{1}{1}\,\LPA{1}{2}\\
&\,+\frac{25}{12}\,\LPA{1}{4}\,\zeta_3+\frac{1}{210}\,\LPA{1}{7}-\frac{1}{4}\,\LSMA{4}\,\LSMA{2}\,\LSPA{1}-\frac{5}{16}\,\LSMA{2}\,\LSMA{0}\,\LSPB{3}{1}+\frac{1}{4}\,\LSMA{2}\,\LSMC{2}{1}{1}\,\LSPA{1}\\
&\,+\frac{1}{4}\,\LSMA{0}\,\LSMB{4}{1}\,\LSPA{1}-\frac{1}{8}\,\LSMA{0}\,\LSMB{2}{1}\,\LSPA{3}-\frac{1}{4}\,\LSMA{0}\,\LSMD{2}{1}{1}{1}\,\LSPA{1}+\frac{3}{2}\,\LSPA{1}\,\zeta_3^2-\frac{125}{8}\,\LSPA{7}\\
&\,+\frac{1}{2}\,\LSPC{4}{1}{2}+\frac{11}{4}\,\LSPC{4}{2}{1}+\frac{3}{4}\,\LSPC{3}{3}{1}-\frac{1}{2}\,\LSPE{2}{1}{2}{1}{1}-\frac{3}{2}\,\LSPE{2}{2}{1}{1}{1}+\frac{25}{4}\,\zeta_7+5\,\LSPC{5}{1}{1}
\esp\eeq
\beq\bsp\nonumber
&\,-4\,\LSPE{3}{1}{1}{1}{1}+\frac{1}{4}\,\LSPC{2}{2}{1}\,\LPA{1}{2}
+a_0\Bigg\{-\frac{1309}{4}\,\LSPA{1}\,\LMA{2}{2}\,\LMA{0}{2}\\
&\,\qquad-\frac{8535}{4}\,\LSPA{3}\,\LMA{0}{2}\,\LPA{1}{2}+\frac{235}{4}\,\LMA{0}{2}\,\LPA{1}{5}+\frac{4617}{16}\,\LMA{0}{4}\,\LPA{1}{3}\\
&\,\qquad-\frac{32027}{24}\,\LSMB{2}{1}\,\LSPA{1}\,\LMA{0}{3}-\frac{11415}{8}\,\LSPA{5}\,\LMA{0}{2}-\frac{310}{9}\,\LSPA{1}\,\LMA{0}{6}\\
&\,\qquad+\frac{15225}{64}\,\LSPA{3}\,\LMA{0}{4}+\frac{24279}{4}\,\LSPC{3}{1}{1}\,\LMA{0}{2}-\frac{823}{2}\,\LSMA{0}\,\LSMB{2}{1}\,\LPA{1}{3}\\
&\,\qquad+\frac{2235}{2}\,\LSPA{5}\,\LPA{1}{2}-\frac{365}{4}\,\LSPA{3}\,\LPA{1}{4}+205\,\LMA{2}{2}\,\LPA{1}{3}+1911\,\LSPA{3}\,\LMA{2}{2}\\
\phantom{g_9^{(4)}(w,\ws) }
&\,\qquad+2130\,\LSPC{2}{2}{1}\,\LMA{0}{2}-2623\,\LMA{0}{2}\,\zeta_5+992\,\LSPA{1}\,\LMB{2}{1}{2}+63\,\LSPC{3}{1}{1}\,\LPA{1}{2}\\
&\,\qquad-288\,\LSPC{2}{2}{1}\,\LPA{1}{2}+2396\,\LPA{1}{4}\,\zeta_3+1830\,\LSPA{1}\,\LPA{3}{2}-1612\,\LSMA{4}\,\LSMA{2}\,\LSPA{1}\\
&\,\qquad+1344\,\LSMA{2}\,\LSMA{0}\,\LSPB{3}{1}-520\,\LSMA{2}\,\LSMC{2}{1}{1}\,\LSPA{1}+11839\,\LSMA{0}\,\LSMB{4}{1}\,\LSPA{1}\\
&\,\qquad+4330\,\LSMA{0}\,\LSMB{3}{2}\,\LSPA{1}+3780\,\LSMA{0}\,\LSMB{2}{1}\,\LSPA{3}+562\,\LSMA{0}\,\LSMD{2}{1}{1}{1}\,\LSPA{1}\\
&\,\qquad+3556\,\LSPA{1}\,\zeta_3^2+2256\,\LSPA{7}-164778\,\LSPC{5}{1}{1}-33216\,\LSPC{4}{1}{2}-89088\,\LSPC{4}{2}{1}\\
&\,\qquad-33912\,\LSPC{3}{3}{1}-12048\,\LSPC{3}{2}{2}-17820\,\LSPE{3}{1}{1}{1}{1}-2928\,\LSPE{2}{1}{2}{1}{1}\\
&\,\qquad-8784\,\LSPE{2}{2}{1}{1}{1}-23796\,\zeta_7\Bigg\}
+b_1\, \zeta_2\, \LMA{2}{2} \LSPA{1} + b_2\,\zeta_2 \, \LMA{0}{2} \LSPA{1} \gFL{2}{1}\\
\,&+b_3\,\zeta_2 \, \gFL{2}{1}\, \gFL{3}{2} +b_4\,\zeta_2\, \gFL{2}{0}\, \gFL{2}{1}+b_5\,\zeta_2\, \hFL{4}{1} \\
&\,+b_6\,\zeta_2\, \hFL{5}{3} +b_7\,\zeta_2\, \gFL{3}{0}+b_8\,\zeta_2\, \gFL{4}{2} +b_9\,\zeta_2\, \gFL{5}{4}\\
&\,+b_{10}\, \zeta_3\, \hFL{4}{2}+b_{11}\, \zeta_3\, \hFL{3}{0}+b_{12}\, \zeta_3 \,[\gFL{2}{1}]^2\\
&\,+b_{13} \,\zeta_3 \,\gFL{4}{3}+b_{14}\, \zeta_3\, \gFL{3}{1}+b_{15}\,\zeta_4\, \gFL{3}{2} +b_{16}\,\zeta_4\, \gFL{2}{0}\\
&\,+b_{17}\, \zeta_3\,\zeta_2\, \gFL{2}{1} +b_{18}\, \zeta_5\, \gFL{2}{1}
\,.
\esp\eeq
In these expressions, $a_i$ for $i=0,\ldots,8$, and $b_j$ for $j=1,\ldots, 18$, denote undetermined rational numbers. The one symbol-level parameter, $a_0$, enters both $g_1^{(4)}$ and $g_0^{(4)}$. We observe that $a_0$ enters these formulae in a complicated way, and that there is no nonzero value of $a_0$ that simplifies the associated large rational numbers. We therefore suspect that $a_0=0$, although we currently have no proof. The remaining parameters account for beyond-the-symbol ambiguities. We will see in the next section that one of these parameters, $b_1$, is not independent of the others.

\subsection{Analytic results for the NNLL correction to the BFKL eigenvalue and the N$^3$LL correction to the impact factor}
Having at our disposal analytic expressions for the four-loop remainder function at NNLLA and N$^3$LLA, we use these results to extract the BFKL eigenvalue and the impact factors to the same accuracy in perturbation theory. We proceed as in Section~\ref{sec:Phi2}, i.e., we use our knowledge of the space of SVHPLs and the corresponding functions in $(\nu,n)$ space to find a function whose inverse Fourier-Mellin transform reproduces the four-loop results we have derived.

Let us start with the computation of the BFKL eigenvalue at NNLLA. 
Expanding eq.~\eqref{eq:MHV_MRK_2} to order $a^4$, we can extract the following relation,
\beq\bsp\label{eq:I[E2]}
\mathcal{I}\left[E^{(2)}_{\nu,n}\right] &\,= 
12\,\left\{\LPA{1}{2}+\pi ^2\right\}g^{(4)}_3(w,\ws)
-8\,\LSPA{1}\,g^{(4)}_2(w,\ws)
+4\,g^{(4)}_1(w,\ws)\\
&\,
-8\,\LSPA{1}\,\pi ^2\,g^{(3)}_2(w,\ws)
+2\,\pi ^2\,g^{(2)}_1(w,\ws)\,\LPA{1}{2}\\
&\,-\mathcal{I}\left[E^{(1)}_{\nu,n}\, \Phi^{(1)}_{\textrm{Reg}}(\nu,n)\right]-\mathcal{I}\left[E_{\nu,n}\, \Phi^{(2)}_{\textrm{Reg}}(\nu,n)\right]\,.
\esp\eeq
The right-hand side of eq.~\eqref{eq:I[E2]} is completely known, up to some rational numbers mostly parametrizing our ignorance of beyond-the-symbol terms in the three- and four-loop coefficient functions at NNLLA. It can be written exclusively in terms of SVHPLs of weight six with eigenvalue $(+,+)$ under $\mathbb{Z}_2\times\mathbb{Z}_2$ transformations. The results of Section~\ref{sec:nu_n} then allow us to write down an ansatz for the NNLLA correction to the BFKL eigenvalue, similar to the ansatz~\eqref{eq:Phi_2_ansatz} we made for the NNLLA correction to the impact factor, but at higher weight. More precisely, we assume that we can write $E^{(2)}_{\nu,n} = \sum_i\alpha_i\,P_i$, where $\alpha_i$ denote rational numbers and $P_i$ runs through all possible monomials of weight five with the correct symmetry properties that we can construct out of the building blocks given in eq.~\eqref{mod_basis}, i.e.,
\beq
P_i\in\left\{\Enun{5}, \zeta_2\,V\,\dEOne, \EnunOne\,N\,\Ffourtilde,\zeta_5,\ldots\right\}\,.
\eeq
The rational coefficients $\alpha_i$ can then be fixed by inserting our ansatz into eq.~\eqref{eq:I[E2]} and performing the inverse Fourier-Mellin transform to $(w,\ws)$ space. We find that there is a unique solution for the $\alpha_i$, and the result for the NNLLA correction to the BFKL eigenvalue then takes the form,
\beq\bsp\label{eq:E2}
E^{(2)}_{\nu,n} &\,= -E^{(1)}_{\nu,n}\,\Phi^{(1)}_{\textrm{Reg}}(\nu,n) -E_{\nu,n}\,\Phi^{(2)}_{\textrm{Reg}}(\nu,n)+\frac{3}{8}\,\dE{2}\,\Enun{2}+\frac{3}{32}\,N^2\,\dE{2}+\frac{1}{8}\,V^2\,\dE{2}\\
&\,-\frac{1}{8}\,V\,\dE{3}+\frac{1}{48}\,\dE{4}+\frac{\pi ^2}{12}\,\dE{2}-\frac{3}{4}\,\dEOne\,V\,\Enun{2}-\frac{5}{16}\,\dEOne\,N^2\,V\\
&\,-\frac{\pi ^2}{4}\,\dEOne\,V+\frac{1}{8}\,\EnunOne\,\dEPOne{2}+\frac{3}{16}\,N^2\,\Enun{3}+\frac{61}{4}\,\Enun{2}\,\zeta_3+\frac{1}{8}\,\Enun{5}+\frac{5 \pi ^2}{6}\,\Enun{3}\\
&\,+\frac{19}{128}\,\EnunOne\,N^4+\frac{5}{16}\,\EnunOne\,N^2\,V^2+\frac{3 \pi ^2}{16}\,\EnunOne\,N^2+\frac{\pi ^2}{4}\,\EnunOne\,V^2+\frac{35}{16}\,N^2\,\zeta_3+\frac{1}{2}\,V^2\,\zeta_3\\
&\,+\frac{11 \pi ^2}{6}\,\zeta_3+10\,\zeta_5 + a_0\,\mathcal{E}_5 + \sum_{i=1}^5a_i\,\zeta_2\,\mathcal{E}_{3,i} + a_6\,\zeta_4\,\mathcal{E}_2 + \sum_{i=7}^8a_i\,\zeta_3\,\mathcal{E}_{1,i}\,,
\esp\eeq
where the quantities $\mathcal{E}_{3,i}$,  $\mathcal{E}_{2}$, and $\mathcal{E}_{1,i}$ capture the beyond-the-symbol ambiguities in $g_1^{(4)}$, and $\mathcal{E}_{5}$ corresponds to the one symbol-level ambiguity. They are given by,
\begin{eqnarray}
\nonumber\mathcal{E}_5 &=& \frac{124}{3}\,N^2\,\dE{2}+\frac{1210}{3}\,V^2\,\dE{2}-\frac{35}{3}\,V\,\dE{3}-\frac{31}{6}\,\dE{4}-\frac{151}{2}\,\dEOne\,N^2\,V\\
&&+\frac{124}{3}\,N^2\,\Enun{3}-\frac{140}{3}\,V^2\,\Enun{3}-\frac{31}{2}\,\EnunOne\,N^4+\frac{10903}{12}\,N^2\,\zeta_3+\frac{13960}{3}\,V^2\,\zeta_3\\
\nonumber&&-62\,\dE{2}\,\Enun{2}+70\,\dEOne\,V\,\Enun{2}-760\,\dEOne\,V^3+248\,\EnunOne\,\dEPOne{2}\\
\nonumber&&+7431\,\Enun{2}\,\zeta_3-97\,\EnunOne\,N^2\,V^2+16072\,\zeta_5\,,\\
\mathcal{E}_{3,1} &=&-\frac{3}{4}\,\EnunOne\,N^2-\dE{2}+5\,\Enun{3}+6\,\EnunOne\,V^2-2\,\EnunOne\,\pi ^2+8\,\zeta_3\,,\\
\mathcal{E}_{3,2} &=&\Enun{3}\,,\\
\mathcal{E}_{3,3} &=&\frac{3}{4}\,\EnunOne\,N^2-3\,\dEOne\,V+3\,\Enun{3}+12\,\zeta_3\,,\\
\mathcal{E}_{3,4} &=&
-\frac{1}{8}\,\dE{2}+\frac{9}{4}\,\dEOne\,V-\frac{3}{4}\,\EnunOne\,N^2-\frac{3}{2}\,\EnunOne\,V^2-\frac{25}{2}\,\zeta_3-2\,\Enun{3}
\,,\\
\mathcal{E}_{3,5} &=&
\frac{3}{8}\,\EnunOne\,N^2-\frac{3}{2}\,\Enun{3}\,,\\
\mathcal{E}_{2} &=& 90\,\EnunOne\,,\\
\mathcal{E}_{1,7} &=& \Enun{2}-\frac{1}{4}\,N^2\,,\\
\mathcal{E}_{1,8} &=&\frac{1}{2}\,\Enun{2}\,.
\end{eqnarray}
We observe that the most complicated piece is $\mathcal{E}_{5}$. It would be absent if our conjecture that $a_0=0$ is correct. Some further comments are in order about eq.~\eqref{eq:E2}:
\begin{enumerate}
\item In ref.~\cite{Fadin2011we} it was argued, based on earlier work~\cite{Lipatov1976zz,Kuraev1976ge,Balitsky1978ic,Fadin2006bj}, that the BFKL eigenvalue should vanish as $(\nu,n)\to0$ to all orders in perturbation theory, i.e., $\omega(0,0)=0$. While this statement depends on how one approaches the limit, the most natural way seems to be to set the discrete variable $n$ to 0 before taking the limit $\nu\to0$.  Indeed in this limit $E_{\nu,n}$ and $E^{(1)}_{\nu,n}$ vanish. However, we find that $E^{(2)}_{\nu,n}$ does not vanish in this limit, but rather it approaches a constant,
\beq\label{eq:E2nuto0}
\lim_{\nu\to0}E^{(2)}_{\nu,0} = -\frac{1}{2}\pi^2\,\zeta_3\,.
\eeq
We stress that the limit is independent of any of the undetermined constants that parameterize the beyond-the-symbol terms in the three- and four-loop coefficients.
While we have confidence in our result for $E^{(2)}_{\nu,n}$ given our assumptions (such as the vanishing of $g_n^{(\ell)}$ and $h_n^{(\ell)}$ as $w\to0$), we have so far no explanation for this observation. 
\item While the $(\nu,n)$-space basis constructed in Section~\ref{sec:nu_n} involves the new functions $\Ffourtilde$, $\Fsixatilde$ and $\Fseventilde$, we find that $E^{(2)}_{\nu,n}$ is free of these functions and can be expressed entirely in terms of $\psi$ functions and rational functions of $\nu$ and $n$. Moreover, the $\psi$ functions arise only in the form of the LLA BFKL eigenvalue and its derivative with respect to $\nu$. We are therefore led to conjecture that, to all loop orders, the BFKL eigenvalue and the impact factor can be expressed as linear combinations of uniform weight of monomials that are even in both $\nu$ and $n$ and are constructed exclusively out of multiple $\zeta$ values\footnote{Note that we can not exclude the appearance of multiple $\zeta$ values at higher weights, as multiple $\zeta$ values are reducible to ordinary $\zeta$ values until weight eight.} and the quantities $N$, $V$, $\EnunOne$ and $D_\nu$ defined in Section~\ref{sec:nu_n}.
\end{enumerate}

We now move on and and extract the impact factor at N$^3$LLA from the four-loop amplitude at the same logarithmic accuracy. Equation~\eqref{eq:MHV_MRK_2} at order $a^4$ yields the following relation for the impact factor at N$^3$LLA,
\beq\bsp\label{eq:I[Phi3]}
\mathcal{I}\left[\Phi^{(3)}_{\textrm{Reg}}(\nu,n)\right] &\, =
-4\,\left\{\LPA{1}{3}+3\,\LSPA{1}\,\pi ^2\right\}\,g^{(4)}_3(w,\ws)
+4\left\{\LPA{1}{2}+\pi ^2\right\}\,g^{(4)}_2(w,\ws)\\
&\,-4\,\LSPA{1}\,g^{(4)}_1(w,\ws)
+4\,g^{(4)}_0(w,\ws)
+8\,\pi ^2\,g^{(3)}_2(w,\ws)\,\LPA{1}{2}\\
&\,
-4\,\LSPA{1}\,\pi ^2\,g^{(3)}_1(w,\ws)
-2\,\pi ^2\left\{\LPA{1}{3}-\frac{\pi ^2}{3}\,\LSPA{1}\right\}\,g^{(2)}_1(w,\ws)\\
&\,
+2\,\pi ^2\,g^{(2)}_0(w,\ws)\,\LPA{1}{2}
+\frac{\pi ^4}{8}\,\LSPA{1}\,\LMA{0}{2}-\frac{\pi ^4}{3}\,\LPA{1}{3}-\frac{73 \pi ^6}{1260}\,\LSPA{1}-2\,\LSPA{1}\,\zeta_3^2\,.
\esp\eeq
In order to determine $\Phi^{(3)}_{\textrm{Reg}}(\nu,n)$, we proceed in the same way as we did for $E^{(2)}_{\nu,n}$, i.e., we write down an ansatz for $\Phi^{(3)}_{\textrm{Reg}}(\nu,n)$ that has the correct transcendentality and symmetry properties and fix the free coefficients by requiring the inverse Fourier-Mellin transform of the ansatz to match the right-hand side of eq.~\eqref{eq:I[Phi3]}. Building upon our conjecture that the impact factor can be expressed purely in terms of $\psi$ functions and rational functions of $\nu$ and $n$, we construct a restricted ansatz\footnote{We have constructed the full basis of functions in $(\nu,n)$ space through weight six and the explicit map to $(w,\ws)$ functions of weight seven. It is therefore not necessary for us to restrict our ansatz in this way. It is, however, sufficient, and computationally simpler to do so.} that is a linear combination just of monomials of $\zeta$ values and $N$, $V$, $D_\nu$ and $\EnunOne$. Just like in the case of $E^{(2)}_{\nu,n}$, we find that there is a unique solution for the coefficients in the ansatz, thus giving further support to our conjecture. Furthermore, we are forced along the way to fix one of the beyond-the-symbol parameters appearing in $g_0^{(4)}$,
\beq
b_1=-\frac{15}{8}\,a_1-\frac{3}{16}\,a_2-\frac{3}{32}\,a_4+\frac{9}{16}\,a_5+\frac{1}{64}\,b_3+\frac{1}{8}\,b_4-\frac{3}{16}\,b_5-\frac{1}{32}\,b_6+\frac{1}{4}\,b_7+\frac{3}{32}\,b_8+\frac{3}{16}\,.
\eeq
The final result for the impact factor at N$^3$LLA then takes the form,
\begin{align}
\Phi^{(3)}_{\textrm{Reg}}(\nu,n) & ={1\over 3}\left[\Phi^{(1)}_{\textrm{Reg}}(\nu,n)\right]^3-E^{(2)}_{\nu,n}\,\EnunOne -\Phi^{(2)}_{\textrm{Reg}}(\nu,n)\,\Enun{2}
-\frac{1}{24}\,\dEP{2}{2}\\
\nonumber&+\frac{1}{4}\,\dEOne\,V\,\dE{2}-\frac{1}{24}\,\dEOne\,\dE{3}+\frac{1}{8}\,\dE{2}\,\Enun{3}-\frac{3}{32}\,\EnunOne\,N^2\,\dE{2}\\
\nonumber&-\frac{37 \pi ^2}{96}\,\EnunOne\,\dE{2}-\frac{1}{24}\,\dE{2}\,\zeta_3-\frac{1}{4}\,\dEOne\,V\,\Enun{3}+\frac{3}{16}\,\dEOne\,\EnunOne\,N^2\,V\\
\nonumber&+\frac{11 \pi ^2}{24}\,\dEOne\,\EnunOne\,V+\frac{9}{4}\,\dEOne\,V\,\zeta_3+\frac{1}{16}\,\dEPOne{2}\,\Enun{2}-\frac{3}{64}\,N^2\,\dEPOne{2}
\end{align}
\begin{align}
\nonumber
\phantom{\Phi^{(3)}_{\textrm{Reg}}(\nu,n)}&-\frac{1}{8}\,V^2\,\dEPOne{2}+\frac{3 \pi ^2}{32}\,\dEPOne{2}+\frac{37}{256}\,N^4\,\Enun{2}+\frac{5}{32}\,N^2\,V^2\,\Enun{2}\\
\nonumber&-\frac{23 \pi ^2}{128}\,N^2\,\Enun{2}-\frac{21 \pi ^2}{32}\,V^2\,\Enun{2}+\frac{161}{12}\,\Enun{3}\,\zeta_3+\frac{7}{48}\,\Enun{6}+\frac{\pi ^2}{3}\,\Enun{4}-\frac{\pi ^4}{72}\,\Enun{2}\\
\nonumber&+\frac{7}{16}\,\EnunOne\,N^2\,\zeta_3-\frac{13 \pi ^2}{2}\,\EnunOne\,\zeta_3-\frac{45}{1024}\,N^6-\frac{41}{128}\,N^4\,V^2+\frac{5 \pi ^2}{512}\,N^4-\frac{3}{16}\,N^2\,V^4\\
\nonumber&-\frac{5 \pi ^2}{128}\,N^2\,V^2+\frac{\pi ^4}{24}\,N^2+\frac{\pi ^4}{8}\,V^2+\frac{5}{2}\,\zeta_3^2-\frac{311 \pi ^6}{11340}+3\,\EnunOne\,V^2\,\zeta_3+10\,\EnunOne\,\zeta_5\\
\nonumber&+\frac{15}{64}\,N^2\,\Enun{4}+a_0\,\mathcal{P}_6+\sum_{i=1}^5a_i\,\zeta_2\,\mathcal{P}_{a,4,i}+a_6\,\zeta_4\,\mathcal{P}_{a,2}
+\sum_{i=7}^8a_i\,\zeta_3\,\mathcal{P}_{a,3,i}\\
\nonumber&
+\sum_{i=2}^9b_i\,\zeta_2\,\mathcal{P}_{b,4,i}
+\sum_{i=10}^{14}b_i\,\zeta_3\,\mathcal{P}_{b,3,i}
+\sum_{i=15}^{16}b_i\,\zeta_4\,\mathcal{P}_{b,2,i}
+b_{17}\,\zeta_2\zeta_3\,\mathcal{P}_{b,1,1}
+b_{18}\,\zeta_5\,\mathcal{P}_{b,1,2}\,,
\end{align}
where $\mathcal{P}_{i,j,\ldots}$ parametrize the beyond-the-symbol terms in the four-loop coefficient functions, and $\mathcal{P}_6$ parameterizes the one symbol-level ambiguity,
\begin{eqnarray}
\mathcal{P}_6&=& \frac{105}{2}\,\dEP{2}{2}-\frac{152}{3}\,\EnunOne\,N^2\,\dE{2}-\frac{2690}{3}\,\EnunOne\,V^2\,\dE{2}+\frac{595}{3}\,\EnunOne\,V\,\dE{3}
\nonumber\\
&&-\frac{7}{6}\,\EnunOne\,\dE{4}-\frac{10455}{2}\,\dE{2}\,\zeta_3+\frac{249}{8}\,N^2\,\dEPOne{2}+\frac{2655}{2}\,V^2\,\dEPOne{2}\\
\nonumber&&+\frac{103}{16}\,N^4\,\Enun{2}+\frac{317}{4}\,N^2\,V^2\,\Enun{2}+\frac{197}{24}\,N^2\,\Enun{4}+\frac{515}{6}\,V^2\,\Enun{4}+\frac{61793}{6}\,\EnunOne\,N^2\,\zeta_3\\
\nonumber&&+\frac{13777}{3}\,\EnunOne\,V^2\,\zeta_3+\frac{111}{128}\,N^6+\frac{345}{32}\,N^4\,V^2-385\,\dEOne\,V\,\dE{2}-30\,\dEOne\,\dE{3}\\
\nonumber&&+16\,\dE{2}\,\Enun{3}-420\,\dEOne\,V\,\Enun{3}+7\,\dEOne\,\EnunOne\,N^2\,V-760\,\dEOne\,\EnunOne\,V^3\\
\nonumber&&-22606\,\dEOne\,V\,\zeta_3-34\,\dEPOne{2}\,\Enun{2}+1140\,V^4\,\Enun{2}+15231\,\Enun{3}\,\zeta_3+6548\,\EnunOne\,\zeta_5\\
\nonumber&&+46992\,\zeta_3^2\,,\\
\mathcal{P}_{a,4,1}&=&
\frac{5}{8}\,\EnunOne\,\dE{2}-\frac{3}{2}\,\dEOne\,\EnunOne\,V+\frac{33}{8}\,\dEPOne{2}-\frac{183}{32}\,N^2\,\Enun{2}\\
&&\nonumber-\frac{129}{8}\,V^2\,\Enun{2}-\frac{5}{4}\,\Enun{4}+\frac{3}{128}\,N^4+\frac{171}{32}\,N^2\,V^2+\frac{\pi ^2}{4}\,N^2+\pi ^2\,\Enun{2}-68\,\EnunOne\,\zeta_3\,,\\
\mathcal{P}_{a,4,2}&=&-\frac{3}{16}\,\EnunOne\,\dE{2}+\frac{3}{4}\,\dEOne\,\EnunOne\,V+\frac{7}{16}\,\dEPOne{2}-\frac{51}{64}\,N^2\,\Enun{2}\\
\nonumber&&-\frac{33}{16}\,V^2\,\Enun{2}-\frac{1}{4}\,\Enun{4}-\frac{7}{256}\,N^4+\frac{19}{64}\,N^2\,V^2-12\,\EnunOne\,\zeta_3\,,\\
\mathcal{P}_{a,4,3}&=&-\frac{3}{2}\,\EnunOne\,\dE{2}+\frac{9}{4}\,\dEPOne{2}-\frac{3}{2}\,N^2\,\Enun{2}-\frac{9}{2}\,V^2\,\Enun{2}-\frac{3}{4}\,\Enun{4}-\frac{9}{64}\,N^4\\
\nonumber&&+\frac{9}{8}\,N^2\,V^2+6\,\dEOne\,\EnunOne\,V-48\,\EnunOne\,\zeta_3\,,
\end{eqnarray}
\begin{eqnarray}
\mathcal{P}_{a,4,4}&=&\frac{49}{32}\,\EnunOne\,\dE{2}-\frac{27}{8}\,\dEOne\,\EnunOne\,V-\frac{45}{32}\,\dEPOne{2}+\frac{117}{128}\,N^2\,\Enun{2}\\
\nonumber&&+\frac{111}{32}\,V^2\,\Enun{2}+\frac{1}{2}\,\Enun{4}+\frac{73}{2}\,\EnunOne\,\zeta_3+\frac{69}{512}\,N^4-\frac{21}{128}\,N^2\,V^2\,,\\
\mathcal{P}_{a,4,5}&=&-\frac{3}{16}\,\EnunOne\,\dE{2}-\frac{3}{4}\,\dEOne\,\EnunOne\,V-\frac{15}{16}\,\dEPOne{2}+\frac{105}{64}\,N^2\,\Enun{2}\\
\nonumber&&+\frac{63}{16}\,V^2\,\Enun{2}+\frac{3}{8}\,\Enun{4}+\frac{3}{256}\,N^4-\frac{69}{64}\,N^2\,V^2+18\,\EnunOne\,\zeta_3\,,\\
\mathcal{P}_{a,2}&=&{-\frac{45}{4}\,N^2-45\,\Enun{2}}\,,
\\
\mathcal{P}_{a,3,7}&=&\frac{1}{6}\,\dE{2}-\frac{1}{3}\,\Enun{3}-\frac{4}{3}\,\zeta_3-\EnunOne\,V^2\,,\\
\mathcal{P}_{a,3,8}&=&-\frac{1}{24}\,\dE{2}+\frac{1}{4}\,\dEOne\,V-\frac{1}{6}\,\Enun{3}-\frac{1}{8}\,\EnunOne\,N^2-\frac{1}{2}\,\EnunOne\,V^2-\frac{13}{6}\,\zeta_3\,,\\
\mathcal{P}_{b,4,2}&=&\frac{3}{4}\,N^2\,\Enun{2}+\frac{3}{16}\,N^4+\frac{21}{4}\,N^2\,V^2+3\,\EnunOne\,\dE{2}+12\,\dEOne\,\EnunOne\,V\\
\nonumber&&+3\,\dEPOne{2}+9\,V^2\,\Enun{2}\,,\\
\mathcal{P}_{b,4,3}&=&\frac{7}{192}\,\EnunOne\,\dE{2}-\frac{7}{16}\,\dEOne\,\EnunOne\,V-\frac{9}{64}\,\dEPOne{2}+\frac{33}{256}\,N^2\,\Enun{2}\\
\nonumber&&+\frac{19}{64}\,V^2\,\Enun{2}+\frac{5}{24}\,\Enun{4}+\frac{37}{12}\,\EnunOne\,\zeta_3+\frac{9}{1024}\,N^4-\frac{1}{256}\,N^2\,V^2\,,\\
\mathcal{P}_{b,4,4}&=&-\frac{5}{24}\,\EnunOne\,\dE{2}-\frac{1}{2}\,\dEOne\,\EnunOne\,V-\frac{3}{8}\,\dEPOne{2}+\frac{9}{32}\,N^2\,\Enun{2}\\
\nonumber&&+\frac{7}{8}\,V^2\,\Enun{2}+\frac{5}{12}\,\Enun{4}+\frac{14}{3}\,\EnunOne\,\zeta_3+\frac{3}{128}\,N^4+\frac{11}{32}\,N^2\,V^2\,,\\
\mathcal{P}_{b,4,5}&=&\frac{3}{16}\,\EnunOne\,\dE{2}+\frac{1}{2}\,\dEOne\,\EnunOne\,V+\frac{1}{2}\,\dEPOne{2}-\frac{31}{64}\,N^2\,\Enun{2}\\
&&\nonumber-\frac{27}{16}\,V^2\,\Enun{2}-\frac{9}{16}\,\Enun{4}+\frac{\pi ^2}{8}\,\Enun{2}-\frac{1}{128}\,N^4-\frac{3}{64}\,N^2\,V^2+\frac{\pi ^2}{32}\,N^2-8\,\EnunOne\,\zeta_3\,,\\
\mathcal{P}_{b,4,6}&=&-\frac{5}{96}\,\EnunOne\,\dE{2}+\frac{1}{2}\,\dEOne\,\EnunOne\,V+\frac{17}{96}\,\dEPOne{2}-\frac{25}{128}\,N^2\,\Enun{2}\\
\nonumber&&-\frac{15}{32}\,V^2\,\Enun{2}-\frac{11}{48}\,\Enun{4}-\frac{49}{12}\,\EnunOne\,\zeta_3-\frac{17}{1536}\,N^4+\frac{11}{384}\,N^2\,V^2\,,\\
\mathcal{P}_{b,4,7}&=&\Phi^{(2)}_{\textrm{Reg}}(\nu,n)-\frac{2}{3}\,\EnunOne\,\dE{2}-\frac{3}{8}\,\dEPOne{2}+\frac{1}{4}\,N^2\,\Enun{2}+\frac{7}{4}\,V^2\,\Enun{2}\\
&&\nonumber-\frac{\pi ^2}{2}\,\Enun{2}+\frac{1}{3}\,\EnunOne\,\zeta_3-\frac{5}{128}\,N^4-\frac{7}{8}\,N^2\,V^2+\frac{\pi ^2}{48}\,N^2+\frac{\pi ^2}{4}\,V^2-\frac{11 \pi ^4}{180}\\
\nonumber&&+\frac{5}{24}\,\Enun{4}+\dEOne\,\EnunOne\,V\,,\\
\mathcal{P}_{b,4,8}&=&-\frac{5}{32}\,\EnunOne\,\dE{2}+\frac{1}{8}\,\dEOne\,\EnunOne\,V-\frac{7}{32}\,\dEPOne{2}+\frac{27}{128}\,N^2\,\Enun{2}\\
\nonumber&&+\frac{33}{32}\,V^2\,\Enun{2}+\frac{3}{8}\,\Enun{4}-\frac{\pi ^2}{4}\,\Enun{2}+\frac{7}{512}\,N^4-\frac{19}{128}\,N^2\,V^2+3\,\EnunOne\,\zeta_3\,,
\end{eqnarray}
\begin{eqnarray}
\mathcal{P}_{b,4,9}&=&\frac{1}{24}\,\Enun{4}\,,\\
\mathcal{P}_{b,3,10}&=&-\frac{1}{48}\,\dE{2}+\frac{3}{8}\,\dEOne\,V-\frac{1}{3}\,\Enun{3}-\frac{1}{8}\,\EnunOne\,N^2-\frac{1}{4}\,\EnunOne\,V^2-\frac{25}{12}\,\zeta_3\,,\\
\mathcal{P}_{b,3,11}&=&\frac{1}{16}\,\EnunOne\,N^2-\frac{1}{4}\,\Enun{3}\,,\\
\mathcal{P}_{b,3,12}&=&-\frac{1}{2}\,\dEOne\,V+\frac{1}{2}\,\Enun{3}+\frac{1}{8}\,\EnunOne\,N^2+2\,\zeta_3\,,\\
\mathcal{P}_{b,3,13}&=&\frac{1}{6}\,\Enun{3}\,,\\
\mathcal{P}_{b,3,14}&=&-\frac{1}{6}\,\dE{2}+\frac{5}{6}\,\Enun{3}-\frac{1}{8}\,\EnunOne\,N^2-\frac{\pi ^2}{3}\,\EnunOne+\frac{4}{3}\,\zeta_3+\EnunOne\,V^2\,,\\
\mathcal{P}_{b,2,15}&=&\frac{1}{2}\,\Enun{2}\,,\\
\mathcal{P}_{b,2,16}&=&\Enun{2}-\frac{1}{4}\,N^2\,,\\
\mathcal{P}_{b,1,1}&=&\EnunOne\,,\\
\mathcal{P}_{b,1,2}&=&\EnunOne\,.
\end{eqnarray}
 Again, the undetermined function at symbol level, $\mathcal{P}_6$, is the most complicated term, but it would be absent if $a_0=0$.

Finally, we remark that the $\nu\to0$ behavior of
$\Phi^{(\ell)}_{\textrm{Reg}}(\nu,n)$ is nonvanishing, and even singular
for $\ell=2$ and 3.  Taking the limit after setting $n=0$, as in the case of
$E_{\nu,n}^{(2)}$, we find that the constant term is given
in terms of the cusp anomalous dimension,
\bea
\label{eq:Phi1nuto0}
\lim_{\nu\to0}\Phi^{(1)}_{\textrm{Reg}}(\nu,0)
&\sim& \frac{\gamma_K^{(2)}}{4} \,+\,{\cal O}(\nu^4) \,, \\
\label{eq:Phi2nuto0}
\lim_{\nu\to0}\Phi^{(2)}_{\textrm{Reg}}(\nu,0)
&\sim& \frac{\pi^2}{4\,\nu^2} + \frac{\gamma_K^{(3)}}{4} 
\,+\,{\cal O}(\nu^2)\,, \\
\label{eq:Phi3nuto0}
\lim_{\nu\to0}\Phi^{(3)}_{\textrm{Reg}}(\nu,0)
&\sim& - \frac{\pi^4}{8\,\nu^2} + \frac{\gamma_K^{(4)}}{4} 
\,+\,{\cal O}(\nu^2)\,.
\eea
This fact is presumably related to the appearance of $\gamma_K(a)$ in
the factors $\omega_{ab}$ and $\delta$, which carry logarithmic
dependence on $|w|$ as $w\to0$.  It may play a role in understanding
the failure of $E^{(2)}_{\nu,0}$ to vanish as $\nu\to0$
in eq.~\eqref{eq:E2nuto0}.


\section{Conclusions and Outlook}
\label{sec:Concl}

In this article we exposed the structure of the multi-Regge limit
of six-gluon scattering in planar $\cN=4$ super-Yang-Mills theory
in terms of the single-valued harmonic polylogarithms introduced by Brown.
Given the finite basis of such functions, it is extremely simple to
determine any quantity that is defined by a power series expansion
around the origin of the $(w,\ws)$ plane.  Two examples which we could
evaluate with no ambiguity are the LL and NLL terms in the multi-Regge
limit of the MHV amplitude.  We could carry this exercise out through
transcendental weight 10, and we presented the analytic formulae
explicitly through six loops in Section~\ref{sec:R6_LLA_NLLA}.
The NMHV amplitudes also fit into the same mathematical framework,
as we saw in Section~\ref{sec:NMHV}:  An integro-differential
operator that generates the NMHV LLA terms from
the MHV LLA ones~\cite{Lipatov2012gk} has a very natural action on
the SVHPLs, making it simple to generate NMHV LLA results
to high order as well.  A clear avenue for future investigation
utilizing the SVHPLs is the NMHV six-point amplitude at 
next-to-leading-logarithm and beyond.  

A second thrust of this article was to understand the Fourier-Mellin
transform from $(w,w^*)$ to $(\nu,n)$ variables.  In practice,
we constructed this map in the reverse direction:  We built an
ansatz out of various elements: harmonic sums and specific rational
combinations of $\nu$ and $n$.  We then implemented the inverse Fourier-Mellin
transform as a truncated sum, or power series around the origin of the
$(w,\ws)$ plane, and matched to the basis of SVHPLs.
We thereby identified specific combinations of the elements as building
blocks from which to generate the full set of SVHPL Fourier-Mellin transforms.
We have executed this procedure completely through weight six in the $(\nu,n)$
space, corresponding to weight seven in the $(w,\ws)$ space.
In generalizing the procedure to yet higher weight, we expect the procedure to be
much the same. Beginning with a linear combination of weight $(p-2)$ HPLs in a single variable $x$, perform
a Mellin transformation to produce weight $(p-1)$ harmonic sums such as
$\psi$, $F_4$, $F_{6a}$, etc.  For suitable combinations of these elements, the inverse Fourier-Mellin transform
will generate weight $p$ SVHPLs in the complex conjugate pair $(w,\ws)$. 
The step of determining which combinations of elements correspond to the
SVHPLs was carried out empirically in this paper.  It would be
interesting to investigate further the mathematical properties of
these building blocks.

Using our understanding of the Fourier-Mellin transform, we could
explicitly evaluate the NNLL MHV impact factor $\Phi^{(2)}_{\textrm{Reg}}(\nu,n)$
which derives from a knowledge of the three-loop remainder function
in the MRK limit~\cite{Dixon2011pw,Fadin2011we}.  We then went
on to four loops, using a computation of the four-loop
symbol~\cite{fourloop} in conjunction with additional constraints
from the multi-Regge limit to determine the MRK symbol up to one
free parameter $a_0$ (which we suspect is zero).  We matched this symbol
to the symbols of the SVHPLs in order to determine the complete
four-loop remainder function in MRK, up to a number of beyond-the-symbol
constants.  This data, in particular $g_1^{(4)}$ and $g_0^{(4)}$, then
led to the NNLL BFKL eigenvalue $E_{\nu,n}^{(2)}$ and
N$^3$LL impact factor $\Phi^{(3)}_{\textrm{Reg}}(\nu,n)$.
These quantities also contain the various beyond-the-symbol
constants.  Clearly the higher-loop NNLL MRK terms
can be determined just as we did at LL and NLL, using the master
formula~\eqref{eq:MHV_MRK} and the SVHPL basis.
However, it would also be worthwhile to understand what constraint
can fix $a_0$, and the host of beyond-the-symbol constants, since they
will afflict all of these terms.  This task may require backing away
somewhat from the multi-Regge limit, or utilizing coproduct information
in some way.

We also remind the reader that we found that the NNLL BFKL eigenvalue
$E_{\nu,n}^{(2)}$ does not vanish as $\nu\to0$, taking the limit after 
setting $n=0$.  This behavior is in contrast to what happens in 
the LL and NLL case.  It also goes against the expectations in
ref.~\cite{Fadin2011we}, and thus calls for further study.

Although the structure of QCD amplitudes in the multi-Regge limit is more
complicated than those of planar $\cN=4$ super-Yang-Mills theory, one can
still hope that the understanding of the Fourier-Mellin $(\nu,n)$ space
that we have developed here may prove useful in the QCD context.

Finally, we remark that the SVHPLs are very likely to be applicable
to another current problem in $\cN=4$ super-Yang-Mills theory, namely
the determination of correlation functions for four off-shell operators.
Conformal invariance implies that these quantities depend on two separate
cross ratios.  The natural arguments of the polylogarithms that appear
at low loop order, after a change of variables from the original
cross ratios, are again a complex pair $(w,\ws)$ (or $(z,\bar{z})$).
The same single-valued conditions apply here as well.
For example, the one-loop off-shell box integral
that enters the correlation function is proportional to
$L_2^-(z,\bar{z})/(z-\bar{z})$.  We expect that the SVHPL framework
will allow great progress to be made in this arena, just as it has
to the study of the multi-Regge limit.


\section*{Acknowledgments}

We thank Vittorio Del Duca for useful discussions, and Johannes Bl\"{u}mlein
and Francis Brown for helpful comments on the manuscript.
This research was supported by the US Department of Energy under
contract DE--AC02--76SF00515 and by the ERC grant ``IterQCD''.


\appendix
\section{Single-valued harmonic polylogarithms}
\label{app:svhpl}
\subsection{Expression of the $L^\pm$ functions in terms of ordinary HPLs}
In this appendix we present the expressions for the $\mathbb{Z}_2\times\mathbb{Z}_2$ eigenfunctions $L_w^\pm(z)$
defined in eq.~\eqref{eq:Lplusminus} as linear combinations of ordinary HPLs of the form $H_{w_1}(z)\,H_{w_2}(\zp)$ up to weight 5. All expressions up to weight 6 are attached as ancillary files in computer-readable format. We give results only for the Lyndon words, as all other cases can be reduced to the latter.  In the following, we use the condensed notation~(\ref{eq:Hwcompact}) for the HPL arguments $z$ and $\zp$ to improve the readability of the formulas.

\subsection{Lyndon words of weight 1}
\bea
\LSMA{0} &=& H_0+\Hb_0 = \log|z|^2\,, \\
\LSPA{1} &=& H_1+\Hb_1 +\frac{1}{2} H_0+\frac{1}{2} \Hb_0
 = -\log|1-z|^2 + \frac{1}{2}\log|z|^2\,,
\eea


\subsection{Lyndon words of weight 2}

\beq\bsp
\LSMA{2}&\,=\frac{1}{4}\big[ -2\,H_{1,0}+2\,\Hb_{1,0}+2\,H_0\,\Hb_1-2\,\Hb_0\,H_1+2\,H_2-2\,\Hb_2 \big]\\
&\,=\text{Li}_2(z)-\text{Li}_2(\zp)+\frac{1}{2}\,\log|z|^2 \, (\log (1-z)-\log (1-\zp))\,,
\esp\eeq


\subsection{Lyndon words of weight 3}

\begin{eqnarray}
\LSPA{3}&=&\frac{1}{4}\big[ 2\,H_0\,\Hb_{0,0}+2\,H_0\,\Hb_{1,0}+2\,\Hb_0\,H_{0,0}+2\,\Hb_0\,H_{1,0}+2\,H_1\,\Hb_{0,0}+2\,\Hb_1\,H_{0,0}\\
\nonumber&&+2\,H_{0,0,0}+2\,H_{1,0,0}+2\,\Hb_{0,0,0}+2\,\Hb_{1,0,0}+2\,H_3+2\,\Hb_3 \big]\\
\nonumber&=&\text{Li}_3(z)+\text{Li}_3(\zp) -\frac{1}{2}\log|z|^2 \big[\text{Li}_2(\zp)+\text{Li}_2(z)\big]
-\frac{1}{4}\log^2|z|^2 \log |1-z|^2 + \frac{1}{12} \log ^3|z|^2\,,\\
\LSMB{2}{1}&=&\frac{1}{4}\big[ H_0\,\Hb_{1,0}+\Hb_0\,H_{1,0}+H_1\,\Hb_{0,0}+\Hb_1\,H_{0,0}+2\,H_0\,\Hb_{0,0}+2\,H_0\,\Hb_{1,1}\\
\nonumber&&+2\,\Hb_0\,H_{0,0}+2\,\Hb_0\,H_{1,1}+H_{1,0,0}+2\,H_{0,0,0}+2\,H_{2,0}+2\,H_{2,1}+2\,H_{1,1,0}\\
\nonumber&&+\Hb_{1,0,0}+2\,\Hb_{0,0,0}+2\,\Hb_{2,0}+2\,\Hb_{2,1}+2\,\Hb_{1,1,0}+2\,H_0\,\Hb_2+2\,\Hb_0\,H_2\\
\nonumber&&+2\,H_1\,\Hb_2+2\,\Hb_1\,H_2+H_3+\Hb_3-4\,\zeta_3 \big]\\
\nonumber&=&
-\text{Li}_3(1-z)-\text{Li}_3(1-\zp)-\frac{1}{2}\big[\text{Li}_3(z)+\text{Li}_3(\zp)\big]
+\frac{1}{4}\log|z|^2\big[ \text{Li}_2(z) +\text{Li}_2(\zp) \big]\\
\nonumber&&-\frac{1}{2}\log|1-z|^2 \big[\text{Li}_2(z)+ \text{Li}_2(\zp)\big]
-\frac{1}{8}\log^2|z|^2\log|1-z|^2
+\frac{1}{12} \log ^3|z|^2\\
\nonumber&&-\frac{1}{4}\log{z\over\zp} \big[\log ^2(1-z)-\log ^2(1-\zp)]
+\zeta_2\, \log|1-z|^2
+\zeta_3
\,,
\end{eqnarray}


\subsection{Lyndon words of weight 4}

\begin{eqnarray}
\LSPB{3}{1}&=&\frac{1}{4}\big[ H_0\,\Hb_{2,0}+H_0\,\Hb_{1,0,0}-\Hb_0\,H_{2,0}-\Hb_0\,H_{1,0,0}-H_1\,\Hb_{0,0,0}+\Hb_1\,H_{0,0,0}\\
\nonumber&&+H_{0,0}\,\Hb_2+H_{0,0}\,\Hb_{1,0}-\Hb_{0,0}\,H_2-\Hb_{0,0}\,H_{1,0}+2\,H_0\,\Hb_{1,1,0}-2\,\Hb_0\,H_{1,1,0}\\
\nonumber&&+2\,H_{0,0}\,\Hb_{1,1}-2\,\Hb_{0,0}\,H_{1,1}+H_{3,0}-H_{2,0,0}-H_{1,0,0,0}+2\,H_{3,1}-2\,H_{1,1,0,0}\\
\nonumber&&+\Hb_{2,0,0}+\Hb_{1,0,0,0}-\Hb_{3,0}-2\,\Hb_{3,1}+2\,\Hb_{1,1,0,0}-H_0\,\Hb_3+\Hb_0\,H_3-2\,H_1\,\Hb_3\\
\nonumber&&+2\,\Hb_1\,H_3+4\,H_1\,\zeta_3+H_4-4\,\Hb_1\,\zeta_3-\Hb_4 \big]\,,\\
\LSMA{4}&=&\frac{1}{4}\big[ 2\,H_0\,\Hb_{1,0,0}-2\,\Hb_0\,H_{1,0,0}-2\,H_1\,\Hb_{0,0,0}+2\,\Hb_1\,H_{0,0,0}+2\,H_{0,0}\,\Hb_{1,0}\\
\nonumber&&-2\,\Hb_{0,0}\,H_{1,0}-2\,H_{1,0,0,0}+2\,\Hb_{1,0,0,0}+2\,H_4-2\,\Hb_4 \big]\,,
\end{eqnarray}


\begin{eqnarray}
\LSMC{2}{1}{1}&=&\frac{1}{4}\big[ H_0\,\Hb_{1,0,0}+H_0\,\Hb_{1,2}+H_0\,\Hb_{1,1,0}-\Hb_0\,H_{1,0,0}-\Hb_0\,H_{1,2}-\Hb_0\,H_{1,1,0}\\
\nonumber&&-H_1\,\Hb_{0,0,0}-H_1\,\Hb_{2,0}+\Hb_1\,H_{0,0,0}+\Hb_1\,H_{2,0}+H_{0,0}\,\Hb_{1,0}+H_{0,0}\,\Hb_{1,1}\\
\nonumber&&-\Hb_{0,0}\,H_{1,0}-\Hb_{0,0}\,H_{1,1}+H_2\,\Hb_{1,0}-\Hb_2\,H_{1,0}+2\,H_0\,\Hb_{1,1,1}-2\,\Hb_0\,H_{1,1,1}\\
\nonumber&&-2\,H_1\,\Hb_{2,1}+2\,\Hb_1\,H_{2,1}+2\,H_2\,\Hb_{1,1}-2\,\Hb_2\,H_{1,1}+H_{3,1}+H_{2,2}\\
\nonumber&&-H_{1,0,0,0}-H_{1,2,0}-H_{1,1,0,0}+2\,H_{2,1,1}-2\,H_{1,1,1,0}+\Hb_{1,0,0,0}+\Hb_{1,2,0}+\Hb_{1,1,0,0}\\
\nonumber&&-\Hb_{3,1}-\Hb_{2,2}-2\,\Hb_{2,1,1}+2\,\Hb_{1,1,1,0}-H_1\,\Hb_3+\Hb_1\,H_3+2\,H_1\,\zeta_3+H_4\\
\nonumber&&-2\,\Hb_1\,\zeta_3-\Hb_4 \big]\,,
\end{eqnarray}


\subsection{Lyndon words of weight 5}

\begin{eqnarray}
\LSPA{5}&=&\frac{1}{4}\big[ 2\,H_0\,\Hb_{0,0,0,0}+2\,H_0\,\Hb_{1,0,0,0}+2\,\Hb_0\,H_{0,0,0,0}+2\,\Hb_0\,H_{1,0,0,0}+2\,H_1\,\Hb_{0,0,0,0}\\
\nonumber&&+2\,\Hb_1\,H_{0,0,0,0}+2\,H_{0,0}\,\Hb_{0,0,0}+2\,H_{0,0}\,\Hb_{1,0,0}+2\,\Hb_{0,0}\,H_{0,0,0}+2\,\Hb_{0,0}\,H_{1,0,0}\\
\nonumber&&+2\,H_{1,0}\,\Hb_{0,0,0}+2\,\Hb_{1,0}\,H_{0,0,0}+2\,H_{0,0,0,0,0}+2\,H_{1,0,0,0,0}+2\,\Hb_{0,0,0,0,0}+2\,\Hb_{1,0,0,0,0}\\
\nonumber&&+2\,H_5+2\,\Hb_5 \big]\,,\\
\LSPC{3}{1}{1}&=&\frac{1}{4}\big[ H_5+\Hb_5+H_{4,0}+\Hb_{4,0}+H_{4,1}+\Hb_{4,1}+H_{3,2}+\Hb_{3,2}+H_{3,1,0}+\Hb_{3,1,0}\\
\nonumber&&+H_{2,0,0,0}+\Hb_{2,0,0,0}+H_{2,1,0,0}+\Hb_{2,1,0,0}+H_{1,0,0,0,0}+\Hb_{1,0,0,0,0}+H_{1,2,0,0}\\
\nonumber&&+\Hb_{1,2,0,0}+H_{1,1,0,0,0}+\Hb_{1,1,0,0,0}+2\,H_{0,0,0,0,0}+2\,\Hb_{0,0,0,0,0}+2\,H_{3,0,0}+2\,\Hb_{3,0,0}\\
\nonumber&&+2\,H_{3,1,1}+2\,\Hb_{3,1,1}+2\,H_{1,1,1,0,0}+2\,\Hb_{1,1,1,0,0}+4\,\zeta_5+H_0\,\Hb_4+H_0\,\Hb_{3,1}+H_0\,\Hb_{2,0,0}\\
\nonumber&&+H_0\,\Hb_{2,1,0}+H_0\,\Hb_{1,0,0,0}+H_0\,\Hb_{1,2,0}+H_0\,\Hb_{1,1,0,0}+\Hb_0\,H_4+\Hb_0\,H_{3,1}+\Hb_0\,H_{2,0,0}\\
\nonumber&&+\Hb_0\,H_{2,1,0}+\Hb_0\,H_{1,0,0,0}+\Hb_0\,H_{1,2,0}+\Hb_0\,H_{1,1,0,0}+H_1\,\Hb_{0,0,0,0}+H_1\,\Hb_4+H_1\,\Hb_{3,0}\\
\nonumber&&+\Hb_1\,H_{0,0,0,0}+\Hb_1\,H_4+\Hb_1\,H_{3,0}+H_{0,0}\,\Hb_{2,0}+H_{0,0}\,\Hb_{2,1}+H_{0,0}\,\Hb_{1,0,0}+H_{0,0}\,\Hb_{1,2}\\
\nonumber&&+H_{0,0}\,\Hb_{1,1,0}+\Hb_{0,0}\,H_{2,0}+\Hb_{0,0}\,H_{2,1}+\Hb_{0,0}\,H_{1,0,0}+\Hb_{0,0}\,H_{1,2}+\Hb_{0,0}\,H_{1,1,0}\\
\nonumber&&+H_2\,\Hb_{0,0,0}+H_2\,\Hb_3+\Hb_2\,H_{0,0,0}+\Hb_2\,H_3+H_{1,0}\,\Hb_{0,0,0}+H_{1,0}\,\Hb_3+\Hb_{1,0}\,H_{0,0,0}\\
\nonumber&&+\Hb_{1,0}\,H_3+H_{1,1}\,\Hb_{0,0,0}+\Hb_{1,1}\,H_{0,0,0}+2\,H_0\,\Hb_{0,0,0,0}+2\,H_0\,\Hb_{3,0}+2\,H_0\,\Hb_{1,1,1,0}\\
\nonumber&&+2\,\Hb_0\,H_{0,0,0,0}+2\,\Hb_0\,H_{3,0}+2\,\Hb_0\,H_{1,1,1,0}+2\,H_1\,\Hb_{3,1}+2\,\Hb_1\,H_{3,1}+2\,H_{0,0}\,\Hb_{0,0,0}\\
\nonumber&&+2\,H_{0,0}\,\Hb_3+2\,H_{0,0}\,\Hb_{1,1,1}+2\,\Hb_{0,0}\,H_{0,0,0}+2\,\Hb_{0,0}\,H_3+2\,\Hb_{0,0}\,H_{1,1,1}-2\,H_2\,\zeta_3\\
\nonumber&&-2\,\Hb_2\,\zeta_3-2\,H_{1,0}\,\zeta_3-2\,\Hb_{1,0}\,\zeta_3+2\,H_{1,1}\,\Hb_3+2\,\Hb_{1,1}\,H_3-4\,H_{1,1}\,\zeta_3-4\,\Hb_{1,1}\,\zeta_3\\
\nonumber&&-2\,H_0\,\Hb_1\,\zeta_3-2\,\Hb_0\,H_1\,\zeta_3 \big]\,,\\
\LSPC{2}{2}{1}&=&\frac{1}{4}\big[ H_5+\Hb_5+H_{4,1}+\Hb_{4,1}+H_{2,3}+\Hb_{2,3}+H_{1,0,0,0,0}+\Hb_{1,0,0,0,0}+H_{1,3,0}\\
\nonumber&&+\Hb_{1,3,0}+H_{1,1,0,0,0}+\Hb_{1,1,0,0,0}+2\,H_{0,0,0,0,0}+2\,\Hb_{0,0,0,0,0}+2\,H_{4,0}+2\,\Hb_{4,0}\\
\nonumber&&+2\,H_{2,0,0,0}+2\,\Hb_{2,0,0,0}+2\,H_{2,2,0}+2\,\Hb_{2,2,0}+2\,H_{2,2,1}+2\,\Hb_{2,2,1}+2\,H_{1,1,2,0}\\
\nonumber&&+2\,\Hb_{1,1,2,0}-6\,\zeta_5+H_0\,\Hb_{1,0,0,0}+H_0\,\Hb_{1,3}+H_0\,\Hb_{1,1,0,0}+\Hb_0\,H_{1,0,0,0}+\Hb_0\,H_{1,3}\\
\nonumber&&+\Hb_0\,H_{1,1,0,0}+H_1\,\Hb_{0,0,0,0}+H_1\,\Hb_4+H_1\,\Hb_{2,0,0}+\Hb_1\,H_{0,0,0,0}+\Hb_1\,H_4+\Hb_1\,H_{2,0,0}\\
\nonumber&&+H_{0,0}\,\Hb_{1,0,0}+H_{0,0}\,\Hb_{1,1,0}+\Hb_{0,0}\,H_{1,0,0}+\Hb_{0,0}\,H_{1,1,0}+H_2\,\Hb_{1,0,0}+\Hb_2\,H_{1,0,0}\\
\nonumber&&+H_{1,0}\,\Hb_{0,0,0}+H_{1,0}\,\Hb_{2,0}+\Hb_{1,0}\,H_{0,0,0}+\Hb_{1,0}\,H_{2,0}+H_{1,1}\,\Hb_{0,0,0}+\Hb_{1,1}\,H_{0,0,0}\\
\nonumber&&+2\,H_0\,\Hb_{0,0,0,0}+2\,H_0\,\Hb_4+2\,H_0\,\Hb_{2,0,0}+2\,H_0\,\Hb_{2,2}+2\,H_0\,\Hb_{1,1,2}+2\,\Hb_0\,H_{0,0,0,0}\\
\nonumber&&+2\,\Hb_0\,H_4+2\,\Hb_0\,H_{2,0,0}+2\,\Hb_0\,H_{2,2}+2\,\Hb_0\,H_{1,1,2}+2\,H_1\,\Hb_{2,2}+2\,\Hb_1\,H_{2,2}\\
\nonumber&&+2\,H_{0,0}\,\Hb_{0,0,0}+2\,H_{0,0}\,\Hb_{2,0}+2\,\Hb_{0,0}\,H_{0,0,0}+2\,\Hb_{0,0}\,H_{2,0}+2\,H_2\,\Hb_{0,0,0}+2\,H_2\,\Hb_{2,0}\\
\nonumber&&+2\,H_2\,\Hb_{1,1,0}+2\,H_2\,\zeta_3+2\,\Hb_2\,H_{0,0,0}+2\,\Hb_2\,H_{2,0}+2\,\Hb_2\,H_{1,1,0}+2\,\Hb_2\,\zeta_3\\
\nonumber&&+2\,H_{1,1}\,\Hb_{2,0}+2\,\Hb_{1,1}\,H_{2,0}-4\,H_{0,0}\,\zeta_3-4\,\Hb_{0,0}\,\zeta_3+4\,H_{1,0}\,\zeta_3+4\,\Hb_{1,0}\,\zeta_3\\
\nonumber&&+8\,H_{1,1}\,\zeta_3+8\,\Hb_{1,1}\,\zeta_3-4\,H_0\,\Hb_0\,\zeta_3+4\,H_0\,\Hb_1\,\zeta_3+4\,\Hb_0\,H_1\,\zeta_3 \big]\,,
\end{eqnarray}
\begin{eqnarray}
\LSMB{4}{1}&=&\frac{1}{4}\big[ H_0\,\Hb_{2,0,0}+H_0\,\Hb_{1,0,0,0}+\Hb_0\,H_{2,0,0}+\Hb_0\,H_{1,0,0,0}+H_1\,\Hb_{0,0,0,0}+\Hb_1\,H_{0,0,0,0}\\
\nonumber&&+H_{0,0}\,\Hb_{2,0}+H_{0,0}\,\Hb_{1,0,0}+\Hb_{0,0}\,H_{2,0}+\Hb_{0,0}\,H_{1,0,0}+H_2\,\Hb_{0,0,0}+\Hb_2\,H_{0,0,0}\\
\nonumber&&+H_{1,0}\,\Hb_{0,0,0}+\Hb_{1,0}\,H_{0,0,0}+2\,H_0\,\Hb_{0,0,0,0}+2\,H_0\,\Hb_{1,1,0,0}+2\,\Hb_0\,H_{0,0,0,0}\\
\nonumber&&+2\,\Hb_0\,H_{1,1,0,0}+2\,H_{0,0}\,\Hb_{0,0,0}+2\,H_{0,0}\,\Hb_{1,1,0}+2\,\Hb_{0,0}\,H_{0,0,0}+2\,\Hb_{0,0}\,H_{1,1,0}\\
\nonumber&&+2\,H_{1,1}\,\Hb_{0,0,0}+2\,\Hb_{1,1}\,H_{0,0,0}-4\,H_{0,0}\,\zeta_3-4\,H_{1,0}\,\zeta_3+H_{4,0}+H_{2,0,0,0}\\
\nonumber&&+H_{1,0,0,0,0}+2\,H_{0,0,0,0,0}+2\,H_{4,1}+2\,H_{1,1,0,0,0}-4\,\Hb_{0,0}\,\zeta_3-4\,\Hb_{1,0}\,\zeta_3+\Hb_{4,0}\\
\nonumber&&+\Hb_{2,0,0,0}+\Hb_{1,0,0,0,0}+2\,\Hb_{0,0,0,0,0}+2\,\Hb_{4,1}+2\,\Hb_{1,1,0,0,0}-4\,H_0\,\Hb_0\,\zeta_3\\
\nonumber&&-4\,H_0\,\Hb_1\,\zeta_3-4\,\Hb_0\,H_1\,\zeta_3+H_0\,\Hb_4+\Hb_0\,H_4+2\,H_1\,\Hb_4+2\,\Hb_1\,H_4\\
\nonumber&&+H_5+\Hb_5-4\,\zeta_5 \big]\,,\\
\LSMB{3}{2}&=&\frac{1}{4}\big[ H_0\,\Hb_{1,0,0,0}+\Hb_0\,H_{1,0,0,0}+H_1\,\Hb_{0,0,0,0}+\Hb_1\,H_{0,0,0,0}+H_{0,0}\,\Hb_{1,0,0}\\
\nonumber&&+\Hb_{0,0}\,H_{1,0,0}+H_{1,0}\,\Hb_{0,0,0}+\Hb_{1,0}\,H_{0,0,0}+2\,H_0\,\Hb_{0,0,0,0}+2\,H_0\,\Hb_{3,0}+2\,H_0\,\Hb_{1,2,0}\\
\nonumber&&+2\,\Hb_0\,H_{0,0,0,0}+2\,\Hb_0\,H_{3,0}+2\,\Hb_0\,H_{1,2,0}+2\,H_1\,\Hb_{3,0}+2\,\Hb_1\,H_{3,0}+2\,H_{0,0}\,\Hb_{0,0,0}\\
\nonumber&&+2\,H_{0,0}\,\Hb_3+2\,H_{0,0}\,\Hb_{1,2}+2\,\Hb_{0,0}\,H_{0,0,0}+2\,\Hb_{0,0}\,H_3+2\,\Hb_{0,0}\,H_{1,2}+2\,H_{1,0}\,\Hb_3\\
\nonumber&&+2\,\Hb_{1,0}\,H_3+8\,H_{0,0}\,\zeta_3+8\,H_{1,0}\,\zeta_3+H_{1,0,0,0,0}+2\,H_{0,0,0,0,0}+2\,H_{3,0,0}+2\,H_{3,2}\\
\nonumber&&+2\,H_{1,2,0,0}+8\,\Hb_{0,0}\,\zeta_3+8\,\Hb_{1,0}\,\zeta_3+\Hb_{1,0,0,0,0}+2\,\Hb_{0,0,0,0,0}+2\,\Hb_{3,0,0}+2\,\Hb_{3,2}\\
\nonumber&&+2\,\Hb_{1,2,0,0}+8\,H_0\,\Hb_0\,\zeta_3+8\,H_0\,\Hb_1\,\zeta_3+8\,\Hb_0\,H_1\,\zeta_3+H_5+\Hb_5+16\,\zeta_5 \big]\,,\\
\LSMD{2}{1}{1}{1}&=&\frac{1}{4}\big[ H_5+\Hb_5+H_{4,1}+\Hb_{4,1}+H_{3,2}+\Hb_{3,2}+H_{3,1,1}+\Hb_{3,1,1}+H_{2,3}+\Hb_{2,3}\\
\nonumber&&+H_{2,2,1}+\Hb_{2,2,1}+H_{2,1,2}+\Hb_{2,1,2}+H_{1,0,0,0,0}+\Hb_{1,0,0,0,0}+H_{1,3,0}+\Hb_{1,3,0}\\
\nonumber&&+H_{1,2,0,0}+\Hb_{1,2,0,0}+H_{1,2,1,0}+\Hb_{1,2,1,0}+H_{1,1,0,0,0}+\Hb_{1,1,0,0,0}+H_{1,1,2,0}+\Hb_{1,1,2,0}\\
\nonumber&&+H_{1,1,1,0,0}+\Hb_{1,1,1,0,0}+2\,H_{0,0,0,0,0}+2\,\Hb_{0,0,0,0,0}+2\,H_{4,0}+2\,\Hb_{4,0}+2\,H_{3,0,0}\\
\nonumber&&+2\,\Hb_{3,0,0}+2\,H_{3,1,0}+2\,\Hb_{3,1,0}+2\,H_{2,0,0,0}+2\,\Hb_{2,0,0,0}+2\,H_{2,2,0}+2\,\Hb_{2,2,0}\\
\nonumber&&+2\,H_{2,1,0,0}+2\,\Hb_{2,1,0,0}+2\,H_{2,1,1,0}+2\,\Hb_{2,1,1,0}+2\,H_{2,1,1,1}+2\,\Hb_{2,1,1,1}+2\,H_{1,1,1,1,0}\\
\nonumber&&+2\,\Hb_{1,1,1,1,0}-4\,\zeta_5+H_0\,\Hb_{1,0,0,0}+H_0\,\Hb_{1,3}+H_0\,\Hb_{1,2,0}+H_0\,\Hb_{1,2,1}+H_0\,\Hb_{1,1,0,0}\\
\nonumber&&+H_0\,\Hb_{1,1,2}+H_0\,\Hb_{1,1,1,0}+\Hb_0\,H_{1,0,0,0}+\Hb_0\,H_{1,3}+\Hb_0\,H_{1,2,0}+\Hb_0\,H_{1,2,1}\\
\nonumber&&+\Hb_0\,H_{1,1,0,0}+\Hb_0\,H_{1,1,2}+\Hb_0\,H_{1,1,1,0}+H_1\,\Hb_{0,0,0,0}+H_1\,\Hb_4+H_1\,\Hb_{3,0}+H_1\,\Hb_{3,1}\\
\nonumber&&+H_1\,\Hb_{2,0,0}+H_1\,\Hb_{2,2}+H_1\,\Hb_{2,1,0}+\Hb_1\,H_{0,0,0,0}+\Hb_1\,H_4+\Hb_1\,H_{3,0}+\Hb_1\,H_{3,1}\\
\nonumber&&+\Hb_1\,H_{2,0,0}+\Hb_1\,H_{2,2}+\Hb_1\,H_{2,1,0}+H_{0,0}\,\Hb_{1,0,0}+H_{0,0}\,\Hb_{1,2}+H_{0,0}\,\Hb_{1,1,0}\\
\nonumber&&+H_{0,0}\,\Hb_{1,1,1}+\Hb_{0,0}\,H_{1,0,0}+\Hb_{0,0}\,H_{1,2}+\Hb_{0,0}\,H_{1,1,0}+\Hb_{0,0}\,H_{1,1,1}+H_2\,\Hb_{1,0,0}\\
\nonumber&&+H_2\,\Hb_{1,2}+H_2\,\Hb_{1,1,0}+\Hb_2\,H_{1,0,0}+\Hb_2\,H_{1,2}+\Hb_2\,H_{1,1,0}+H_{1,0}\,\Hb_{0,0,0}+H_{1,0}\,\Hb_3\\
\nonumber&&+H_{1,0}\,\Hb_{2,0}+H_{1,0}\,\Hb_{2,1}+\Hb_{1,0}\,H_{0,0,0}+\Hb_{1,0}\,H_3+\Hb_{1,0}\,H_{2,0}+\Hb_{1,0}\,H_{2,1}+H_{1,1}\,\Hb_{0,0,0}\\
\nonumber&&+H_{1,1}\,\Hb_3+H_{1,1}\,\Hb_{2,0}+\Hb_{1,1}\,H_{0,0,0}+\Hb_{1,1}\,H_3+\Hb_{1,1}\,H_{2,0}+2\,H_0\,\Hb_{0,0,0,0}+2\,H_0\,\Hb_4
\end{eqnarray}
\begin{eqnarray}
\nonumber&&+2\,H_0\,\Hb_{3,0}+2\,H_0\,\Hb_{3,1}+2\,H_0\,\Hb_{2,0,0}+2\,H_0\,\Hb_{2,2}+2\,H_0\,\Hb_{2,1,0}+2\,H_0\,\Hb_{2,1,1}\\
\nonumber&&+2\,H_0\,\Hb_{1,1,1,1}+2\,\Hb_0\,H_{0,0,0,0}+2\,\Hb_0\,H_4+2\,\Hb_0\,H_{3,0}+2\,\Hb_0\,H_{3,1}+2\,\Hb_0\,H_{2,0,0}\\
\nonumber&&+2\,\Hb_0\,H_{2,2}+2\,\Hb_0\,H_{2,1,0}+2\,\Hb_0\,H_{2,1,1}+2\,\Hb_0\,H_{1,1,1,1}+2\,H_1\,\Hb_{2,1,1}+2\,\Hb_1\,H_{2,1,1}\\
\nonumber&&+2\,H_{0,0}\,\Hb_{0,0,0}+2\,H_{0,0}\,\Hb_3+2\,H_{0,0}\,\Hb_{2,0}+2\,H_{0,0}\,\Hb_{2,1}+2\,\Hb_{0,0}\,H_{0,0,0}+2\,\Hb_{0,0}\,H_3\\
\nonumber&&+2\,\Hb_{0,0}\,H_{2,0}+2\,\Hb_{0,0}\,H_{2,1}+2\,H_2\,\Hb_{0,0,0}+2\,H_2\,\Hb_3+2\,H_2\,\Hb_{2,0}+2\,H_2\,\Hb_{2,1}\\
\nonumber&&+2\,H_2\,\Hb_{1,1,1}+2\,\Hb_2\,H_{0,0,0}+2\,\Hb_2\,H_3+2\,\Hb_2\,H_{2,0}+2\,\Hb_2\,H_{2,1}+2\,\Hb_2\,H_{1,1,1}\\
\nonumber&&+2\,H_{1,1}\,\Hb_{2,1}-2\,H_{1,1}\,\zeta_3+2\,\Hb_{1,1}\,H_{2,1}-2\,\Hb_{1,1}\,\zeta_3 \big]\,.
\end{eqnarray}


\subsection{Expression of Brown's SVHPLs in terms of the $L^\pm$ functions}

In this appendix we present the expression of Brown's SVHPLs corresponding to Lyndon words in terms of the $\mathbb{Z}_2\times\mathbb{Z}_2$ eigenfunctions $L_w^\pm(z)$.
\beq\bsp
L_0 &\,= \LSMA{0}\,,\\
L_1 &\,= \LSPA{1}-\frac{1}{2}\,\LSMA{0}\,,\\
L_2 &\,= \LSMA{2}\,,\\
L_3 &\,= \LSPA{3}-\frac{1}{12}\,\LMA{0}{3}\,,\\
L_{2,1} &\,= -\frac{1}{4}\,\LSPA{1}\,\LMA{0}{2}+\frac{1}{2}\,\LSPA{3}+\LSMB{2}{1}+\zeta_3\,,\\
L_4 &\,= \LSMA{4}\,,\\
L_{3,1} &\,= -\frac{1}{4}\,\LSMA{2}\,\LMA{0}{2}+\LSMA{4}+\LSPB{3}{1}\,,\\
L_{2,1,1} &\,= -\frac{1}{4}\,\LSMA{2}\,\LSMA{0}\,\LSPA{1}+\frac{1}{2}\,\LSPB{3}{1}+\LSMC{2}{1}{1}\,,\\
L_5 &\,= \LSPA{5}-\frac{1}{240}\,\LMA{0}{5}\,,\\
L_{4,1} &\,= \frac{1}{48}\,\LSPA{1}\,\LMA{0}{4}-\frac{1}{4}\,\LSPA{3}\,\LMA{0}{2}+\frac{1}{2}\,\LMA{0}{2}\,\zeta_3+\frac{3}{2}\,\LSPA{5}+\LSMB{4}{1}+\zeta_5\,,\\
L_{3,2} &\,= -\frac{1}{16}\,\LSPA{1}\,\LMA{0}{4}+\frac{1}{2}\,\LSPA{3}\,\LMA{0}{2}-\frac{7}{2}\,\LSPA{5}-\LMA{0}{2}\,\zeta_3+\LSMB{3}{2}-4\,\zeta_5\,,\\
L_{3,1,1} &\,= \frac{1}{16}\,\LMA{0}{3}\,\LPA{1}{2}-\frac{1}{4}\,\LSMB{2}{1}\,\LMA{0}{2}+\frac{7}{960}\,\LMA{0}{5}-\frac{1}{4}\,\LSMA{0}\,\LSPA{1}\,\LSPA{3}+\frac{1}{2}\,\LSMA{0}\,\LSPA{1}\,\zeta_3\\
&\,+\LSMB{4}{1}+\LSPC{3}{1}{1}\,,\\
L_{2,2,1} &\,= -\frac{3}{16}\,\LMA{0}{3}\,\LPA{1}{2}+\frac{1}{2}\,\LSMB{2}{1}\,\LMA{0}{2}-\frac{13}{960}\,\LMA{0}{5}+\frac{3}{4}\,\LSMA{0}\,\LSPA{1}\,\LSPA{3}-\frac{1}{2}\,\LSMA{0}\,\LSPA{1}\,\zeta_3\\
&\,-\frac{7}{2}\,\LSMB{4}{1}-\frac{1}{2}\,\LSMB{3}{2}+\LSPC{2}{2}{1}\,,\\
L_{2,1,1,1} &\,= \frac{1}{48}\,\LMA{0}{2}\,\LPA{1}{3}-\frac{1}{192}\,\LSPA{1}\,\LMA{0}{4}+\frac{1}{16}\,\LSPA{3}\,\LMA{0}{2}-\frac{1}{8}\,\LMA{0}{2}\,\zeta_3-\frac{1}{4}\,\LSMA{0}\,\LSMB{2}{1}\,\LSPA{1}\\
&\,-\frac{1}{4}\,\LSPA{5}+\frac{1}{2}\,\LSPC{3}{1}{1}+\frac{1}{2}\,\zeta_5+\LSMD{2}{1}{1}{1}\,,\\
L_6 &\,= \LSMA{6}\,,\\
L_{5,1} &\,= -\frac{1}{4}\,\LSMA{4}\,\LMA{0}{2}+\frac{1}{48}\,\LSMA{2}\,\LMA{0}{4}+2\,\LSMA{6}+\LSPB{5}{1}\,,\\
L_{4,2} &\,= \frac{3}{4}\,\LSMA{4}\,\LMA{0}{2}-\frac{1}{12}\,\LSMA{2}\,\LMA{0}{4}-\frac{11}{2}\,\LSMA{6}+\LSPB{4}{2}\,,
\esp\eeq
\beq\bsp
L_{4,1,1} &\,= \frac{1}{16}\,\LSMA{2}\,\LSPA{1}\,\LMA{0}{3}-\frac{1}{4}\,\LSPB{3}{1}\,\LMA{0}{2}-\frac{1}{4}\,\LSMA{4}\,\LSMA{0}\,\LSPA{1}+\frac{1}{2}\,\LSMA{2}\,\LSMA{0}\,\zeta_3+\frac{3}{2}\,\LSPB{5}{1}+\LSMC{4}{1}{1}\,,\\
L_{3,2,1} &\,= -\frac{3}{16}\,\LSMA{2}\,\LSPA{1}\,\LMA{0}{3}+\frac{1}{2}\,\LSPB{3}{1}\,\LMA{0}{2}+\frac{3}{4}\,\LSMA{4}\,\LSMA{0}\,\LSPA{1}-\frac{1}{2}\,\LSMA{2}\,\LSMA{0}\,\zeta_3-\frac{7}{2}\,\LSPB{5}{1}+\LSMC{3}{2}{1}\,,\\
L_{3,1,2} &\,= -\frac{1}{4}\,\LSMA{2}\,\LSMA{0}\,\LSPA{3}-\frac{3}{2}\,\LSMA{2}\,\LSMA{0}\,\zeta_3+\LSMC{3}{1}{2}+3\,\LSPB{5}{1}+\LSPB{4}{2}\,,\\
L_{3,1,1,1} &\,= \frac{1}{16}\,\LSMA{2}\,\LMA{0}{2}\,\LPA{1}{2}+\frac{1}{4}\,\LSMA{4}\,\LMA{0}{2}-\frac{5}{192}\,\LSMA{2}\,\LMA{0}{4}-\frac{1}{4}\,\LSMC{2}{1}{1}\,\LMA{0}{2}-\frac{1}{4}\,\LSMA{0}\,\LSPA{1}\,\LSPB{3}{1}\\
&\,-\LSMA{6}+\LSMC{4}{1}{1}+\LSPD{3}{1}{1}{1}\,,\\
L_{2,2,1,1} &\,= -\frac{1}{4}\,\LSMA{2}\,\LMA{0}{2}\,\LPA{1}{2}-\frac{3}{4}\,\LSMA{4}\,\LMA{0}{2}+\frac{1}{12}\,\LSMA{2}\,\LMA{0}{4}+\frac{3}{4}\,\LSMC{2}{1}{1}\,\LMA{0}{2}+\frac{11}{4}\,\LSMA{6}\\
&\,+\frac{1}{4}\,\LSMA{2}\,\LSPA{1}\,\LSPA{3}-\frac{1}{2}\,\LSMA{2}\,\LSPA{1}\,\zeta_3+\frac{3}{4}\,\LSMA{0}\,\LSPA{1}\,\LSPB{3}{1}-\frac{1}{2}\,\LSMC{3}{1}{2}-5\,\LSMC{4}{1}{1}-\LSMC{3}{2}{1}+\LSPD{2}{2}{1}{1}\,,\\
L_{2,1,1,1,1} &\,= -\frac{5}{192}\,\LSMA{2}\,\LSPA{1}\,\LMA{0}{3}+\frac{1}{16}\,\LSPB{3}{1}\,\LMA{0}{2}+\frac{1}{48}\,\LSMA{2}\,\LSMA{0}\,\LPA{1}{3}+\frac{1}{8}\,\LSMA{4}\,\LSMA{0}\,\LSPA{1}-\frac{1}{4}\,\LSMA{2}\,\LSMA{0}\,\zeta_3\\
&\,-\frac{1}{4}\,\LSMA{0}\,\LSMC{2}{1}{1}\,\LSPA{1}-\frac{1}{4}\,\LSPB{5}{1}+\frac{1}{2}\,\LSPD{3}{1}{1}{1}+\LSME{2}{1}{1}{1}{1}\,.
\esp\eeq

\section{Analytic continuation of harmonic sums}
\label{app:blumlein}
In this section we review the analytic continuation of multiple harmonic sums and the structural relations between them, as presented by Bl\"{u}mlein~\cite{Blumlein2009ta}. Multiple harmonic sums are defined by,
\beq
S_{a_1,\cdots,a_n}(N) = \sum_{k_1=1}^{N}\sum_{k_2=1}^{k_1} \cdots \sum_{k_n=1}^{k_{n-1}} \frac{\sgn(a_1)^{k_1}}{k_1^{|a_1|}}\cdots \frac{\sgn(a_n)^{k_n}}{k_n^{|a_n|}}\,,
\eeq
where the $a_k$ are positive or negative integers, and $N$ is a positive integer.  For the cases in which we are interested, they are similar to the Euler-Zagier sums~\eqref{eq:Euler-Zagier}, except that the summation range differs slightly. They are related to Mellin transforms of real functions or distributions $f(x)$,
\beq\label{eq:sum_int}
S_{a_1,\ldots,a_n}(N)  = \int_0^1 dx\; x^N\,f_{a_1,\ldots,a_n} =  {\bf M}[f_{a_1,\ldots,a_n}(x)](N)\,.
\eeq
Typically $f(x)$ are HPLs weighted by factors of $1/(1\pm x)$. To avoid singularities at $x=1$, it is often useful to consider the $+$-distribution,
\beq
{\bf M}[(f(x))_+](N) = \int_0^1 dx\,\left(x^N-1\right)\,f(x) \, .
\eeq
The weight $|w|$ of the harmonic sum is given by $|w|=\sum_{k=1}^n |a_k|$. The number of harmonic sums of weight $w$ is equal to $2\cdot3^{|w|-1}$, but not all of them are independent. For example, they obey shuffle relations~\cite{Hoffman2004bf}. It is natural to ask whether these are the only relations they satisfy. In fact, it is known that in the special case $N\to\infty$, in which the sums reduce to multiple zeta values, many new relations emerge~\cite{Broadhurst1996az,Broadhurst1996kc,Borwein1999js,Blumlein2009cf}.
In ref.~\cite{Blumlein2009ta}, an analytic continuation of the harmonic sums was considered. It is defined by the integral representation, eq.~\eqref{eq:sum_int}, where $N$ is allowed to take complex values. This allows for two new operations---differentiation and evaluation at fractional arguments---which generate new structural relations among the harmonic sums.

In the present work, harmonic sums with negative indices do not appear, so we will assume that $a_k>0$. This assumption provides a considerable simplification.  The derivative relations allow for the extraction of logarithmic factors,
\beq
{\bf M}[\log^l(x) f(x)] (N) = \frac{d^l}{d N^l} {\bf M}[f(x)](N)\,,
\eeq
which explains why the derivatives of the building blocks in Section~\ref{sec:nu_n} generate SVHPLs. In ref.~\cite{Blumlein2009ta}, all available relations are imposed, and the following are the irreducible functions through weight five:\\
\\
\underline{weight 1}
\beq
S_1(N)=\psi(N+1)+\gamma_E = {\bf M}\left[\left(\frac{1}{x-1}\right)_+\right](N)
\eeq
\\
\underline{weight 3}
\beq
F_4(N)= {\bf M}\left[\left(\frac{\Li_2(x)}{1-x}\right)_+\right](N)\\
\eeq
\\
\underline{weight 4}
\beq\bsp
F_{6a}(N)&= {\bf M}\left[\left(\frac{\Li_3(x)}{1-x}\right)_+\right](N)\\
F_{7}(N)&= {\bf M}\left[\left(\frac{S_{1,2}(x)}{x-1}\right)_+\right](N)\\
\esp\eeq
\\
\underline{weight 5}
\beq\bsp
F_{9}(N)&= {\bf M}\left[\left(\frac{\Li_4(x)}{x-1}\right)_+\right](N)\\
F_{11}(N)&= {\bf M}\left[\left(\frac{S_{2,2}(x)}{x-1}\right)_+\right](N)\\
F_{13}(N)&= {\bf M}\left[\left(\frac{\Li_2^2(x)}{x-1}\right)_+\right](N)\\
F_{17}(N)&= {\bf M}\left[\left(\frac{S_{1,3}(x)}{x-1}\right)_+\right](N)
\esp\eeq
There are no irreducible basis functions of weight two. These functions are meromorphic with poles at the negative integers. To use these functions in the integral transform~\eqref{eq:integral_transform}, we need the expansions near the poles. Actually, we only need the expansions around zero, since the expansions around any integer can be obtained from them using the recursion relations of ref.~\cite{Blumlein2009ta},
\beq\bsp
\psi^{(n)}(1+z) &= \psi^{(n)}(z) + (-1)^n \frac{n!}{z^{n+1}}\\
F_4(z) &= F_4(z-1)-\frac{1}{z}\left[\zeta_2-\frac{S_1(z)}{z}\right]\\
F_{6a}(z) &= F_{6a}(z-1) -\frac{\zeta_3}{z} + \frac{1}{z^2} \left[\zeta_2 -\frac{S_1(z)}{z}\right]\\
F_7(z) &= F_7(z-1) +\frac{\zeta_3}{z} - \frac{1}{2 z^2} \left[S_1^2(z)+S_2(z)\right]\\
F_9(z) &= F_{9}(z-1) +\frac{\zeta_4}{z} - \frac{\zeta_3}{z^2}+ \frac{\zeta_2}{z^3}  - \frac{1}{z^4} S_1(z)\\
F_{11}(z) &= F_{11}(z-1) + \frac{\zeta_4}{4 z} - \frac{\zeta_3}{z^2}+ \frac{1}{2 z^3} \left[S_1^2(z) +  S_2(z) \right]\\
F_{13}(z) &= F_{13}(z-1) + \frac{\zeta_2^2}{z} - \frac{4\zeta_3}{z^2}- \frac{2 \zeta_2}{z^2} S_1(z) + \frac{2 S_{2,1}(z)}{z^2} + \frac{2}{z^3} \left[S_1^2(z)+S_2(z)\right] \\
F_{17}(z) &= F_{17}(z-1) + \frac{\zeta_4}{z} - \frac{1}{6 z^2}\left[S_1^3(z) + 3 S_1(z) S_2(z) + 2 S_3(z)\right]\,.
\esp\eeq
The expansions around zero can be obtained from the integral representations. We find that, for $\delta\to0$, the expansions can all be expressed simply in terms of multiple zeta values,
\beq\bsp
S_{1}(\delta) &= -\sum_{n=1}^{\infty}(-\delta)^{n}\zeta_{n+1} \,, \\
F_{4}(\delta) &= \sum_{n=1}^{\infty}(-\delta)^{n}\zeta_{n+1,2} \,, \\
F_{6a}(\delta) &= \sum_{n=1}^{\infty}(-\delta)^{n}\zeta_{n+1,3} \,, \\
F_{7}(\delta) &= -\sum_{n=1}^{\infty}(-\delta)^{n}\zeta_{n+1,2,1} \,, \\
F_{9}(\delta) &= -\sum_{n=1}^{\infty}(-\delta)^{n}\zeta_{n+1,4} \,, \\
F_{11}(\delta) &= -\sum_{n=1}^{\infty}(-\delta)^{n}\zeta_{n+1,3,1} \,, \\
F_{13}(\delta) &= -\sum_{n=1}^{\infty}(-\delta)^{n}\left( 2\zeta_{n+1,2,2}+4\zeta_{n+1,3,1}\right) \,, \\
F_{17}(\delta) &= -\sum_{n=1}^{\infty}(-\delta)^{n} \zeta_{n+1,2,1,1}\,.
\esp\eeq
These single-variable functions can be assembled to form two-variable functions of $\nu$ and $n$, such that their inverse Fourier-Mellin transforms produce sums of SVHPLs. This construction is not unique, because other building blocks could be added. We choose to define the two-variable functions as,
\beq\bsp
\tilde{F}_4 \,=\,&\sgn(n) \left\{F_4\Big(i\nu+\frac{|n|}{2}\Big) + F_4\Big(-i\nu+\frac{|n|}{2}\Big) - \frac{1}{4}\dE{2}-\frac{1}{8}N^2 E_{\nu,n}-\frac{1}{2}V^2 E_{\nu,n}\right.\\
&\left.\phantom{\sgn(n)}+\frac{1}{2}\,\Big(\psi_{-}+V\Big)\dEOne + \zeta_2 E_{\nu,n} -4\,\zeta_3\right\} + N \left\{\frac{1}{2}\,V \psi_{-} + \frac{1}{2}\zeta_2 \right\} \,, \\
\tilde{F}_{6a}\,=\,&\sgn(n)\left\{F_{6a}\Big(i\nu+\frac{|n|}{2}\Big) - F_{6a}\Big(-i\nu+\frac{|n|}{2}\Big) - \frac{1}{12}\,\dE{3} -\frac{3}{8}\, N^2\,V E_{\nu,n}-\frac{1}{2}\,V^3 E_{\nu,n} \right. \\
&\left.\phantom{\sgn(n)}+\frac{1}{4}\,\Big(\psi_{-}+V\Big)\dE{2} +\zeta_2\,\dEOne+\zeta_3\,\psi_{-} \right\}+N \left\{\frac{1}{16}\,\left(N^2+12\,V^2\right)\psi_{-} + \zeta_2 V\right\} \,, \\
\tilde{F}_{7}\,=\,&  F_{7}\Big(i\nu+\frac{|n|}{2}\Big) - F_{7}\Big(-i\nu+\frac{|n|}{2}\Big) -\frac{1}{2}\,\tilde{F}_{6a}+\frac{1}{2}\,V \tilde{F}_4-\Big[\frac{1}{8}\,(\psi_{-})^2 - \frac{1}{4}\,\psi'_{+}+\frac{1}{2}\,\zeta_2\Big]\dEOne \\
&+\Big[\frac{1}{2}\,\tilde{F}_4+\frac{1}{16}\,N^2\,E_{\nu,n}+\frac{1}{4}\,V^2\,E_{\nu,n}-\frac{1}{4}\,V\,\dEOne + \frac{1}{8}\,\dE{2}-\zeta_3\Big]\psi_{-} + 5\,V\zeta_3 \\
&+\sgn(n) N \left\{-\frac{1}{8}\,V\,E_{\nu,n}^2-\frac{1}{2}\,V^3-\frac{3}{32}\,VN^2-\Big[\frac{1}{8}\,(\psi_{-})^2-\frac{1}{4}\,\psi'_{+}+\frac{1}{2}\,\zeta_2\Big]\,V \right\} \,,
\esp\eeq
where
\beq\bsp
\psi_{-}&\equiv\psi\Big(1+i\nu+\frac{|n|}{2}\Big)-\psi\Big(1-i\nu+\frac{|n|}{2}\Big),\\
\psi'_{+}&\equiv\psi'\Big(1+i\nu+\frac{|n|}{2}\Big)+\psi'\Big(1-i\nu+\frac{|n|}{2}\Big)\, .
\esp\eeq


\subsection{The basis in $(\nu,n)$ space in terms of single-valued HPLs}
In this appendix we present the analytic expressions for the basis of $\mathbb{Z}_2\times\mathbb{Z}_2$ eigenfunctions in $(\nu,n)$ space in terms of single-valued HPLs in $(w,\ws)$ space up to weight five. The $\mathbb{Z}_2\times\mathbb{Z}_2$ acts on $(w,\ws)$ space via conjugation and inversion, while it acts on $(\nu,n)$ space via $[n\leftrightarrow-n]$ and $[\nu\leftrightarrow-\nu, n\leftrightarrow-n]$.  The eigenvalue under $\mathbb{Z}_2\times\mathbb{Z}_2$ in $(w,\ws)$ space will be referred to as \emph{parity}.

\paragraph{Basis of weight 1 with parity $(+,+)$:} 
\beq\bsp
\mathcal{I}\left[ 1\right] &\, = 2\,\LSPA{1}\,.
\esp\eeq


\paragraph{Basis of weight 1 with parity $(+,-)$:} 

\beq\bsp
\mathcal{I}\left[ \delta_{0,n}\right] &\, = \LSMA{0}\,.
\esp\eeq


\paragraph{Basis of weight 2 with parity $(+,+)$:} 

\begin{eqnarray}
\mathcal{I}\left[ \EnunOne\right] &=& \LPA{1}{2}-\frac{1}{4}\,\LMA{0}{2}\,,\\
\mathcal{I}\left[\delta_{0,n}/(i\nu)\right] & =& \frac{1}{2}\,\LMA{0}{2}\,.
\end{eqnarray}


\paragraph{Basis of weight 2 with parity $(+,-)$:} 

\beq\bsp
\mathcal{I}\left[ V\right] &\, = -\LSMA{0}\,\LSPA{1}\,.
\esp\eeq


\paragraph{Basis of weight 2 with parity $(-,-)$:} 

\beq\bsp
\mathcal{I}\left[ N\right] &\, = 4\,\LSMA{2}\,.
\esp\eeq


\paragraph{Basis of weight 3 with parity $(+,+)$:} 
\begin{eqnarray}
\mathcal{I}\left[ \Enun{2}\right] &=& \frac{2}{3}\,\LPA{1}{3}-\LSPA{3}\,,\\
\mathcal{I}\left[ N^2\right] &=& 12\,\LSPA{3}-2\,\LSPA{1}\,\LMA{0}{2}\,,\\
\mathcal{I}\left[ V^2\right] &=& \frac{1}{2}\,\LSPA{1}\,\LMA{0}{2}-\LSPA{3}\,.
\end{eqnarray}

\paragraph{Basis of weight 3 with parity $(+,-)$:} 
\begin{eqnarray}
\mathcal{I}\left[ V\,\EnunOne\right] & =& \frac{1}{6}\,\LMA{0}{3}-2\,\LSMB{2}{1}\,,\\
\mathcal{I}\left[ \dEOne\right] & =& -\frac{1}{12}\,\LMA{0}{3}-\LSMA{0}\,\LPA{1}{2}+4\,\LSMB{2}{1}\,,\\
\mathcal{I}\left[ \kd{2}\right] & =& \frac{1}{6}\,\LMA{0}{3}\,.
\end{eqnarray}

\paragraph{Basis of weight 3 with parity $(-,+)$:} 

\beq\bsp
\mathcal{I}\left[ N\,V\right] = -\LSMA{2}\,\LSMA{0}\,.
\esp\eeq

\paragraph{Basis of weight 3 with parity $(-,-)$:} 
\beq\bsp
\mathcal{I}\left[N\,\EnunOne\right] &\, = 2\,\LSMA{2}\,\LSPA{1}\,.
\esp\eeq


\paragraph{Basis of weight 4 with parity $(+,+)$:} 
\begin{eqnarray}
\mathcal{I}\left[ \Enun{3}\right] & = &\frac{1}{2}\,\LMA{2}{2}+\frac{1}{2}\,\LMA{0}{2}\,\LPA{1}{2}+\frac{7}{96}\,\LMA{0}{4}+\frac{1}{2}\,\LPA{1}{4}-\frac{3}{2}\,\LSMA{0}\,\LSMB{2}{1}\\
&&\null\nonumber-\frac{5}{2}\,\LSPA{1}\,\LSPA{3}-3\,\LSPA{1}\,\zeta_3\,,\\
\mathcal{I}\left[N^2\,\EnunOne\right] & =& \frac{1}{12}\,\LMA{0}{4}+2\,\LMA{2}{2}-2\,\LSMA{0}\,\LSMB{2}{1}+2\,\LSPA{1}\,\LSPA{3}-4\,\LSPA{1}\,\zeta_3\,,\\
\mathcal{I}\left[V^2\,\EnunOne\right] & =& -\frac{1}{2}\,\LMA{2}{2}-\frac{1}{4}\,\LMA{0}{2}\,\LPA{1}{2}-\frac{1}{12}\,\LMA{0}{4}+\frac{3}{2}\,\LSMA{0}\,\LSMB{2}{1}+\frac{1}{2}\,\LSPA{1}\,\LSPA{3}\\
&&\null\nonumber-\LSPA{1}\,\zeta_3\,,\\
\mathcal{I}\left[V\,\dEOne\right] & = &\frac{3}{4}\,\LMA{0}{2}\,\LPA{1}{2}+\frac{1}{16}\,\LMA{0}{4}+\LMA{2}{2}-2\,\LSMA{0}\,\LSMB{2}{1}-2\,\LSPA{1}\,\LSPA{3}+4\,\LSPA{1}\,\zeta_3\,,~~\\
\mathcal{I}\left[ \dE{2}\right] & =& -\frac{1}{2}\,\LMA{0}{2}\,\LPA{1}{2}-\frac{1}{24}\,\LMA{0}{4}-2\,\LMA{2}{2}+4\,\LSPA{1}\,\LSPA{3}-8\,\LSPA{1}\,\zeta_3\,,\\
\mathcal{I}\left[ \kd{3}\right] & =& \frac{1}{24}\,\LMA{0}{4}\,.
\end{eqnarray}

\paragraph{Basis of weight 4 with parity $(+,-)$:} 
\begin{eqnarray}
\mathcal{I}\left[ V\,\Enun{2}\right] & =& \frac{1}{8}\,\LSPA{1}\,\LMA{0}{3}+\frac{1}{6}\,\LSMA{0}\,\LPA{1}{3}-\LSMA{0}\,\zeta_3-2\,\LSMB{2}{1}\,\LSPA{1}\,,\\
\mathcal{I}\left[ N^2\,V\right] & =& \frac{1}{3}\,\LSPA{1}\,\LMA{0}{3}-2\,\LSMA{0}\,\LSPA{3}\,,\\
\mathcal{I}\left[ V^3\right] & =& \frac{1}{2}\,\LSMA{0}\,\LSPA{3}-\frac{1}{6}\,\LSPA{1}\,\LMA{0}{3}\,,\\
\mathcal{I}\left[\EnunOne\,\dEOne\right] & =& -\frac{1}{8}\,\LSPA{1}\,\LMA{0}{3}-\frac{1}{2}\,\LSMA{0}\,\LPA{1}{3}+\frac{1}{2}\,\LSMA{0}\,\LSPA{3}+\LSMA{0}\,\zeta_3+2\,\LSMB{2}{1}\,\LSPA{1}\,,
\end{eqnarray}

\paragraph{Basis of weight 4 with parity $(-,+)$:} 

\begin{eqnarray}
\mathcal{I}\left[N\,V\,\EnunOne\right] & =& -2\,\LSPB{3}{1}\,,\\
\mathcal{I}\left[N\,\dEOne\right] & =& 8\,\LSPB{3}{1}-2\,\LSMA{2}\,\LSMA{0}\,\LSPA{1}\,.
\end{eqnarray}

\paragraph{Basis of weight 4 with parity $(-,-)$:}

\begin{eqnarray}
\mathcal{I}\left[ \Ffourtilde\right] & =& -\frac{1}{4}\,\LSMA{2}\,\LMA{0}{2}+\LSMA{2}\,\LPA{1}{2}+4\,\LSMA{4}-6\,\LSMC{2}{1}{1}\,,\\
\mathcal{I}\left[ N\,\Enun{2}\right] & =& \frac{1}{2}\,\LSMA{2}\,\LMA{0}{2}-6\,\LSMA{4}+8\,\LSMC{2}{1}{1}\,,\\
\mathcal{I}\left[ N^3\right] & =& 40\,\LSMA{4}-6\,\LSMA{2}\,\LMA{0}{2}\,,\\
\mathcal{I}\left[ N\,V^2\right] & =& \frac{1}{2}\,\LSMA{2}\,\LMA{0}{2}-2\,\LSMA{4}\,.
\end{eqnarray}


\paragraph{Basis of weight 5 with parity $(+,+)$:} 
\begin{eqnarray}
\mathcal{I}\left[\Enun{4}\right] &=& \frac{17}{96}\,\LSPA{1}\,\LMA{0}{4}-\frac{5}{4}\,\LSPA{3}\,\LMA{0}{2}+\frac{2}{5}\,\LPA{1}{5}+\frac{43}{4}\,\LSPA{5}+\LMA{0}{2}\,\LPA{1}{3}+4\,\LMA{0}{2}\,\zeta_3\\
&&\null\nonumber-4\,\LSPA{3}\,\LPA{1}{2}-8\,\LPA{1}{2}\,\zeta_3-4\,\LSMA{0}\,\LSMB{2}{1}\,\LSPA{1}+12\,\LSPC{3}{1}{1}+8\,\LSPC{2}{2}{1}\,,\\
\mathcal{I}\left[N^2\,\Enun{2}\right] &=& \frac{1}{3}\,\LMA{0}{2}\,\LPA{1}{3}-\frac{1}{24}\,\LSPA{1}\,\LMA{0}{4}+4\,\LSPA{1}\,\LMA{2}{2}+3\,\LSPA{3}\,\LMA{0}{2}-8\,\LMA{0}{2}\,\zeta_3\\
&&\null\nonumber-25\,\LSPA{5}-24\,\LSPC{3}{1}{1}-16\,\LSPC{2}{2}{1}\,,\\
\mathcal{I}\left[N^4\right] &=& \frac{13}{6}\,\LSPA{1}\,\LMA{0}{4}-20\,\LSPA{3}\,\LMA{0}{2}+140\,\LSPA{5}\,,\\
\mathcal{I}\left[V^2\,\Enun{2}\right] &=& -\frac{1}{12}\,\LMA{0}{2}\,\LPA{1}{3}-\frac{13}{96}\,\LSPA{1}\,\LMA{0}{4}+\frac{1}{4}\,\LSPA{3}\,\LMA{0}{2}-\frac{1}{4}\,\LSPA{5}-\LSPA{1}\,\LMA{2}{2}\\
&&\null\nonumber+2\,\LMA{0}{2}\,\zeta_3+10\,\LSPC{3}{1}{1}+4\,\LSPC{2}{2}{1}-4\,\zeta_5\,,\\
\mathcal{I}\left[N^2\,V^2\right] &=& -\frac{1}{8}\,\LSPA{1}\,\LMA{0}{4}+\LSPA{3}\,\LMA{0}{2}-5\,\LSPA{5}\,,\\
\mathcal{I}\left[V^4\right] &=& \frac{5}{96}\,\LSPA{1}\,\LMA{0}{4}-\frac{1}{4}\,\LSPA{3}\,\LMA{0}{2}+\frac{3}{4}\,\LSPA{5}\,,\\
\mathcal{I}\left[V\,\EnunOne\,\dEOne\right]&=& \frac{7}{48}\,\LSPA{1}\,\LMA{0}{4}-\frac{3}{4}\,\LSPA{3}\,\LMA{0}{2}-\frac{3}{2}\,\LMA{0}{2}\,\zeta_3+\frac{7}{2}\,\LSPA{5}+\LSPA{1}\,\LMA{2}{2}+\LSMA{0}\,\LSMB{2}{1}\,\LSPA{1}\\
&&\null\nonumber-12\,\LSPC{3}{1}{1}-4\,\LSPC{2}{2}{1}+6\,\zeta_5\,,\\
\mathcal{I}\left[\dEPOne{2}\right] &=& \frac{3}{2}\,\LMA{0}{2}\,\LPA{1}{3}-\frac{1}{3}\,\LSPA{1}\,\LMA{0}{4}-2\,\LSPA{1}\,\LMA{2}{2}+2\,\LSPA{3}\,\LMA{0}{2}+2\,\LMA{0}{2}\,\zeta_3\\
&&\null\nonumber-4\,\LSPA{3}\,\LPA{1}{2}+8\,\LPA{1}{2}\,\zeta_3-8\,\LSMA{0}\,\LSMB{2}{1}\,\LSPA{1}-9\,\LSPA{5}+48\,\LSPC{3}{1}{1}+16\,\LSPC{2}{2}{1}-24\,\zeta_5\,,~~\\
\mathcal{I}\left[\EnunOne\,\dE{2}\right] &=& \frac{1}{6}\,\LSPA{1}\,\LMA{0}{4}-\LMA{0}{2}\,\LPA{1}{3}-\LSPA{3}\,\LMA{0}{2}+4\,\LSPA{3}\,\LPA{1}{2}-8\,\LPA{1}{2}\,\zeta_3\\
&&\null\nonumber+4\,\LSMA{0}\,\LSMB{2}{1}\,\LSPA{1}+2\,\LSPA{5}-24\,\LSPC{3}{1}{1}-8\,\LSPC{2}{2}{1}+12\,\zeta_5 \,,\\
\label{eq:NF4tilde}
\mathcal{I}\left[N\,\Ffourtilde\right] &=& \
\frac{1}{12}\,\LSPA{1}\,\LMA{0}{4}-\frac{7}{4}\,\LSPA{3}\,\LMA{0}{2}+\
\frac{7}{2}\,\LMA{0}{2}\,\zeta_3-\LSPA{1}\,\LMA{2}{2}-\LSMA{0}\,\LSMB{2}{1}\,\LSPA{1}\\
&&\null\nonumber+15\,\LSPA{5}+12\,\LSPC{3}{1}{1}+8\,\LSPC{2}{2}{1}\,.
\end{eqnarray}


\paragraph{Basis of weight 5 with parity $(+,-)$:} 
\begin{eqnarray}
\mathcal{I}\left[\Fseventilde\right] &=& \frac{5}{8}\,\LSMA{0}\,\LMA{2}{2}-\frac{11}{48}\,\LMA{0}{3}\,\LPA{1}{2}+\frac{1}{4}\,\LSMB{2}{1}\,\LMA{0}{2}+\frac{59}{3840}\,\LMA{0}{5}\\
&&\null\nonumber+\frac{5}{48}\,\LSMA{0}\,\LPA{1}{4}+\frac{3}{2}\,\LSMA{0}\,\LSPA{1}\,\LSPA{3}-\frac{7}{2}\,\LSMB{3}{2}-\LSMB{2}{1}\,\LPA{1}{2}-8\,\LSMA{0}\,\LSPA{1}\,\zeta_3\\
&&\null\nonumber-10\,\LSMB{4}{1}+7\,\LSMD{2}{1}{1}{1}\,,\\
\mathcal{I}\left[V\,\Enun{3}\right] &=& \frac{1}{2}\,\LSMA{0}\,\LMA{2}{2}+\frac{3}{16}\,\LMA{0}{3}\,\LPA{1}{2}+\frac{3}{4}\,\LSMB{2}{1}\,\LMA{0}{2}-\frac{1}{192}\,\LMA{0}{5}\\
&&\null\nonumber-\frac{1}{4}\,\LSMA{0}\,\LSPA{1}\,\LSPA{3}+\frac{9}{2}\,\LSMA{0}\,\LSPA{1}\,\zeta_3-\frac{9}{2}\,\LSMB{3}{2}-6\,\LSMB{4}{1}-12\,\LSMD{2}{1}{1}{1}\,,
\end{eqnarray}
\begin{eqnarray}
\mathcal{I}\left[N^2\,V\,\EnunOne\right] &=& -\frac{1}{4}\,\LMA{0}{3}\,\LPA{1}{2}-\frac{1}{48}\,\LMA{0}{5}+\LSMB{2}{1}\,\LMA{0}{2}+\LSMA{0}\,\LSPA{1}\,\LSPA{3}-2\,\LSMA{0}\,\LSPA{1}\,\zeta_3\\
&&\null\nonumber-8\,\LSMB{4}{1}-2\,\LSMB{3}{2}\,,\\
\mathcal{I}\left[V^3\,\EnunOne\right] &=& \frac{3}{16}\,\LMA{0}{3}\,\LPA{1}{2}-\frac{3}{4}\,\LSMB{2}{1}\,\LMA{0}{2}+\frac{23}{960}\,\LMA{0}{5}-\frac{3}{4}\,\LSMA{0}\,\LSPA{1}\,\LSPA{3}\\
&&\null\nonumber+\frac{3}{2}\,\LSMA{0}\,\LSPA{1}\,\zeta_3+\frac{3}{2}\,\LSMB{3}{2}+4\,\LSMB{4}{1}\,,\\
\mathcal{I}\left[\Enun{2}\,\dEOne\right] &=& -\frac{1}{2}\,\LSMA{0}\,\LMA{2}{2}-\frac{7}{24}\,\LMA{0}{3}\,\LPA{1}{2}-\frac{1}{48}\,\LMA{0}{5}-\frac{1}{6}\,\LSMA{0}\,\LPA{1}{4}+\LSMA{0}\,\LSPA{1}\,\LSPA{3}~~~~~~\\
&&\null\nonumber-2\,\LSMA{0}\,\LSPA{1}\,\zeta_3+4\,\LSMB{4}{1}+3\,\LSMB{3}{2}+8\,\LSMD{2}{1}{1}{1}\,,\\
\mathcal{I}\left[N^2\,\dEOne\right] &=& \frac{3}{2}\,\LMA{0}{3}\,\LPA{1}{2}+\frac{1}{24}\,\LMA{0}{5}-2\,\LSMA{0}\,\LMA{2}{2}-4\,\LSMB{2}{1}\,\LMA{0}{2}-8\,\LSMA{0}\,\LSPA{1}\,\LSPA{3}\\
&&\null\nonumber+16\,\LSMA{0}\,\LSPA{1}\,\zeta_3+48\,\LSMB{4}{1}+12\,\LSMB{3}{2}\,,\\
\mathcal{I}\left[V^2\,\dEOne\right] &=& \frac{1}{2}\,\LSMA{0}\,\LMA{2}{2}-\frac{3}{8}\,\LMA{0}{3}\,\LPA{1}{2}-\frac{1}{480}\,\LMA{0}{5}+\LSMB{2}{1}\,\LMA{0}{2}+2\,\LSMA{0}\,\LSPA{1}\,\LSPA{3}\\&&\null\nonumber-4\,\LSMA{0}\,\LSPA{1}\,\zeta_3-12\,\LSMB{4}{1}-5\,\LSMB{3}{2}\,,\\
\mathcal{I}\left[V\,\dE{2}\right] &=& -\frac{1}{15}\,\LMA{0}{5}-2\,\LSMA{0}\,\LMA{2}{2}-2\,\LSMA{0}\,\LSPA{1}\,\LSPA{3}+4\,\LSMA{0}\,\LSPA{1}\,\zeta_3+24\,\LSMB{4}{1}\\
&&\null\nonumber+12\,\LSMB{3}{2}\,,\\
\mathcal{I}\left[\dE{3}\right] &=& \frac{1}{2}\,\LMA{0}{3}\,\LPA{1}{2}+\frac{7}{40}\,\LMA{0}{5}+6\,\LSMA{0}\,\LMA{2}{2}-48\,\LSMB{4}{1}-24\,\LSMB{3}{2}\,,\\
\mathcal{I}\left[\kd{4}\right] &=& \frac{1}{120}\,\LMA{0}{5}\,.
\end{eqnarray}


\paragraph{Basis of weight 5 with parity $(-,+)$:} 
\begin{eqnarray}
\mathcal{I}\left[\Fsixatilde\right] &=&
\frac{1}{12}\,\LSMA{2}\,\LMA{0}{3}-\LSMA{4}\,\LSMA{0}+\LSMA{2}\,\LSMB{2}{1}-\LSPA{1}\,\LSPB{3}{1}\,,\\
\mathcal{I}\left[V\,\Ffourtilde\right] &=&
\frac{1}{48}\,\LSMA{2}\,\LMA{0}{3}-\frac{1}{2}\,\LSMA{2}\,\LSMA{0}\,\LPA{1}{2}-\frac{3}{4}\,\LSMA{4}\,\LSMA{0}-\LSMA{2}\,\LSMB{2}{1}+3\,\LSMA{0}\,\LSMC{2}{1}{1}~~~~~~\\
&&\null\nonumber+\LSPA{1}\,\LSPB{3}{1}\,,\\
\mathcal{I}\left[N\,V\,\Enun{2}\right] &=&
-\frac{1}{48}\,\LSMA{2}\,\LMA{0}{3}+\frac{1}{2}\,\LSMA{2}\,\LSMA{0}\,\LPA{1}{2}+\frac{3}{4}\,\LSMA{4}\,\LSMA{0}-2\,\LSMA{0}\,\LSMC{2}{1}{1}\\
&&\null\nonumber-2\,\LSPA{1}\,\LSPB{3}{1}\,,\\
\mathcal{I}\left[N^3\,V\right] &=& \frac{3}{4}\,\LSMA{2}\,\LMA{0}{3}-5\,\LSMA{4}\,\LSMA{0}\,,\\
\mathcal{I}\left[N\,V^3\right] &=&
\frac{3}{4}\,\LSMA{4}\,\LSMA{0}-\frac{7}{48}\,\LSMA{2}\,\LMA{0}{3}\,,\\
\mathcal{I}\left[N\,\EnunOne\,\dEOne\right] &=&
-\frac{5}{24}\,\LSMA{2}\,\LMA{0}{3}+\frac{3}{2}\,\LSMA{4}\,\LSMA{0}-\LSMA{2}\,\LSMA{0}\,\LPA{1}{2}+4\,\LSPA{1}\,\LSPB{3}{1}\,.
\end{eqnarray}


\paragraph{Basis of weight 5 with parity $(-,-)$:} 
\begin{eqnarray}
\mathcal{I}\left[N\,\Enun{3}\right] &=& \frac{5}{8}\,\LSMA{2}\,\LSPA{1}\,\LMA{0}{2}-\frac{15}{2}\,\LSMA{4}\,\LSPA{1}-\frac{1}{2}\,\LSMA{2}\,\LSPA{3}-\LSMA{2}\,\LPA{1}{3}+12\,\LSMC{2}{1}{1}\,\LSPA{1}\,,~~~~~\\
\mathcal{I}\left[\EnunOne\,N^3\right] &=& -\frac{1}{2}\,\LSMA{2}\,\LSPA{1}\,\LMA{0}{2}+2\,\LSMA{4}\,\LSPA{1}+6\,\LSMA{2}\,\LSPA{3}-16\,\LSMA{2}\,\zeta_3-4\,\LSMA{0}\,\LSPB{3}{1}\,,\\
\mathcal{I}\left[\EnunOne\,N\,V^2\right] &=& -\frac{1}{8}\,\LSMA{2}\,\LSPA{1}\,\LMA{0}{2}+\frac{1}{2}\,\LSMA{4}\,\LSPA{1}-\frac{1}{2}\,\LSMA{2}\,\LSPA{3}+\LSMA{0}\,\LSPB{3}{1}\,,\\
\mathcal{I}\left[N\,V\,\dEOne\right] &=& \frac{3}{4}\,\LSMA{2}\,\LSPA{1}\,\LMA{0}{2}-3\,\LSMA{4}\,\LSPA{1}+\LSMA{2}\,\LSPA{3}+4\,\LSMA{2}\,\zeta_3-2\,\LSMA{0}\,\LSPB{3}{1}\,,\\
\mathcal{I}\left[N\,\dE{2} \right]&=& -\LSMA{2}\,\LSPA{1}\,\LMA{0}{2}+12\,\LSMA{4}\,\LSPA{1}-4\,\LSMA{2}\,\LSPA{3}-16\,\LSMA{2}\,\zeta_3\,,\\
\mathcal{I}\left[\EnunOne\,\Ffourtilde\right] &=& \frac{1}{8}\,\LSMA{2}\,\LSPA{1}\,\LMA{0}{2}+\frac{2}{3}\,\LSMA{2}\,\LPA{1}{3}+\frac{1}{2}\,\LSMA{4}\,\LSPA{1}+\frac{1}{2}\,\LSMA{2}\,\LSPA{3}-\frac{1}{2}\,\LSMA{0}\,\LSPB{3}{1}\\
&&\null\nonumber-2\,\LSMA{2}\,\zeta_3-4\,\LSMC{2}{1}{1}\,\LSPA{1}\,.
\end{eqnarray}


\end{document}